\documentclass[acmsmall]{acmart}
\renewcommand\footnotetextcopyrightpermission[1]{}
\acmJournal{PACMPL}
\acmVolume{1}
\acmNumber{CONF} 
\acmArticle{1}
\acmYear{2018}
\acmMonth{1}
\acmDOI{} 
\startPage{1}

\setcopyright{none}

\usepackage{booktabs}   
\usepackage{subcaption} 

\usepackage[normalem]{ulem}

\usepackage{amsmath}
\usepackage{listings}
\usepackage{mathpartir}
\usepackage{syntax}
\usepackage{amsthm}
\usepackage{amsfonts}
\usepackage{stmaryrd}
\usepackage{xspace}
\usepackage{flushend}
\usepackage{placeins}
\usepackage{algpseudocode}
\usepackage[noabbrev,capitalize,nameinlink]{cleveref}
\usepackage{wrapfig}

\newcommand{\code}[1]{\texttt{#1}}

\newcommand{\TOOL}{Leto\xspace}

\newenvironment{wraplst}
               {\hspace*{1.8em}\begin{minipage}{0.94\linewidth}}
               {\end{minipage}}

\newenvironment{wrappedwraplst}
               {\begin{flushright}\begin{minipage}{0.9\linewidth}}
               {\end{minipage}\end{flushright}}

\newenvironment{wraplst*}
               {\hspace*{1.4em}\begin{minipage}{0.953\linewidth}}
               {\end{minipage}}

\newenvironment{wrappedwraplst*}
               {\begin{flushright}\begin{minipage}{0.91\linewidth}}
               {\end{minipage}\end{flushright}}

\newcommand{\algoref}{sao2013self,hoemmen2011fault,oboril2011numerical,ABFT,shantharam2012fault,bronevetsky2008soft,roy1994algorithm,roy1996algorithm,sao2016self}

\newcommand{\softref}{li2016processor,reisSwift05,santini2017effectiveness,wei2012blockwatch,yim2011hauberk}

\newcommand{\misc}{han2013approximate,sullivan2012truncated,sullivan2013truncated}

\newcommand{\concern}{saggese2005microprocessor,shivakumar2002modeling,mitra2005robust,rajaraman2006seat,omana2003model,zhang2006soft,mitra2006combinational,buchner1997comparison,wang2011soft,rao2007computing,johnston2000scaling,ando20031,haque2010hard,gomez2014gpgpus,gu2004error,mukherjee2003systematic,wang2004characterizing,weaver2004techniques,amarasinghe09,chen2012efficient,yim2014characterization,kurd2010westmere,siewiorek2017reliable,malaiya1981reliability,malaiya1979survey,gschwind2011softbeam,reinhardt2000transient,gomaa2003transient}

\newcommand{\worstTime}{158.77\xspace}

\begin{document}

\title[]{Verifying Programs Under Custom Application-Specific Execution Models}

\authorsaddresses{}

\author{Brett Boston}
\affiliation{
  \institution{Massachusetts Institute of Technology}
}

\author{Zoe Gong}
\affiliation{
  \institution{Massachusetts Institute of Technology}
}

\author{Michael Carbin}
\affiliation{
  \institution{Massachusetts Institute of Technology}
}

\date{}

\begin{abstract}

Researchers have recently designed a number of application-specific fault tolerance mechanisms that enable applications to either be naturally resilient to errors or include additional detection and correction steps that can bring the overall execution of an application back into an envelope for which an acceptable execution is eventually guaranteed. A major challenge to building an application that leverages these mechanisms, however, is to verify that the implementation satisfies the basic invariants that these mechanisms require---given a model of how faults may manifest during the application's execution.

To this end we present \TOOL, an SMT based automatic verification system that enables developers to verify their applications with respect to a {\em first-class} execution model specification. Namely, \TOOL enables software and platform developers to programmatically specify the execution semantics of the underlying hardware system as well as verify assertions about the behavior of the application's resulting execution.  In this paper, we present the {\TOOL} programming language and its corresponding verification system. We also demonstrate {\TOOL} on several applications that leverage application-specific fault tolerance mechanisms.

\end{abstract}

\maketitle
\thispagestyle{plain}

\vspace{-0.5ex}
\section{Introduction}
\label{sec:introduction}

Due to the aggressive scaling of technology sizes in modern
computer processor fabrication, modern processors have become more vulnerable
to errors that result from natural variations in processor manufacturing,
natural variations in transistor reliability as processors physically age over
time, and natural variations in these processors' operating environments (e.g.,
temperature variation and cosmic/environmental
radiation)~\cite{Borkar05,mitra2005robust, shivakumar2002modeling,
kurd2010westmere,mitra2006combinational,mukherjee2003systematic,
yim2014characterization, johnston2000scaling, amarasinghe09}.

\vspace{-.15cm}
\paragraph{\bf Challenges.} These systems encounter {\em faults}---anomalies
in the underlying physical device---that produce {\em errors}---unanticipated
or incorrect values that are visible to the program. Simple fault models
include bit flips in the output of arithmetic, logical, and memory operations.
These faults can be {\em transient}---occurring nondeterministically with
the device eventually returning to correct behavior---or
{\em permanent}---with the device never returning to correct behavior.
A key challenge for building applications for these platforms is that reasoning
about the reliability of these programs requires reasoning about the
operation of the underlying execution model and its impact on the application's
behavior.

\vspace{-1ex}
\subsection{Application-Specific Fault Tolerance}

In response to this increased error vulnerability, researchers have begun to
expand on historical results for algorithm-based fault tolerance
~\cite{\algoref}, alternatively {\em application-specific fault tolerance}, to
identify new opportunities for low-overhead mechanisms that can steer an
application's execution to produce {\em acceptable} results: results that are
within some tolerance of the result expected from a fully reliable execution.
For example, application-specific fault tolerance techniques for linear algebra
produce lightweight checksums that developers can use to validate that the
computation produced the correct results.  For some applications, these
checksums are exact, enabling the exact error detection capabilities of
dual-modular redundancy but with lower overhead.  However, for other
applications, these checksums either are not known to exist or, at best,
compromise on their error coverage.

Other techniques include selective $n$-modular redundancy in which a developer
either manually or with the support of a dynamic fault-injection
tool identifies instructions or regions of code that do not need to be
protected for the application to produce an acceptable result---as determined
by an empirical evaluation~\cite{carbin10, approxilyzer, thomas16, vulfi16}.
Another class of techniques is fault-tolerant algorithms that through the
addition of algorithm-specific checking and correction code are tolerant to
faults~\cite{hoemmen2011fault, sao2013self, du2012, sao2016self}.

\vspace{-.15cm}
\paragraph{\bf Challenges.} A major barrier to implementing these techniques is
that their results either rely on empirical guarantees or---for
self-stabilizing algorithms---hinge on the assumption that the fault model of
the underlying computing substrate matches the assumptions of the algorithmic
formalization.

\subsection{Verified Application-Specific Fault Tolerance}

To address these challenges we present {\TOOL}, an SMT based, automatic
verification system that supports reasoning about unreliably executed programs.
{\TOOL} enables a
developer to verify their application-specific fault tolerance
mechanism by providing tools to 1) programmatically specify the behavior
of the computing substrate's fault model and 2) verify
{\em relational} assertions that relate the behavior of the unreliably executed
program to that of the reliable execution.   Specifically, {\TOOL} automatically
weaves the behavior of the underlying hardware system---as given by a
specification---into the execution of the main program. In addition, {\TOOL}'s
program logic enables a developer to specify relational assertions that, for
example, constrain the difference in results of the unreliable execution of the
program from that of its reliable execution.

\subsubsection{First-Class Execution Models}

{\TOOL} permits developers to programmatically specify a stateful execution
model.
\cref{fig:seu} presents a Leto specification for a first-class {\em
single-event upset} (SEU) execution model with specifications for
multiplication.
An SEU model is a common {\em fault model} that application developers in the
area of fault tolerance use to model the execution behavior of an application
such that they can provide a variety of fault tolerance
mechanisms~\cite{chen2008error, yim2010measurement}.
The underlying assumption is that faults in a
system (e.g., those due to cosmic radiation) occur with a probability such that
at most one fault will occur during the execution of an application.

\begin{wrapfigure}[10]{l}{0.43\linewidth}
\begin{wrappedwraplst*}
  \vspace{-3ex}
\begin{lstlisting}
bool upset = false; ~\label{seu:upset}~

operator *(real x1, real x2) ~\label{seu:mult1}~
   ensures (result == x1 * x2);

operator *(real x1, real x2) ~\label{seu:mult2}~
   when (!upset)
   modifies (upset)
   ensures (upset);
\end{lstlisting}
\end{wrappedwraplst*}
\vspace{-.22cm}
\caption{Unbounded SEU Execution Model}
\label{fig:seu}
\vspace{-.22cm}
\end{wrapfigure}

The model exports two versions of the multiplication operator.
\cref{seu:mult1} specifies the standard reliable implementation of
multiplication. The model denotes this fact with its \texttt{ensures} clause
which asserts what must be true of the model's state and outputs after
execution of the operation. This operation specifically constrains the value of
\texttt{result}---which represents the result of the operation---to equal
\(\code{x1 * x2}\) where \code{*} has standard multiplication semantics.

\cref{seu:mult2} specifies an unreliable implementation of the multiplication
operator.
This implementation does not place any constraints on \code{result} and
therefore permits unbounded errors.

\vspace{-.15cm}
\paragraph{\bf Stateful.} Because the model is an SEU model it
must track whether a fault has occurred so that the model exposes at most
one fault to the application.  This model is therefore \textit{stateful} and to
achieve this semantics, the model includes a boolean valued state
variable---\texttt{upset} (\cref{seu:upset})---that records whether or not a
fault has already occurred during the execution of the program. The model
additionally predicates the availability of its unreliable operations by a {\em
guard}.  Specifically, an operation's guard is the optionally-specified boolean
expression that occurs after the \texttt{when} keyword.  The \texttt{when}
clause for the unreliable version requires \texttt{!upset} indicating that the
unreliable version is only enabled if a fault has yet to occur.
Leto models the dynamic execution of the application such that the execution
exposes only enabled operation implementations at a given program point.

\subsubsection{Acceptability Properties.}

Leto enables developers to automatically verify the basic goal of an
application-specific fault tolerance mechanism: ensure that the resulting
application satisfies its {\em acceptability properties}~\cite{pldi12}, such as
its {\em safety} and {\em accuracy}.

\paragraph{\bf Safety Properties: } standard properties of the execution of
the application that must be true of a single execution of the application.
Such properties include, for example, memory safety and the assertion that the
application returns results that are within some range. For example, a
computation of a distance metric must---regardless of the accuracy
of its results---return a value that is non-negative. In {\TOOL}, a developer
specifies safety properties with the standard \texttt{assert} statements
typically seen in verification systems.

\paragraph{\bf Accuracy Properties: } properties of the unreliable or {\em
relaxed} execution of the application that relate its behavior and results to
that of a reliably executed version. Accuracy properties are {\em relational}
in that they relate values of the state of the program between its two semantic
interpretations. For example, the assertion \code{abs(x<o> - x<r>) <
epsilon} in {\TOOL} specifies that the difference in value of \texttt{x}
between the program's original, reliable execution (denoted by \code{x<o>}) and
relaxed execution (denoted by \code{x<r>}) is at most \code{epsilon}.

\paragraph{\bf Execution-Specific Properties.} Given a first-class execution
model, Leto also enables developers to refer to the state of the execution. For
example, in many self-stabilizing iterative algorithms, the proof of
convergence for the algorithm in the presence of faults requires reasoning
about three cases: 1) the portion of the execution in which no fault has
occurred, 2) the iteration on which a fault occurs (assuming an SEU model), and
3) the portion of the execution after the fault.
Leto enables developers to verify such properties by exposing the state of the
fault model into the program logic.

\paragraph{\bf Asymmetric  Relational Verification.} {\TOOL} provides and
implements an Asymmetric Relational Hoare Logic~\cite{pldi12} as its core
program logic.  An Asymmetric Relational Hoare Logic is a variant of the
standard Hoare Logic that natively refers to the values of variables between
two executions of the program.  {\TOOL}'s use of a relational program logic
serves two goals: 1) it gives a semantics to accuracy properties and 2) it
enables tractable verification of safety properties.  For example, proving the
memory safety of an application outright can be challenging for many
applications. However, application-specific fault tolerance mechanisms can
typically be designed and deployed such that it is possible to verify that for
any given array access or memory access, errors in the application do not {\em
interfere} with the accessed address. Such properties are typically easier to
verify for a protected application than verifying the safety of the memory
access outright.  {\TOOL} therefore enables developers to tractably verify a
strong {\em relative} safety guarantee: if the original application satisfies
all of the specified safety properties, then relaxed executions of the
application with its deployed application-specific fault tolerance mechanisms
also satisfy these safety properties.


\subsection{Contributions}

This paper presents the following contributions:

\vspace{-.15cm}
\paragraph{\bf First-Class Execution Models.} We present a language for
specifying execution models that provides stateful, input-dependent selection
of each operation's implementation.
We demonstrate numerous sample execution models.

\vspace{-.15cm}
\paragraph{\bf Execution Model Refinement}
We present a language construct for refining models into new models, permitting
developers to create submodels that satisfy the specification of their
supermodels.
We describe the process by which refinement creates an interface between
hardware vendors and software developers.

\vspace{-.15cm} \paragraph{\bf Programming Language and Semantics.}  We present
language constructs that enable the developers to specify assertions that
refer to the state of the execution model. These constructs enable a developer
to, for example, specify the precise properties that self-stabilizing
applications require to verify high-level convergence properties.

\vspace{-.15cm}
\paragraph{\bf Program Logic and Verification Algorithm.} {\TOOL}'s program
logic enables developers to lower the overhead of verifying a standard safety
property by enabling techniques that demonstrate the non-interference between the
application's faults and the validity of a property. {\TOOL}'s verification
algorithm additionally automates this process through the inclusion of loop
invariant inference .

\vspace{-.15cm} \paragraph{\bf Case Studies.} We evaluate {\TOOL} on several
self-correcting algorithms (Jacobi, Self-stabilizing Conjugate Gradient,
Self-stabilizing Steepest Descent, and Self-correcting Connected Components)
and demonstrate that it is possible to verify the key invariants required to prove
that these algorithms' self-stability guarantees hold for their
implementations. We consider execution models that capture a range of
substrates, including emerging hardware systems that bound potential error,
emerging hardware security vulnerabilities (Rowhammer~\cite{rowhammer}), as well as
standard fault modelling assumptions that expose unbounded errors to the
application.

{\TOOL}'s contributions enable developers to specify and verify the rich
properties seen in applications with application-specific fault tolerance
mechanisms.

\section{Example Execution Models}
\begin{wrapfigure}[11]{l}{0.55\linewidth}
  \vspace{-3ex}
\begin{wrappedwraplst}
\begin{lstlisting}
vector<real> product(uint N,
                     vector<real> x(N),
                     vector<real> y(N))
{
   vector<real> result(N);

   for (uint i = 0; i < N; ++i) ~\label{vecprod:loop}~
   {
      result[i] = x[i] * y[i];
   }
   return result;
}
\end{lstlisting}
\end{wrappedwraplst}
  \vspace{-3ex}
\caption{Vector Product}
\label{fig:vecprod}
\end{wrapfigure}

\cref{fig:vecprod} presents an implementation of a vector-vector product in
\TOOL.
It defines a function \code{product} that takes a size parameter \code{N}, a
vector \code{x} of size \code{N}, a vector \code{y} of size \code{N}, and
returns the element-wise product of \code{x} and \code{y}.

In \cref{sec:add-seu} through \cref{sec:multicycle-model} we introduce
several execution models.
We highlight the key features of \TOOL that these models use and demonstrate
them on variants of the vector product function.
We use this specific example because vector products are a component of many
numerical algorithms and exist in three of our four benchmarks as an
intermediate calculation in matrix-vector products.

In \cref{sec:refinement} we present a technique in \TOOL for refining loosely
defined models into stricter models.
This technique ensures that these refined submodels satisfy the specifications
of their respective supermodels.

\subsection{\bf Additive Single-Event Upset Execution Model}
\label{sec:add-seu}

The additive single-event upset execution model in \cref{fig:add-seu}
exports two versions of the multiplication operator.
\cref{aseu:reliable} specifies the standard reliable implementation of
multiplication. The model denotes this fact with its \texttt{ensures} clause
which asserts what must be true of the model's state and outputs after
execution of the operation. This operation specifically constrains the value of
\texttt{result}---which represents the result of the operation---to equal
\code{x1 * x2}.

\begin{wrapfigure}[11]{l}{0.52\linewidth}
\begin{wrappedwraplst}
\begin{lstlisting}
const real eps = ...; ~\label{aseu:err}~
bool upset = false; ~\label{aseu:upset}~

operator*(real x1, real x2) ~\label{aseu:reliable}~
   ensures (result == x1 * x2);

operator*(real x1, real x2) ~\label{aseu:unrel1}~
   when !upset
   modifies (upset)
   ensures upset &&
      x1*x2-eps < result < x1*x2+eps;
\end{lstlisting}
\end{wrappedwraplst}
\vspace{-2ex}
  \caption{Additive SEU Execution Model}
\label{fig:add-seu}
\end{wrapfigure}

\paragraph{\bf Bounded Error.}
\TOOL enables developers to place bounds on the results of binary operations by
using inequalities over the \code{result} variable.
In addition, \cref{aseu:unrel1} specifies an unreliable implementation for
multiplication.
The semantics of this unreliable operator guarantees that even in
the presence of an error, the result is within \code{eps} of the original
result where \code{eps} is a programmer specified constant in the model on
\cref{aseu:err}.

~\\

\vspace{-4ex}
\subsubsection{Vector Product Under Additive SEU}
\cref{fig:vecprod-seu} presents a vector product implementation
annotated to verify under the additive SEU execution model.
All of the relaxation in this implementation occurs in the loop
(\cref{vseu:loop}) where the execution model may corrupt the value of
\code{result[i]}.
On \cref{vseu:assert} the implementation uses a relational assertion
(\code{assert_r}) to verify that \code{model.eps} bounds the impact of these
errors (\cref{vseu:merr}).
The notation \code{model.eps} is a first-class reference to the state of the
execution model at that point in the program.

\begin{figure}[b]
\begin{wraplst}
  \begin{lstlisting}
property_r bounded_diff(vector<real> x, uint N) : ~\label{vseu:prop}~
   ~$\forall$~(uint i)((i < N<r>) -> (abs(x<o>[i] - x<r>[i]) < model.eps)); ~\label{vseu:merr}~

requires_r eq(N) && eq(x) && eq(y) ~\label{vseu:pre}~
vector<real> product(uint N, vector<real> x(N), vector<real> y(N))
{
   vector<real> result(N);

   for (uint i = 0; i < N; ++i) ~\label{vseu:loop}~
      invariant_r bounded_diff(result, i)
   {
      result[i] = x[i] *. y[i]; ~\label{vseu:mult}~
   }

   assert_r(bounded_diff(result, N)); ~\label{vseu:assert}~

   return result;
}
\end{lstlisting}
\end{wraplst}
  \vspace{-1ex}
  \caption{Vector Product Under Additive SEU}
  \label{fig:vecprod-seu}
  \vspace{-3ex}
\end{figure}

Verifying this vector product implementation relies on \TOOL's specification
capabilities to establish bounds on the error in the product.
{\TOOL}'s two features that diverge from traditional programming languages
are that developers can specify that some operations in the program may execute
with an alternative semantics and---as consequence---write {\em relational}
assertions that relate values between the standard, {\em original execution}
and the alternative {\em relaxed execution} of the program.  

\paragraph{\bf Relaxed Execution.} {\TOOL} exports custom operations by
enabling developers to specify that an operation may execute according to the
execution model specification (versus a standard implementation)  by appending
a dot to the operation as in the operation `\texttt{*.}' (\cref{vseu:mult}).

\paragraph{\bf Bounded Error}
Many self healing iterative algorithms for solving linear systems of equations
experience increases in solve times proportional to the magnitude of errors
they experience during execution.
For example, the Jacobi iterative method has the property that the change in
the number of iterations to converge after an error is bounded logarithmically
by the magnitude of the error.
Thus, by bounding the magnitude of errors, a developer using the Jacobi method
can derive a static bound on the maximum impact  errors can have on convergence
time.

To verify that this vector product implementation has bounded error the
developer uses the \code{assert\_r} statement on \cref{vseu:assert} to assert
\code{bounded\_diff(result, N)}.
\code{bounded\_diff} is a \emph{relational property} application.
A property is a hygienic macro to enable code reuse within loop invariants and
assertions.
We define the \code{bounded\_diff} property on \cref{vseu:prop}.
This property takes a vector \code{x} and a size \code{N} and enforces that for
every index \code{i}, \code{model.eps} bounds the error in \code{x[i]}.
\code{x<o>[i]~-~x<r>[i]} computes the error in \code{x} where \code{x<o>} refers
to the value of \code{x} in the standard, original execution of the program
and \code{x<r>} refers to the value of \code{x} in the relaxed
execution.
To facilitate the verification of this condition we also include it in the loop
invariant on \cref{vseu:loop}.

Another necessary component to verifying this assertion is that \code{N},
\code{x}, and \code{y} have the same values in both the original and relaxed
executions.
The implementation enforces this property through the relational function
precondition (\code{requires_r}) on \cref{vseu:pre}.
\TOOL expands terms of the form \code{eq(x)} to \code{x<o> == x<r>}.

\subsubsection{Verification Algorithm}
Leto provides an automated verification algorithm that performs {\em
relational} forward symbolic execution to discharge assertions in the program.
Namely, Leto traverses the program, building a logical characterization of the
state of the program at each point and verifies that the resulting logical
formula ensures that a given \texttt{assert} or \texttt{assert_r} statement is
valid.   This approach also works in concert with the
developer's specification annotations; these include both function
preconditions and loop invariants. Leto also provides support for automatic
loop invariant inference, which can lower the annotation burden of the
developer by automatically inferring additional loop invariants.
In the vector product under additive SEU program \TOOL infers \code{eq(N)},
\code{eq(x)}, \code{eq(y)} and \code{i <= N}, which is necessary to demonstrate
that all of the vector accesses are in bounds.

\subsection{Switchable Rowhammer Model}
\label{sec:rowhammer-model}

\begin{wrapfigure}[11]{l}{0.43\linewidth}
  \vspace{-3ex}
\begin{wrappedwraplst}
\begin{lstlisting}
const uint eps = ...;
bool reliable = false;

@region(unreliable) ~\label{be:reliable}~
write(uint dest, uint src)
   ensures (dest == src);

@region(unreliable) ~\label{be:relaxed}~
write(uint dest, uint src)
   when (!reliable)
   ensures (src + eps < dest);
\end{lstlisting}
\end{wrappedwraplst}
\vspace{-.2cm}
\caption{Switchable Rowhammer Model}
\label{fig:big-err}
\vspace{-.2cm}
\end{wrapfigure}

We present the switchable rowhammer execution model in \cref{fig:big-err}.
This execution model simulates a Rowhammer attack that allows an attacker to
selectively flip bits in DRAM by issuing frequent reads on DRAM
rows~\cite{rowhammer}.
 
Unlike the models we have presented thus far, this one enables developers to
model memory errors.
Additionally, it enables switchability, permitting the program to selectively
disable errors to emulate selective Rowhammer protection techniques
\cite{aweke2016anvil}.

\paragraph{\bf Memory Regions.}
In addition to binary operators, \TOOL permits the specification of read and
write behavior.
Read and write specifications contain an additional \code{@region} annotation
allowing developers to partition their program variables into multiple memory
regions with differing read and write characteristics.

Both operator specifications in \cref{fig:big-err} govern writes to variables
stored in the \code{unreliable} memory region.
When \TOOL encounters an expression of the form \code{v = e} where \code{v}
is in the memory region \code{unreliable}, \TOOL substitutes occurrences of
\code{dest} in the model with \code{v} and occurrences of \code{src} with
\code{e}.
\cref{be:reliable} specifies a reliable write operator while \cref{be:relaxed}
specifies a faulty write operator.
The faulty write operator permits errors that are larger than
\code{eps}, the result being that the system stores an erroneous value in the
variable represented by
\code{dest} and subsequent reads of that variable return the
erroneous value.

\paragraph{\bf Switchability.}
The \code{reliable} flag  models the fact that it is possible to selectively
enable Rowhammer protection techniques \cite{aweke2016anvil} to trade
performance for reliability. Specifically if \code{reliable} is set to
\code{true}, then the model does not generate errors.  In addition to
Rowhammer, this model can simulate scenarios that selectively use ECC-protected
caches~\cite{ kim2007multibit, yoon2009ecc, energyecc} in conjunction with
traditional caches.  

\subsubsection{Vector Product Under Switchable Rowhammer}
\begin{figure}
\begin{wraplst}
\begin{lstlisting}
property_r large_error(vector<uint> x, uint N) : ~\label{rseu:prop}~
   ~$\forall$~(uint i)(i < N<r> -> (eq(x[i]) || model.eps * model.eps < x<r>[i]));

requires ~$\forall$~(uint i)(x[i] < model.eps && y[i] < model.eps) ~\label{rseu:pre}~
requires_r eq(N) && eq(x) && eq(y)
vector<uint> product(uint N, vector<uint> x(N), vector<uint> y(N))
{
   @region(unreliable) vector<uint> result(N); ~\label{rseu:decl}~

   for (uint i = 0; i < N; ++i) ~\label{rseu:loop}~
      invariant_r large_error(result, i)
   {
      result[i] = x[i] * y[i]; ~\label{rseu:fwrite}~
   }

   assert_r(large_error(result, N)); ~\label{rseu:assert}~

   return result;
}
\end{lstlisting}
\end{wraplst}
  \vspace{-1ex}
  \caption{Vector Product Under Switchable Rowhammer}
  \label{fig:rseu}
\end{figure}

\cref{fig:rseu} presents a vector product implementation annotated
to verify under the switchable rowhammer model.
The relaxation in this implementation occurs in the loop (\cref{rseu:loop})
where faulty writes may corrupt the value of \code{result[i]}.
On \cref{rseu:assert} the implementation verifies that the impact of errors is
larger than \code{model.eps~*~model.eps}.

\paragraph{\bf Memory Regions.}
\cref{rseu:decl} places the \code{result} vector in the \code{unreliable}
memory region.
Thus, \TOOL will use the execution model specification of the \code{write}
operator on \cref{rseu:fwrite}.

\paragraph{\bf Detectable Errors.}
In some systems, large errors are always detectable as they produce invalid
results that an application can easily check for.
For example, in Self-Correcting Connected Components (SC-CC)
\cite{sao2016self}---an iterative algorithm for computing the connected
subgraphs in a graph---a large error in the output vector produces references
to nonexistent graph nodes.
Thus, SC-CC detects and corrects all large errors by scanning through the
output vector after each iteration for invalid entries and recomputing
erroneous elements reliably.

The assertion on \cref{rseu:assert} uses the \code{large\_error} property on
\cref{rseu:prop} to enforce that the impact of errors is sufficiency large to
produce invalid values in \code{result}.
The \code{large\_error} property takes a vector \code{x} and a size \code{N} and
ensures that for every index \code{i}, \code{x[i]} is equal across both
executions, or \code{x<r>[i]} contains a trivially invalid value.

To support this, the function precondition on \cref{rseu:pre} mandates that
every element of \code{x} and \code{y} is bounded between 0 and
\code{model.eps}.
This ensures that any non-erroneous value in \code{result<r>} is bounded
between 0 and \code{model.eps * model.eps}.
Therefore, if a program using this vector product function scans for elements
that are out of this bound it will detect all errors.
Such a program could then correct these errors by setting the \code{reliable}
flag in the switchable rowhammer model to \code{true} and recomputing the
erroneous elements.

\subsection{Multicycle Error Model}
\label{sec:multicycle-model}

\cref{fig:multicycle} presents a multicycle error execution model.
A multicycle error is an error state in which multiple consecutive instructions
experience errors \cite{inoue2011high}.
This implementation permits a single multicycle error and tracks the state of this
error through the use of model variables \code{stuck} and \code{length}.
The \code{stuck} flag represents whether or not the system is currently
experiencing a multicycle fault while the \code{length} variable indicates
how many instructions the fault will continue for.
We leave the \code{length} variable unbound, permitting the multicycle error to
persist for an arbitrary number of operations.
This implementation is compatible with the vector product implementation from
the additive SEU example (\cref{fig:add-seu}) and requires no modifications to
verify.

\cref{multi:reliable} describes a reliable multiplication implementation that
the model may use before the fault occurs (\code{!stuck}) and after it ends
(\code{length == 0}).

\begin{wrapfigure}[14]{l}{0.52\textwidth}
  \vspace{-3ex}
\begin{wrappedwraplst}
\begin{lstlisting}
const real eps = ...;
bool stuck = false;
uint length;

@label(reliable) ~\label{multi:label1}~
operator*(real x1, real x2) ~\label{multi:reliable}~
   when !stuck || length == 0
   ensures result == x1 * x2;

@label(unreliable) ~\label{multi:label2}~
operator*(real x1, real x2) ~\label{multi:faulting1}~
   when length > 0
   modifies (stuck, length)
   ensures stuck &&
      length == old(length) - 1 && ~\label{multi:old}~
      x1*x2-eps < result < x1*x2+eps;
\end{lstlisting}
\end{wrappedwraplst}
  \vspace{-2ex}
\caption{Multicycle Error Execution Model}
\label{fig:multicycle}
\end{wrapfigure}

\cref{multi:faulting1} encodes an operator that the model may use during, or to
begin a multicycle error.
The model may substitute these operators so long as a multicycle error has not
occurred and resolved before the current instruction (\code{length > 0}).
This operator sets \code{stuck}, decrements \code{length}, and constrains
\code{result} to be within \code{eps} of the original result.

Together, these two operators ensure that at some point the system may be
stuck experiencing faults on all multiplications, but after \code{length}
multiplications it will unstick and execution will be reliable.

\paragraph{\bf Temporal Variable References}
To support more complex models than one can express with boolean flags, \TOOL
allows developers to differentiate between variable state before and after each
operation.
By default, all variables in \code{ensures} clauses refer to their state after
the operator executes.
However, by wrapping the variable in \code{old()} the wrapped variable will
instead refer to its state before the operator executes.
Temporal variable references through the \code{old} keyword is in the spirit of
similar constructs in ESC/Java~\cite{flanagan2001houdini},
Spec\#~\cite{specsharp}, and many other verification systems.
The multicycle error model uses this feature to track the length of a
multicycle event on \cref{multi:old} by asserting that the \code{length}
variable after the multiplication is one less than its value prior to the
multiplication.

\paragraph{\bf Named Operators}
To enable developers to \emph{refine} models, \TOOL supports optional labels on
operators.
Refinement is the process by which developers may create submodels that satisfy
the specification of supermodels.
It enables developers to construct a lattice of such models such that programs
verified under an abstract execution model will also verify under specialized
versions of that model.
\cref{multi:label1} gives the \code{reliable} label to the first operator while
\cref{multi:label2} gives the \code{unreliable} label to the second operator.

\subsubsection{Vector Product Under Multicycle}
\TOOL verifies the vector product implementation from the additive SEU example
(\cref{fig:vecprod-seu}) under this model with no modifications to the
implementation thus demonstrating \TOOL's high automatic reasoning power,
as well as the modular characteristic of \TOOL's execution models.

\subsection{Refinement}
\label{sec:refinement}

To enable an parallel development process in which developers may successively
build their hardware, models, and programs in tandem, \TOOL supports
execution model \emph{refinement}.
Refinement enables developers to construct a lattice of models such that more
precise submodels satisfy the specification of less precise supermodels.

By verifying refinement, \TOOL guarantees that programs verified under an
abstract execution model will also verify under specialized versions of that
model.
Therefore, refinement separates model elaboration from program verification.
That is, it provides an interface between hardware vendors and software
developers.
Software developers may verify their programs under loosely defined execution
models, while hardware vendors may provide detailed models that are true to
the underlying semantics of their hardware.
As long as software developers use \TOOL to verify that these precise models
refine their loose models, \TOOL guarantees that their programs will run as
expected on the hardware vendor's product.

\cref{fig:refine} presents an SEU model refined from the multicycle model.
\cref{refine:refine} indicates that this model is a refinement of the
multicycle model.
\TOOL supports multiple refinement, allowing submodels to refine any number of
supermodels.
\cref{refine:import1} and \cref{refine:import2} import the reliable and
unreliable operators from the multicycle model by name.
This makes these operators available to the submodel.
Developers may add additional operators by indicating that they
refine a named operator in each supermodel.
\TOOL checks that this submodel operator refines each supermodel operator by
verifying that the \code{when} clause of the submodel operator logically
implies the \code{when} clause of each supermodel operator and the
\code{ensures} clause of the submodel operator logically implies the
\code{ensures} clause of each supermodel operator.
Every additional operator in a submodel must explicitly refine some named
operator in each supermodel.
When using multiple refinement, any imported operators from one model must also
explicitly refine an operator in every other supermodel.

\begin{wrapfigure}[7]{l}{0.40\linewidth}
\vspace{-3ex}
\begin{wrappedwraplst*}
\begin{lstlisting}
refines multicycle; ~\label{refine:refine}~
import multicycle.reliable; ~\label{refine:import1}~
import multicycle.unreliable; ~\label{refine:import2}~

multicycle.length = 1; ~\label{refine:length}~
\end{lstlisting}
\end{wrappedwraplst*}
\vspace{-2ex}
\caption{Refined SEU Model}
\label{fig:refine}
\vspace{-3ex}
\end{wrapfigure}

\cref{refine:length} additionally sets the \code{length} variable in the
multicycle model to 1.
When refining a model, the submodel implicitly imports all variable
declarations and initializations from the supermodel.
The submodel may declare and initialize variables in its own namespace, but it
may not modify the variable state of the supermodel with one exception:
submodels may initialize uninitialized variables in the supermodel.
We enforce this constraint on model variables because \TOOL programs may
inspect, modify, and assert over model variables so there must be no
statically discernible difference between the submodel and the supermodel state
to a program written with knowledge of supermodel state variables.

Since \TOOL verifies that the refined SEU model refines the multicycle model,
\TOOL guarantees that the vector product implementation in
\cref{fig:vecprod-seu} verifies under the new refined SEU model without any
changes to the vector product implementation.

\newcommand{\judge}{\Downarrow}
\newcommand{\denote}[1]{{\llbracket #1 \rrbracket}}
\newcommand{\fresh}[1]{\textrm{fresh}({#1})}

\newcommand{\step}[2]{{#1} \xrightarrow{\mu} {#2}}
\newcommand{\trans}[2]{{#1} \rightarrow^* {#2}}

\newcommand{\rstep}[2]{{#1} \rightarrow_{r} {#2}}
\newcommand{\rtrans}[2]{{#1} \rightarrow^{*}_{r} {#2}}

\newcommand{\reval}[2]{{#1} \Downarrow_{r} {#2}}
\newcommand{\pstep}[2]{{#1} \rightarrow_{p} {#2}}
\newcommand{\ptrans}[2]{{#1} \rightarrow^{*}_{p} {#2}}
\newcommand{\peval}[2]{{#1} \Downarrow_{p} {#2}}

\newcommand{\inj}[1]{\textit{inj}_t({#1})}

\newcommand{\injo}[1]{\textit{inj}_o({#1})}
\newcommand{\injr}[1]{\textit{inj}_r({#1})}
\newcommand{\eeval}[2]{ {#1} \Downarrow_{r} {#2} }

\newcommand{\Int}{\textrm{Int}_{\textrm{N}}}
\newcommand{\Float}{\textrm{Float}}
\newcommand{\bad}{\text{\textit{bad}}}

\newcommand{\lop}{\text{\textit{lop}}}
\newcommand{\lope}[2]{\ensuremath{{{#1}\;\lop\;{#2}}}}

\newcommand{\iop}{\text{\textit{iop}}}
\newcommand{\iope}[2]{\ensuremath{{{#1}\;\iop\;{#2}}}} 

\newcommand{\cmp}{\text{\textit{cmp}}}
\newcommand{\cmpe}[2]{\ensuremath{{#1}\;\leq\;{#2}}}

\newcommand{\fop}{\text{\textit{fop}}}
\newcommand{\fope}[2]{\ensuremath{{{#1}\;\fop\;{#2}}}}

\newcommand{\true}{\text{\textit{true}}}
\newcommand{\false}{\text{\textit{false}}}

\newcommand{\mywrite}[2][]{\text{\ensuremath{\texttt{write}_{\ell}}~{#2}{#1}}}
\newcommand{\myskip}{\text{\texttt{skip}}}
\newcommand{\assign}[2]{{\text{\ensuremath{{#1} \; \texttt{=} \; {#2}}}}}

\newcommand{\fail}[1]{\text{\ensuremath{\texttt{fail} \; {#1}}}}
\newcommand{\assert}[1]{\text{\ensuremath{\texttt{assert} \; {#1}}}}
\newcommand{\relassert}[1]{\text{\ensuremath{\texttt{assert\_r} \; {#1}}}}
\newcommand{\assume}[1]{\text{\ensuremath{\texttt{assume} \; {#1}}}}
\newcommand{\ifthen}[3]{\text{\ensuremath{\texttt{if} \; {#1} \; {#2} \; {#3}}}}
\newcommand{\seq}[2]{\text{\ensuremath{{#1} \; \texttt{;} \; {#2}}}}
\newcommand{\while}[2]{\text{\ensuremath{\texttt{while} \; {#1} \; {#2}}}}
\newcommand{\whileR}[3]{\text{\ensuremath{\texttt{while} \; {#1} \; {#2} \; {#3}}}}
\newcommand{\whileIR}[4]{\text{\ensuremath{\texttt{while} \; {#1} \; {#2} \; {#3} \; {#4}}}}
\newcommand{\return}{\text{\texttt{return}}}
\newcommand{\returnx}[1]{\text{\ensuremath{\texttt{return} \; #1}}}
\newcommand{\call}[2]{\text{\ensuremath{#1\texttt{(}#2\texttt{)}}}}

\newcommand{\heap}{h}
\newcommand{\Heap}{H}
\newcommand{\Location}{\mathrm{Loc}}

\renewcommand{\frame}[0]{\sigma}
\newcommand{\Frame}[0]{\Sigma}

\newcommand{\env}{\varepsilon}
\newcommand{\Env}{\mathrm{E}}

\newcommand{\piframe}[1]{\pi_{1}{{#1}}}
\newcommand{\piheap}[1]{\pi_{2}{{#1}}}
\newcommand{\pimodel}[1]{\pi_{3}{{#1}}}

\newcommand{\Exp}{\mathit{Exp}}
\newcommand{\BExp}{\mathit{BExp}}
\newcommand{\CExp}{\mathit{CExp}}

\newcommand{\reliable}[1]{#1\texttt{<r>}}
\newcommand{\unreliable}[1]{#1\texttt{<u>}}

\newcommand{\tuple}[1]{\langle #1 \rangle}
\newcommand{\envTuple}[4]{\langle {#1}, \; {#2}, \; {#3}, \; {#4} \rangle}
\newcommand{\denv}{\envTuple{\frame}{\heap}{\region}{m}}

\newcommand{\tvar}[1]{{{#1}\texttt{<t>}}}

\newcommand{\origvar}[1]{{{#1}\texttt{<o>}}}
\newcommand{\relvar}[1]{{{#1}\texttt{<r>}}}

\newcommand{\arraytype}[2]{\texttt{array<}{#1}, {#2}\texttt{>}}

\newcommand{\bstep}{\Downarrow}
\newcommand{\bio}{b_{io}}
\newcommand{\bir}{b_{ir}}
\newcommand{\bs}{\bio; \bir}
\newcommand{\sstep}{\rightarrow}
\newcommand{\limp}{\rightarrow}

\newcommand{\region}{\theta}
\newcommand{\Region}{\Theta}

\begin{figure*}[htp]
\begin{small}
    \begin{minipage}{0.55\linewidth}
    \begin{align*}
      c &\in \text{constants},\ x \in \text{variables},\ r \in \text{memory regions}     \\
      \\[-.3cm]
      S &\rightarrow (\code{@region } r)^?\ \tau\ x\ |\ (\code{@region } r)^?\ \tau\ x\ (c^+) \ |\ x = E \\
      &|\ S \ ; \ S\ |\ \code{assume}\ P\ |\ \code{assert_r}\ P_r\ |\ \code{assert}\ P   \\
      &|\code{if}\ (B)\ \{ \; S \; \} \ \code{else}\ \{ \; S \; \}\ |\ \code{while}\ (B)\ I^*\ \{\ S\ \} \\
      \\[-0.3cm]
      \tau &\rightarrow \code{int }|\code{ real }|\code{ bool} \\
      I &\rightarrow \code{invariant}\ P\ |\ \code{invariant_r}\ P_r \\
      \\[-0.3cm]
      M  & \rightarrow \texttt{model} \; \{ \; x^* \;\textit{O}^+ \; \} \qquad
      \textit{O} \rightarrow \texttt{operator} \; \textit{Op} \;  (\tau\ x)^+ \; C^+  \\
      \textit{Op} & \rightarrow \neg \; \mid \; \neg. \; \mid \; \oplus \; \mid \; \prec \; \mid \; \diamond \; \mid \; \texttt{read} \; \mid \texttt{write}\\
      C  & \rightarrow  \texttt{when} \; P \; \mid \; \texttt{ensures} \; P \; \mid \; \code{modifies} \; x^+ \\
    \end{align*}
    \end{minipage}
    \begin{minipage}{0.40\linewidth}
    \begin{align*}
      \diamond &\rightarrow \land\ |\ \land.\ |\ \limp\ |\ \limp. \ |\ \ldots \\
      \prec &\rightarrow <\ |\ <.\ |\ =\ |\ =.\ |\ \ldots \\
      \oplus &\rightarrow +\ |\ +.\ |\ \times\ |\ \times.\ |\  \ldots \\
      \\[-0.3cm]
      E &\rightarrow x\ |\ E\ \oplus E\ |\ \code{model.}x\ |\ x[E] \\
      E_r &\rightarrow x\code{<o>}\ |\ x\code{<r>}\ |\ E_r\ \oplus E_r\ \\
      &|\ x\code{<o>}[E_r]\ |\ x\code{<r>}[E_r]\ |\ E \\
      \\[-0.3cm]
      B &\rightarrow \true\ |\ \false\ |\ E\ \prec\ E \\
        &|\ B\ \diamond\ B\ |\ f\ E^*\ |\ \lnot\ B\ |\ \lnot.\ B \\
      P &\rightarrow \forall\ x\  B\ |\ \exists\ x\ B\   |\ B \\
      P_r &\rightarrow \true\ |\ \false\ |\ E_r\ \prec\ E_r\ | \ P_r\ \diamond\ P_r \\
      & |\ \lnot\ P_r\ |\ \lnot.\ P_r\ |\ \forall\ x\ P_r\ |\ \exists\ x\ P_r\   \\
\end{align*}
\end{minipage}
\end{small}
\vspace{-.6cm}
\caption{Language Syntax}
\label{fig:syntax}
\vspace{-.15cm}
\end{figure*}

\section{Language}

\cref{fig:syntax} presents the core of {\TOOL}'s programming language.
Leto provides a general-purpose imperative language that includes specification
primitives (e.g., \texttt{requires}) in the spirit of
ESC/Java~\cite{flanagan2001houdini}, Boogie~\cite{boogie},
Eiffel~\cite{eiffel}, and Spec\#~\cite{specsharp} to support verifying
applications.

\vspace{-.15cm} \paragraph{\bf Programs.} A program consists of a
sequence of statements \(S\).
Although we also support functions, we elide them here.

\vspace{-.15cm}
\paragraph{\bf Data Types.} The language includes primitive data types ($\tau$) of (signed and unsigned) integers, reals, and booleans as well as vectors/matrices of
these types. A developer can use the $\texttt{@region}\;r$ annotation to state the variable is allocated in a named memory region, $r$, for which reads
and writes may have a custom semantics according to the execution model.

\vspace{-.15cm}
\paragraph{\bf Expressions.}  Leto includes standard numerical operations, comparison, and logical expressions, along with dotted notations (e.g., $\texttt{x} \; \texttt{+.} \; \texttt{y}$) that communicate
that the operation may have a custom semantics as specified in the execution model.

\vspace{-.15cm}
\paragraph{\bf Memory Operations.} Leto supports reads from and writes to variables, including values of both primitive and array/matrix type. Reads and writes to variables allocated
in a designated memory region operate with the semantics as given in the program's execution model.

\vspace{-.15cm}
\paragraph{\bf Assertions and Assumptions.} {\TOOL} also enables developers to
specify assertions and assumptions on the state of the program. {\TOOL}'s
language includes both standard \texttt{assert} statements and \texttt{assume}
statements (with their traditional meaning). Each such statement can use a
quantified boolean expression, $P$, that quantifies over the value of
variable (e.g., the index of an array/matrix). A relational assertion
statement, \texttt{assert_r}, uses a quantified relational boolean expression,
$P_r$, that specifies a relationship between the original and relaxed
executions to verify.

\vspace{-.15cm}
\paragraph{\bf Control flow.}  {\TOOL}'s language includes standard control
constructs, such as sequential composition, \texttt{if} statements,
\texttt{while}, and \texttt{for} statements.
For \texttt{while} and \texttt{for} statements, a developer can specify {\em
loop invariants} to support verification via the syntax \texttt{invariant}
(unary loop invariant) and \texttt{invariant_r} (relational loop invariant). A
loop invariant specifies a property that must be true on entry to the loop, as
well as at the end of each loop iteration. Loop invariants are a key to
verifying applications that contain loops because automatically inferring loop
invariants is undecidable in general. Therefore, a developer may need to
specify additional loop invariants when Leto's loop invariant inference
procedure is insufficient. 

\vspace{-.15cm} \paragraph{\bf Execution Model.}  An execution model $M$
consists of a set of state variables $x^*$ and {\em operation specifications}
(\textit{O*}). Each operation specification (\textit{O}) specifies 1) the
target operator for the specification, 2) a list of variables as parameters to
the specification, and 3) a  set of clauses. A clause is either a \texttt{when}
clause, which guards the execution of the specification with a predicate $P$,
an \texttt{ensures} clause, which establishes a relationship on the output of
the specification given the inputs to the specification and fault model's state
variables, or a \texttt{modifies} clause, that specifies which of the model's
state variables changes as a result of using the operation. Predicates consist
of standard operations over standard expressions with the addition of the
distinguished \texttt{result} variable, which captures the result of the
specification's execution.

\begin{wrapfigure}{l}{0.5\linewidth}
  \begin{small}
    \vspace{-1.5em}
\begin{tabular}{rcl}
  $x$ & $\in$ & $\mathrm{Var}$ \\
  $r$ & $\in$ & $\mathrm{Reg}$ \\
  $n$ & $\in$ & $\Int$ \\ \\
\end{tabular}
\begin{tabular}{rcl}
   $\textit{Exp}$ & $\rightarrow$ & $n \mid \; r \; \oplus \; \textit{r}$   \\
    
  $\mathit{S}$ & $\rightarrow$ & $\assign{\textit{r}}{\textit{Exp}} \; \mid \;\assign{\textit{r}}{x} \;\mid\; \assign{\textit{x}}{r} $ \\
  & $\mid$ & \assert{r} \; $\mid$ \; \assume{r} \\
  & $\mid$ & $\texttt{skip} \; \mid \; \seq{\textit{S}}{\textit{S}} \; \mid \; \ifthen{r}{S}{S}$ \\
  & $\mid$ & $\whileR{r}{P_r^*}{\textit{S}}$ 
\end{tabular}
\vspace{-.15cm}
\caption{Syntax of Lowered Language (Abbreviated)}
\label{fig:syntax-lowered}
\vspace{-.25cm}
  \end{small}
\end{wrapfigure}

\newcommand{\rf}{\rho}

\subsection{Dynamic Semantics} 

We next present an abbreviated dynamic semantics of {\TOOL}'s language.  We
expand on these semantics in \cref{sec:dyn-semantics} and
\autoref{sec:full-semantics}.  We formalize the semantics via a lowered syntax
of the \TOOL language that includes registers (\cref{fig:syntax-lowered}).  We
assume a standard compilation process that translates high-level Leto to this
lowered language.  Additionally, to support our formalization of Leto's program
logic on this representation, we extend the predicates \(P\) and \(P_r\) with
registers into \(P^*\) and \(P^*_r\), respectively.

\subsubsection{Preliminaries}

{\TOOL}'s semantics models an abstract machine that includes a {\em frame}, a
{\em heap}, and an {\em execution model state}.  {\TOOL} allocates memory for
program variables (both scalar and array) in the heap. A frame serves two
roles: 1) a frame maps a program variable to the address of the memory region
allocated for that variable in the heap, and 2) a frame maps a register to its
current value in the program.  The model state stores the values for state
variables within the execution model.

\paragraph{\bf Frames, Heaps, Model States, Environments.} A {\em frame},
\(
\frame \in \Frame =  \mathrm{Var} \cup \mathrm{Reg} \rightarrow \Int
\),
is a finite map from variables
and registers to N-bit integers. A {\em heap},
\(
  \heap \in \Heap = \Location \rightarrow \Int
\),
is a finite map from locations ($n \in \Location \subset \Int$) to N-bit
integer values. A {\em region map},
\(
  \region \in \Region = \Location \rightarrow \mathrm{Region}
\),
is a finite map from locations to memory regions.
A {\em model state},
\(
  m \in M = \mathrm{Var} \rightarrow \Int
\),
is a finite map from model state variables to N-bit integer values. An {\em
environment},
\(
  \env \in \Env = \Frame \times \Heap \times \Region \times M
\),
is a tuple consisting of a frame, a heap, a region map, and a model state. An {\em execution model specification},
\(
  \mu \subseteq \textit{Op} \times \textit{list}(\mathrm{Var}) \times \textit{set}(\mathrm{Var}) \times P \times P
\),
is a relation consisting of tuples of an operation $\textit{op} \in
\textit{Op}$, a list of input variables, a set of modified variables, and two
unary logical predicates representing the when and ensures clauses of the
operation.

\paragraph{\bf Initialization.} For clarity of presentation, we assume a compilation and execution model in which memory locations for program variables are allocated and the corresponding mapping in the frame are done prior to execution of the program (similar in form to C-style declarations).

\subsubsection{Execution Model Semantics}

Here we provide an abbreviated presentation of the execution model relation
\(
  \tuple{m, \textit{op}, (\textit{args})} \judge_{\mu} \tuple{n, \; m'}.
\)
The relation states that given the arguments, $\textit{args}$ to an operation
$op$, evaluation of the operation from the model state $m$ yields a result $n$
and a new model state $m'$ under the execution model specification $\mu$. 
\vspace{.05cm}
\begin{small}
\begin{mathpar}
\inferrule[f-binop]
{
	\mu(\oplus, [x_1, x_2], X, P_w, P_e) \\
  \forall x_m \in X \cdot \fresh{x_m'} \\
	m[x_1 \mapsto n_1][x_2 \mapsto n_2] \models P_w \\
	m'[x_1 \mapsto n_1]
    [x_2 \mapsto n_2]
    [\forall x_m \in X \cdot x_m \mapsto x_m']
    [\textit{result} \mapsto n_3] \models P_e \\
	\mathrm{dom}(m) = \mathrm{dom}(m')
}
{
	\tuple{m, \oplus, (n_1, n_2)} \judge_{\mu} \tuple{n_3, \; m'}
}
\end{mathpar}
\end{small}
\vspace{-.15cm}

The [\textsc{f-binop}] rule specifies the meaning of this relation for binary
operations. This relation states that the value of an operation $\oplus$ given
a tuple of input values $(n_1,
n_2)$ and an execution model state $m$ evaluates to value $n_3$ and a new
model state $m'$. The rule relies on the relation $\mu(\textit{op},
$\textit{vlist}$, X, P_w, P_e)$ which specifies the list of argument names,
$\textit{vlist}$, the set of modified variables \(X\), the {\em precondition}
$P_w$, and the {\em postcondition} $P_e$ for the operation $op$ in the
developer-provided execution model.
The set of modified variables is the union of the \texttt{modifies} clauses in
the operation's specification.
The precondition of an operation is the conjunction of the \texttt{when}
clauses in the operation's specification.
The postcondition of an operation is the conjunction of the \texttt{ensures}
clauses in the operation's specification.

The semantics of the model relation non-deterministically selects an operation
specification, result value, and output model state subject to the constraint
that:
1) the current model state satisfies the precondition (after the inputs to
    the operation have been appropriately assigned into the model state),
2) the output model state satisfies the postcondition (after the inputs,
   modified variables, and result value have been appropriately assigned into
   the model state), and
3) the domains of the input and output state are the same.

Because of the uniformity of the execution model specification, the semantics
for other operations (e.g., reads and writes) is similar with the sole
distinction being the number of arguments passed to the operation. For clarity
of presentation, we elide the presentation of rules for those operations, but
we provide them in \cref{sec:full-semantics}, \cref{fig:ext-model-semantics}.

\subsubsection{Language Semantics}
We next present the non-deterministic small-step transition relation
$\step{\tuple{s, \env}}{\tuple{s', \env'}}$ of a {\TOOL} program. The relation
states that execution of statement $s$ from the environment $\env$ takes one
step yielding the statement $s'$ and environment $\env'$ under the execution
model specification $\mu$. The semantics of the statements are largely similar
to that of traditional approaches except for the ability to the statements to
encounter faults. Broadly, we categorize {\TOOL}'s instructions into four
categories: {\em register instructions}, \emph{assertions}, {\em memory
instructions}, and {\em control flow}.

\vspace{-3ex}
\begin{mathpar}
    \inferrule[assign]
    {
    }
    {
        \step{\tuple{\assign{r}{n}, \denv}}{\tuple{\texttt{skip},
        \envTuple{\frame[r \mapsto n]}{\heap}{\region}{m}}}
    }

    \inferrule[binop]
    {
        n_1 = \frame(r_1) \\
        n_1 = \frame(r_2) \\
        \tuple{m, \oplus, (n_1, n_2)} \judge_{\mu} \tuple{n_3, \; m'} \\
    }
    {
        \step{\tuple{\assign{r}{r_1 \oplus r_2}, \denv}}{\tuple{\assign{r}{n_3}, \envTuple{\frame}{\heap}{\region}{m'}}}
    }
\end{mathpar}

\paragraph{\bf Register Instructions.} The rules [\textsc{assign}] and
[\textsc{binop}] specify the semantics of two of {\TOOL}'s register
manipulation instructions. [\textsc{Assign}] defines the semantics of assigning
an integer value to a register, $\assign{r}{n}$. This has the expected
semantics updating the value of $r$ within the current frame with the value
$n$. Of note is that register assignment executes fully reliably without
faults. 

[\textsc{binop}] specifies the semantics of a register only binary operation,
$\assign{r}{r_1 \; \oplus \; r_2}$. Note that reads of the input registers
execute fully reliably. The result of the operation is $n_3$, which is the
value of the operation given the semantics of that operation's execution  model
when executed from the model state $m$ on parameters $n_1$ and $n_2$. Executing
the execution  model may change the values of the execution  model's state
variables. Therefore, the instruction evaluates to an instruction that assigns
$n_3$ to the destination register and evaluates with a environment that
consists of the unmodified frame, the unmodified heap, and the modified
execution  model state. Note that by virtue of the fact that both the frame and
heap are unmodified, faults in register instructions cannot modify the contents
or organization of memory.

\paragraph{\bf Assertions.}   
The \texttt{assert} and \texttt{assume} statements have standard semantics,
yielding a \texttt{skip} and continuing the execution of the program if their
conditions are satisfied. For either of these statements, if their conditions
evaluate to false, then execution yields \texttt{fail} denoting that execution
has failed and become stuck in error.  
We provide the rules for both \texttt{assert} and \texttt{assume} in
\cref{sec:dyn-semantics}, \cref{fig:main-operational}.

\begin{mathpar}
    \inferrule[read]
    {
        a = \frame(x) \\
        n = h(a) \\
        q = \region(a) \\
        \tuple{m, \code{read}, (n, q)} \judge_{\mu} \tuple{n', \; m'}
    }
    {
        \step{\tuple{\assign{r}{x}, \denv}}{\tuple{\assign{r}{n'},
        \envTuple{\frame}{\heap}{\region}{m'}}}
    }

    \inferrule[write]
    {
         a = \frame(x) \\
         n_{\textit{old}} = h(a) \\
         n_{\textit{new}} = \frame(r) \\
         q = \region(a) \\
        \tuple{m, \code{write}, (n_{\textit{old}}, n_{\textit{new}}, q)} \judge_{\mu} \tuple{n_r, \; m'} \\
    }
    {
        \step{\tuple{\assign{x}{r}, \denv}}
             {\tuple{\texttt{skip},  \envTuple{\frame}
                                              {h[ a \mapsto n_r]}
                                              {\region}
                                              {m'}}}
    }
\end{mathpar}

\paragraph{\bf Memory Instructions.}
The rules [\textsc{read}] and [\textsc{write}] specify the semantics of two of
{\TOOL}'s memory manipulation instructions. [\textsc{read}] defines the
semantics of reading the value of a program variable $x$ from it's
corresponding memory location: $\assign{r}{x}$. The rule fetches the program
variable's memory address from the frame, reads the value of the memory
location $n = h(a)$ and the region the memory location belongs to $q =
\region(a)$ and then executes the execution  model with the program variable's
current value in memory and the memory region it resides in as a parameters.
The execution  model non-deterministically yields a result $n'$ that the rule uses to complete its implementing by issuing an assignment to the register. 

[\textsc{write}] defines the semantics of writing the value of a register to
memory. The rule reads the value of the memory location to record the old value of
the memory location, reads the value of the input register, fetches the region
the memory location corresponds to, and then executes
the execution  model with these values as parameters. The execution  model yields a new
value $n_r$ that the rule then assigns to the value the program variable.

\paragraph{\bf Control Flow.}
The rules for control flow~  have standard
semantics.  An important note is that the semantics of these statements is such
that the transfer of control from one instruction to another always executes
reliably and, therefore, faults do not introduce control flow errors into the
program. This modeling assumption is consistent with standard fault injection
and reliability analysis models~\cite{vulfi16,sampson11}.
We provide the control flow rules in \cref{sec:dyn-semantics},
\cref{fig:main-operational}.

\paragraph{\bf Big-Step Semantics.}  To support the formalization in the
remainder of the paper, we introduce the big-step relation $\tuple{s, \env}
\Downarrow_m v \subseteq S \times \Env \times V$ where $V ::= \Env  \; | \;
\code{fail} \; \Env$ such that $\tuple{s, \env} \Downarrow v$ is the reflexive
transitive closure of $\rightarrow_m$ that yields the environment \(\env\) if
execution ends successfully in a \code{skip} statement or yields the pair
$\code{fail} \; \env$ when the execution ends in a failure. We also introduce
the big-step relation $\tuple{s, \env} \Downarrow v \subseteq S \times \Env
\times V$ where $\tuple{s, \env} \Downarrow v \equiv \tuple{s, \env}
\Downarrow_\rho v$ where $\rho$ denotes a {\em fully reliable} fault model
where the only implementations exposed for each operation are fully reliable
implementations.

\newcommand{\ljudge}[3]{{\vdash_l \{ \; #1 \; \}  \; #2 \; \{ \; #3 \; \}}}

\newcommand{\RExp}[0]{\mathit{RExp}}
\newcommand{\RBExp}[0]{\mathit{RBExp}}
\newcommand{\RP}[0]{\mathit{RP}}

\newcommand{\Relexp}[0]{\mathit{RE}}
\newcommand{\Bexp}[0]{B}
\newcommand{\Relbexp}[0]{B^*}

\newcommand{\bool}[0]{\mathbb{B}}
\newcommand{\pred}[0]{P^*}
\newcommand{\relpred}[0]{\mathit{P^*_r}}

\newcommand{\myexists}[2]{\exists_{#1} \cdot #2}
\newcommand{\preddenote}[2]{\denote{#1}} 
\newcommand{\relpreddenote}[2]{\denote{#1}} 
\newcommand{\pow}[1]{\mathcal{P}(#1)}
\renewcommand{\Exp}{\mathit{E}}

\newcommand{\reljudge}[4][]{{ \ifthenelse{\equal{#1}{}}{}{#1}  \vdash_{r} \{ \; #2 \; \}  \; #3 \; \{ \; #4 \; \}}}

\newcommand{\lockjudge}[4][]{{\ifthenelse{\equal{#1}{}}{}{#1} \vdash \{ \; #2 \; \}  \; #3 \; \{ \; #4 \; \}}}

\section{Program Logic}
\label{sec:logic}

\def \MathparLineskip {\lineskip=0.1cm}

{\TOOL}'s program logic is a relational program logic in that it relates
relaxed executions of the program to its original, reliable execution.  A key
idea behind our development is the separation of the rules into a part that
solely characterizes the reliable execution of the program, (the Left Rules), a
part that solely characterizes the relaxed execution (the Right Rules), and a
part the characterizes the lockstep execution of the reliable and relaxed
execution (the Lockstep Rules). The result is an Asymmetric Relational Hoare
Logic that characterizes the two interpretations of the program.

\subsection{Preliminaries}

\paragraph{\bf Assertion Logic Syntax and Semantics.} \cref{fig:syntax}
presents our language syntax, including the syntax of our assertion language.
Assertions include standard quantified boolean predicates, $\pred{}$, with the
standard semantic function $\denote{\pred{}} \in \pow{\Env}$ that gives the
denotation of $\pred{}$ as the set of environments that satisfy the predicate.
Assertions also include quantified {\em relational} boolean predicates,
$\relpred{}$, with the semantic function $\denote{\relpred{}} \in \pow{\Env
\times \Env}$ that gives $\relpred{}$ the meaning of the set of pairs of
environments that satisfy the predicate. In our standard convention, the first
environment of the pair corresponds to the state of the reliable execution
whereas the second environment corresponds to that of the relaxed execution.

\vspace{-.15cm} 
\paragraph{\bf Auxiliary Definitions.}  To support the formalization in the
remainder of the paper we define the auxiliary notation $\inj{\cdot}$
where $t \in \{o, r\}$ implements an {\em injection} for standard unary
predicates into a relational domain. For $t = o$, the definition injects a
predicate into the domain of the reliable execution of the program whereas when
$t=r$, the definition injects a predicate into the domain of the relaxed
execution.

\begin{figure*}
\centering
\begin{footnotesize}
\begin{mathpar}
\inferrule[assign-l]
{ }
{
   \ljudge{Q^*_r[n / \injo{r} ]}{\assign{r}{n}}{Q^*_r}
}

\inferrule[assign-r]
{ }
{
   \reljudge[\mu]{Q^*_r[n / \injr{r}]}{\assign{r}{n}}{Q^*_r}
}

  \vspace{0.2cm}
\inferrule[binop-l]
{ }
{
   \ljudge{Q^*_r[\injo{r_1 \oplus r_2} / \injo{r}]}{\assign{r}{r_1 \oplus r_2}}{Q^*_r}
}

\inferrule[binop-r]
{
   \fresh{r'} \\
   \exists ([x_1, x_2], X, P^*_w, P^*_e) \in \mu(\oplus) \cdot  \injr{P^*_w[r_1 / x_1][r_2 / x_2]} \\
   Q'^*_r =
            \left(\bigvee_{([x_1, x_2], X, P^*_w, P^*_e) \in \mu(\oplus)}
                  \injr{P^*_w[r_1 / x_1][r_2 / x_2]} \rightarrow
                  \injr{P^*_e[r_1 / x_1][r_2 / x2][\forall x_m \in X \cdot \fresh{x_m'} / x_m][ r' / \textit{result}]} \right)
}
{
   \reljudge[\mu]{Q^*_r[\injr{r'} / \injr{r}] \land Q'^*_r}{\assign{r}{r_1 \oplus r_2}}{Q^*_r}
}

\inferrule[assert-l]
{
}
{
   \ljudge{\true}{\assert{r}}{\injo{r}}
} \qquad
\inferrule[assert-r]
{
}
{
   \reljudge[\mu]{\injr{r}}{\assert{r} }{\injr{r}}
}

\inferrule[assume-l]
{
}
{
  \ljudge{\true}{\assume{r}}{\injo{r}}
} \qquad
\inferrule[assume-r]
{ 
   \reljudge[\mu]{P^*_r}{\assert{r}}{Q^*_r}
}
{
   \reljudge[\mu]{P^*_r}{\assume{r}}{Q^*_r}
}

\end{mathpar}
\end{footnotesize}
\vspace{-.15cm}
\caption{Left and Right Rules for Primitive Statements}
\label{fig:relaxed-logic}
\vspace{-.15cm}
\end{figure*}

\begin{figure*}
\centering
\begin{small}
\begin{mathpar}
 
\boxed{\lockjudge{P^*_r}{s}{Q^*_r}}

\inferrule[split]
{
   \lockjudge[\mu]{P^*_r}{s \sim s}{Q^*_r}
}
{
   \lockjudge[\mu]{P^*_r}{s}{Q^*_r}
}

\inferrule[seq]
{
 \lockjudge[\mu]{P^*_r}{s_1}{R^*_r} \\
 \lockjudge[\mu]{R^*_r}{s_2}{Q^*_r}
}
{
   \lockjudge[\mu]{P^*_r}{\seq{s_1}{s_2}}{Q^*_r}
}

\inferrule[if]
{
   b \equiv \assign{r}{\true} \\
   \lockjudge[\mu]{P^*_r \land \injo{b} \land \injr{b}}{s_1}{Q^*_r} \\
   \lockjudge[\mu]{P^*_r \land \neg \injo{b} \land \injr{b}}{s_2 \sim_r s_1}{Q^*_r} \\
   \lockjudge[\mu]{P^*_r \land \injo{b} \land \neg \injr{b}}{s_1 \sim_r s_2}{Q^*_r} \\
   \lockjudge[\mu]{P^*_r \land \neg \injo{b} \land \neg \injr{b}}{s_2}{Q^*_r}
}
{
   \lockjudge{P^*_r}{\ifthen{r}{s_1}{s_2}}{Q^*_r}
}

\end{mathpar}
\end{small}
\vspace{-.15cm}
\caption{Lockstep Control Flow and Structural Rules}
\label{fig:lockstep-logic}
\vspace{-.25cm}
\end{figure*}

\subsection{Proof Rules}
Figures~\ref{fig:relaxed-logic} and~\ref{fig:lockstep-logic} provide an
abbreviated presentation of the rules of our program logic.
We present the remainder of the rules in \cref{sec:rules}.
We have partitioned the presentation into two parts: 1) the Left Rules and
Right Rules for primitive statements and 2) the Lockstep Rules.

\vspace{-.15cm}
\paragraph{\bf Left Rules.} The Left Rules, which we denote by the judgment
$\ljudge{P^*_r}{s}{Q^*_r}$, characterize the behavior of the reliable execution
of the statement $s$.  The denotation of the judgment is that if $(\env_1,
\env_2) \models P^*_r$,  and $\tuple{s, \env_1} \judge \env_1'$,  then
$(\env_1', \env_2) \models Q^*_r$.  Namely, given a proof in the Left Rules, for
a pair of environments satisfying the precondition of the proof, then if a
reliable execution of $s$ terminates, then the resulting environment pair
satisfies the proof's postcondition.

\vspace{-.15cm}
\paragraph{\bf Right Rules.} The right rules, which we denote by the judgment
$\reljudge[\mu]{P^*_r}{s}{Q^*_r}$, characterize the behavior of the relaxed
execution of $s$ under a fault model specification $\mu$.   The denotation of the
judgment is similar to that of the Left Rules: if  $(\env_1, \env_2) \models
P_r^*$, $\tuple{s, \env_2} \judge_{\mu} \env_2'$,  then $(\env_1, \env_2')
\models Q_r^*$.  Namely, given a proof in the right rules, for a pair of
environments satisfying the proof's precondition, then if execution of $s$
under the fault model specification $\mu$ terminates, then the resulting
environment pair satisfies the proof's postcondition.

\vspace{-.15cm}
\paragraph{\bf Lockstep Rules.} 
The Lockstep Rules together constitute the main
top-level judgment of the logic reasons about relations between the two
semantics as they proceed in lockstep.:  $\lockjudge[\mu]{P_r^*}{s}{Q_r^*}$.
The denotation  is that if $(\env_1, \env_2) \models P_r^*$,
$\tuple{s, \env_1} \judge \env_1'$, and $\tuple{s, \env_2} \judge_{\mu}
\env_2'$, then $(\env_1', \env_2') \models Q_r^*$. 

\subsubsection{Left and Right Rules}

\paragraph{\bf Register Assignment.} The rules [\textsc{assign-l}] and
[\textsc{assign-r}] capture the semantics of the register assignment statement,
$\assign{r}{n}$ in the lowered language. In the reliable execution, the rule
[\textsc{assign-l}] captures the semantics of the assignment statement via the
standard backward characterization of assignment as seen in standard Hoare
logic \cite{hoare}.  The major distinction between a standard presentation and
the presentation here is that the substitution replaces the injected form of
the register $r$ in the postcondition of the statement.  The expression
$\injo{r}$ denotes the value of $r$ in the reliable version of the program. For
the relaxed execution, the rule [\textsc{assign-r}] captures the semantics by
substituting for $\injr{r}$, which denotes the value of $r$ in the relaxed
execution. We note that given these results, assignment is reliable in both the
reliable and relaxed executions with the primary distinction being which
environment is modified (either that corresponding to the reliable execution
or that of the relaxed execution).

\vspace{-.15cm}
\paragraph{\bf Arithmetic Operation.} The rules [\textsc{binop-l}] and
[\textsc{binop-r}]  give the semantics of binary arithmetic operations on
registers: $\assign{r}{r_1 \oplus r_2}$. For the reliable
execution, [\textsc{binop-l}] relies on the backwards characterization of
assignment as seen in [\textsc{assign-l}]  to substitute the value $r$ in the
reliable execution of the program with the value of the arithmetic operation
$\injo{r_1 \oplus r_2}$. For the relaxed execution, [\textsc{binop-r}],
augments the traditional backwards characterization to include the potentially
unreliable execution of the binary operation.

\vspace{-.15cm} 
\paragraph{\bf Assert.} The rules [\textsc{assert-l}] and
[\textsc{assert-r}] give the semantics of assertion statements. There is a
major distinction between the role of assertion statements between the reliable
and relaxed execution of the program. Specifically, while the logic requires
that the condition of an assert statement is verified in the relaxed execution,
the condition of an assert statement in the reliable execution does not need to
be verified; it is instead assumed. The major design point is that {\TOOL}
enables a developer to use a variety of means (e.g., testing, verification, or
code review) to validate an assertion in the original program and transfer that
reasoning to the verification process for the relaxed execution. To achieve
this design, the Left rule for assertions assumes the validity of the assertion
whereas the Right rule asserts. Although the \code{assert} and \code{assume}
have the same semantics in the reliable and relaxed executions, the intentions
of the statements differ.  Specifically, if a developer places an \code{assert}
in the program, the assumption is that they have used other means to evaluate
the validity of that assertion in the reliable execution (potentially including
other verification systems). An \code{assume} statement, however, does not
carry that intention.

\vspace{-.15cm}
\paragraph{\bf Assume.} The rules [\textsc{assume-l}] and [\textsc{assume-r}]
give the semantics of assume statements. The primary distinction for
\code{assume} statements is that while assume statements have their standard
semantics in the reliable execution of the program (no proof obligation is
required), \code{assume} statements do in fact require a proof obligation in
the relaxed semantics. The semantics of an \code{assume} statement in the
relaxed semantics is therefore the same as that of an \code{assert} statement.
The rationale behind this design is that as part of the verification of the
relaxed execution we must verify that faults do not interfere with the
reasoning behind an assumption.

\vspace{-.15cm}
\paragraph{\bf Control Flow.} For clarity of presentation we have elided the
left and right rules control flow because the rules adhere to the standard
formalization as seen in traditional Hoare logic. The only distinction between
these rules and their standard implementation is that they operate over
relational predicates.

\subsubsection{Lockstep Rules}

To support the lock step rules, we first present the [\textsc{stage}] rule,
which joins the Left Rules and Right Rules.
\begin{mathpar}
\inferrule[stage]
{
 \ljudge{P^*_r}{s_1}{R^*_r} \\
 \reljudge[\mu]{R^*_r}{s_2}{Q^*_r}
}
{
   \lockjudge[\mu]{P^*_r}{s_1 \sim s_2}{Q^*_r}
}

\inferrule[inverse-stage]
{
  \reljudge[\mu]{P^*_r}{s_2}{R^*_r} \\
  \ljudge{R^*_r}{s_1}{Q^*_r}
}
{
   \lockjudge[\mu]{P^*_r}{s_1 \sim_r s_2}{Q^*_r}
}
\end{mathpar}
\vspace{-.15cm}
\paragraph{\bf Stage.} The rule [\textsc{stage}] gives a semantics to a pair of
statements $s_1$ and $s_2$ for which the goal is to characterize the behavior
when the reliable execution executes $s_1$ and the relaxed execution executes
$s_2$. The specific composition we have chosen for this rule is to apply the
Left Rules for $s_1$ before applying the Right Rules to $s_2$.  Namely, the
rule first applies the Left Rule for $s_1$, yielding a new predicate $R^*_r$,
before then applying the Right Rule for $s_2$ to $R^*_r$. The rule
[\textsc{split}] provides a rationale for this specific composition.
The rule [\textsc{inverse-stage}] has the opposite semantics, applying the
Right Rule for \(s_2\), yielding a new predicate \(R^*_r\), before then
applying the Left Rule for \(s_1\) to \(R^*_r\).
The rule [\textsc{if}] provides a rationale for this composition during
nonlockstep execution.

\paragraph{\bf Split.} The rule  [\textsc{split}] gives a semantics to
individual statements in the lockstep semantics. The rule relies on the
[\textsc{stage}] rule to apply the left rules for the statement before applying
the right rules. This design forces a specific composition of the rules in
order to achieve more tractable verification. For example, for a statement
{\assert{r}}, this rule will first apply the left rule for assertions, which
can be used to derive $\origvar{r} = true$. Note that this derivation occurs by
assumption as the logic assumes the validity of assertions in the reliable
execution. Next, the rule requires the proof to establish that $\relvar{r} =
true$. If, for example, the predicate $\origvar{r} = \relvar{r}$ is in the
context, then this proof obligation is easily established.

\paragraph{\bf If.} The rule [\textsc{if}] gives the semantics of \code{if}
statements.  The rule considers all cases of the execution of the statement. Specifically, the 
reliable and relaxed executions may proceed in lockstep or they may diverge by
proceeding down different branches. The logic captures this divergence by
leveraging the inverse staging rule to apply the Right Rules for the branch on which the
relaxed execution has taken before applying the Left Rules for the one which
the reliable version has taken.  Again, this forces a specific methodology for
reasoning about the programs in that the logic extracts the full availability
of assertions that may exist on the branch that the relaxed execution takes
before proceeding with the reliable execution.

\subsection{Properties}

{\TOOL}'s program logic ensures two basic properties of {\TOOL} programs: {\em
preservation} and {\em progress}. 
The preservation property states the partial correctness of the logic (but does
not not establish termination---and therefore total correctness).  The
progress property establishes that the relaxed execution of a program verified
with {\TOOL} satisfies all of its \code{assert} and \code{assume}
statements---provided that all reliable executions of the program also satisfy
the program's \code{assert} and \code{assume} statements.  We state these
properties formally below and provide proofs of these theorems in
\cref{sec:proofs}.

\begin{theorem}[Preservation]\ \\
If $\lockjudge[\mu]{P^*_r}{s}{Q^*_r}$ and $ (\env_1, \env_2) \models P^*_r$ 
and $\tuple{s, \env_1} \judge \env_1'$ and $\tuple{s, \env_2} \judge_{\mu}
  \env_2'$,  then $(\env_1', \env_2') \models Q^*_r$
\end{theorem}
{\TOOL}'s preservation property states that given a proof in the program logic
of a program $s$, for all pairs of environments $(\env_1, \env_2)$ that satisfy
the proof's precondition, if the executions of $s$ under both the reliable
semantics and the relaxed semantics terminate in a pair of environments
$(\env_1', \env_2')$, then this pair of environments satisfies the proof's
postcondition.

\begin{theorem}[Progress]\ \\
If $\lockjudge[\mu]{P^*_r}{s}{Q^*_r}$ and $ (\env_1, \env_2) \models P^*_r$ 
and $\tuple{s, \env_1} \judge \env_1'$ and $\tuple{s, \env_2} \judge_{\mu} v_2$, then $\neg \textit{failed}(v_2)$ where $\textit{failed}(\tuple{\code{fail}, \env}) = \textit{true}$
\end{theorem}
{\TOOL}'s progress property states that given a proof in the program logic of a
program $s$, for all pairs of environments $(\env_1, \env_2)$ that satisfy the
proof's precondition, if the reliable execution of $s$ terminates successfully,
then if the relaxed execution of $s$ under $\mu$ terminates, then it does not
terminate in an error.

\section{Implementation}
\TOOL's verification algorithm performs forward symbolic execution to
discharge verification conditions generated by \texttt{assert},
\texttt{assert_t}, \texttt{invariant}, and \texttt{invariant_r} statements in
the program. The algorithm directly implements the Hoare-style relational
program logic from Section~\ref{sec:logic}.
We present a detailed description of the algorithm in \cref{sec:verification}.
We also present a description of our loop invariant inference algorithm in
\cref{sec:houdini}.
The inference algorithm is modeled after Houdini~\cite{flanagan2001houdini},
and its base design is to infer common equivalence relations between programs.

{\TOOL} generates constraints to be solved by Microsoft's Z3 SMT solver
\cite{z3}. Our
system makes use of Z3's real, int, and bool types as well as uninterpreted
functions for arrays/matrices.  As such, our system does not necessarily generate a
set of constraints for which Z3 is complete. The practical impact of this
design is that it is possible for Z3 to be unable to verify valid constraints.
However, we have been able to successfully verify critical fault tolerance
properties for several applications as presented in the following section.

\section{Case Study: Jacobi Iterative Method}
\label{sec:example}

\cref{fig:jacobi} presents an implementation of
the Jacobi iterative method, alternatively Jacobi, in {\TOOL}.
The Jacobi iterative method is an algorithm for solving a system of linear
equations.
Specifically, given a matrix of coefficients \texttt{A} and a vector \texttt{b}
of intercepts, the algorithm computes a solution vector, \texttt{x}, where
$\texttt{A} * \texttt{x} = \texttt{b}$. The algorithm works iteratively by
computing successive approximations of \texttt{x}. For a system of two
equations (where \code{A} is a 2x2 matrix and both \code{b} and \code{x} are of
length two), Jacobi uses the solution vector from the previous iteration, $x^k$,
to produce the solution vector for the current iteration, $x^{k+1}$, using the
following approximation scheme:
\begin{align*}
     x_0^{k + 1}  &= (b_0 - A_{0,1} \cdot x_1^{k}) / A_{0,0} \\
     x_1^{k + 1}  &= (b_1 - A_{1,0} \cdot  x_0^{k}) / A_{1, 1}
\end{align*}
In words, for a given coordinate $x_i$, Jacobi approximates $x_i^{k+1}$, by substituting the values $x_j^{k}$, where $i \not =j$, into the linear equation for $i$, and solving for $x_i^{k+1}$.

Modulo floating-point rounding error, Jacobi converges to the correct \texttt{x} as the number of iterations goes to infinity.

\vspace{-1ex}
\paragraph{\bf Fault Tolerance.} Jacobi is {\em naturally self-stabilizing}.
Specifically, given an execution platform in which faults do not modify the
contents of \texttt{A}, then Jacobi is in a {\em valid state} at the end of
each iteration: if no additional faults occur during its execution, then Jacobi
will converge to the correct solution.

To understand this property intuitively, if an iteration produces an incorrect
solution vector, then the subsequent execution of the computation is equivalent
to having started the computation from scratch with the produced vector as the
initial starting point. Moreover, Jacobi enjoys the nice result that the change
in the number of iterations required to converge from the new starting point is
bounded logarithmically by the magnitude of the error in the solution vector.

Verifying  Jacobi for a given execution platform therefore poses two
challenges: 1) verifying that faults only affect the value of \texttt{x} and 2)
identifying a bound on the number of added iterations in the presence of a
fault.  Note that the latter determination not only serves as important
information for understanding if the implementation will meet the developer's
convergence requirements, but it also serves the practical purpose of setting
the maximum number of iterations such that a faulty execution will produce a
result that is at least as good as a fully reliable execution.

\vspace{-1ex}
\subsection{Multiplicative Single Event Upset Model}

\begin{wrapfigure}[18]{l}{0.52\linewidth}
  \vspace{-1em}
\begin{wrappedwraplst}
\begin{lstlisting}
const real E_REL = ...; ~\label{mseu:erel}~
bool upset = false; ~\label{mseu:upset}~

operator *(real x1, real x2) ~\label{mseu:reliable}~
   ensures (result == x1 * x2);

operator *(real x1, real x2) ~\label{mseu:unrel1}~
   when !upset && (0 < x1 * x2)
   modifies (upset)
   ensures upset &&
      ((1 - E_REL) * x1 * x2 <=
         result <=
         (1 + E_REL) * x1 * x2);

operator *(real x1, real x2) ~\label{mseu:unrel2}~
   when !upset && (x1 * x2 < 0)
   modifies (upset)
   ensures upset &&
           ((1 + E_REL) * x1 * x2 <=
              result <=
              (1 - E_REL) * x1 * x2);
\end{lstlisting}
\end{wrappedwraplst}
\vspace{-.4cm}
  \caption{Multiplicative SEU Execution Model}
\label{fig:mult-seu}
\end{wrapfigure}

We verify Jacobi under the multiplicative SEU model we present in
\cref{fig:mult-seu}.
The model exports three versions of the multiplication operator.
\cref{mseu:reliable} specifies the standard reliable implementation of
multiplication.
\cref{mseu:unrel1} and \cref{mseu:unrel2} each additionally specify an
unreliable implementation for the case where \code{x1 * x2} is positive or negative
respectively.  Each implementation models potentially unreliable
multipliers that are protected by truncated error correction
\cite{sullivan2012truncated, sullivan2013truncated}. Specifically, the
semantics of each unreliable version provides the guarantee that even in the
presence of an error, the result differs by at most \texttt{E_REL*100}\% of the
original result ~\cite{sullivan2012truncated, sullivan2013truncated}, where
\texttt{E_REL} is a constant the model specifies on \cref{mseu:erel}.

\vspace{-1ex}
\subsection{Jacobi Implementation}
\begin{figure}
\begin{wraplst}
\begin{lstlisting}[language=C,keywordstyle=\color{blue}, morekeywords={vector, matrix, invariant, variant, assume, relate, specvar, copy, relational_assume, bool, real, model, relational_assert, spec, requires, function, forall, assert_r},]
uint N; int iters; ~\label{code:decl1}~
matrix<real> A(N,N); vector<real> b(N); vector<real> x(N)~\label{code:decl2}~

for ( ; 0 <= iters; --iters) invariant_r !model.upset -> eq(x) ~\label{code:loop1}~
{
   vector<real> next_x(N);

   for (uint i = 0; i < N; ++i) ~\label{code:loop2}~
      invariant_r !model.upset -> eq(next_x)
      invariant_r bounded_diff(N, next_x)
   {
      real sum = 0;

      for (uint j = 0; j < N; ++j) invariant_r sig(sum) ~\label{code:loop3}~
      {
         if (i != j) { ~\label{code:if-prune}~
            real delta = A[i][j] *. x[j]; ~\label{code:mult}~

            if (E/model.E_REL-E <= abs(delta)) {~\label{code:dyn-check}~
               delta = A[i][j] * x[j]; ~\label{code:fix-mult}~
            }
         }

         sum = sum + delta; ~\label{code:plus}~
      }
      real num = b[i] - sum;
      next_x[i] = num / A[i][i];
   }

   x = next_x;

   assert_r eq(A); ~\label{code:assert-eq}~
   assert_r(bounded_diff(N, x)); ~\label{code:outer-assert}~
}
\end{lstlisting}
\end{wraplst}
\vspace{-.25cm}
  \caption{Jacobi Iterative Method.}
\label{fig:jacobi}
\vspace{-.25cm}
\end{figure}

The overall architecture of the implementation in \cref{fig:jacobi} is
that the outer loop on \cref{code:loop1} computes and stores the solution
vector for the current iteration into \texttt{next_x}. At the end of each
iteration, the implementation updates \texttt{x} by copying \texttt{next_x}
into \texttt{x}. The second loop on \cref{code:loop2} iterates through each
$x_i$ (stored at \texttt{x[i]}), sums the other terms in the $i$th equation
using the thrid loop (\cref{code:loop3}) into \code{sum}, and then computes
\texttt{x[i]} as the value \code{b[i] - sum/A[i][i]}.  We discuss the
definitions of the properties \texttt{sig} and \texttt{bounded\_diff} below.

All of the relaxation in Jacobi occurs in the third loop
(\cref{code:loop3}), where the execution model may corrupt the value of
\code{sum}.  The implementation first performs a relaxed multiplication
(\cref{code:mult}), then dynamically checks that the error in this
multiplication could not have exceeded the statically set bound \code{E} given a
relative error with a maximum percentage deviation from the correct value of
\code{model.E\_REL} (\cref{code:dyn-check}).
The notation \texttt{model.E\_REL} is a first-class reference to the state of
the execution model at that point in the program.
If the value of \code{abs(A[i][j] * x[j])} is less
than \code{E~/~model.E\_REL}, then a relative error cannot exceed the magnitude of
the absolute error bound \code{E} because the quantity \code{A[i][j] * x[j]} is
small.  However, checking this property requires the use of a reliable
multiplication, so we instead perform a relaxed multiplication and
conservatively approximate this property by instead checking that
\code{abs(delta)} is less than \code{E/model.E\_REL - E}.  If it was possible for an
error to exceed this bound, the algorithm repeats the multiplication reliably
(\cref{code:fix-mult}).

\subsection{Specification}
\label{sec:spec}

The verified Jacobi implementation relies on Leto's specification capabilities
to establish self-stability and verify a convergence bound.

\paragraph{\bf Self-Stability.}  To verify that this Jacobi implementation is
self-stabilizing the developer uses the \texttt{assert_r} statement on
~\cref{code:assert-eq} to assert \texttt{eq(A)}, which denotes that \texttt{A}
has the same value in both the original and relaxed execution.
 
This property therefore asserts that faults do not disturb the matrix of
coefficients and therefore precludes any execution models that may disturb the
contents of \texttt{A}.

\paragraph{\bf Convergence Bound.} Jacobi also enjoys a bound on the additional
number of iterations added to its execution given a fault. Specifically,
\(\Delta_c  = \mathcal{O}(\log_T(\frac{1}{(N \times \code{EPS})^2}))\) where
\(\Delta_c\) is the number of additional iterations in the relaxed execution,
\(N\) is the size of the \code{x} vector, \code{EPS} is the maximum
perturbation in each element of  the solution vector to due a fault in an
iteration, and \(T\) is a value between 0 and 1 representing the magnitude of
the non-diagonal elements of \(A\) relative to the magnitude of the
diagonal elements of \(A\). We specify that the maximum perturbation is bounded by \code{EPS}
with the \texttt{assert_r} on ~\cref{code:outer-assert}, which asserts that
\texttt{bounded\_diff(N, x)}.
The property \texttt{bounded\_diff} states that if an upset occurs during an
execution of the outer loop, then the error in each element of \code{x} is
bounded.
For clarity of presentation, we elide the full specification of
\texttt{bounded\_diff}.

\subsection{Verification Approach}

To verify Jacobi, the developer needs to provide a set of loop invariants that
structure the proof.

\vspace{-.15cm}
\paragraph{\bf Outer Loop.} The outermost loop on \cref{code:loop1} has a single
invariant specified by the developer: \code{!model.upset~->~eq(x)}.
Given our target execution model in
\cref{fig:mult-seu}, this invariant therefore states that if a fault has yet to
occur, then \texttt{x} is equivalent between both the original and relaxed
executions. This invariant follows because in the absence of a fault, Jacobi is
a deterministic computation for which any two executions (the original and
relaxed execution) that start from the same state compute the same result. By
default, {\TOOL} models the two executions as starting from the same state.
Therefore, all variables initialized in \cref{code:decl1}
and \cref{code:decl2} have the same values between the two executions.

\vspace{-.15cm}
\paragraph{\bf Middle Loop.} The middle loop on \cref{code:loop2} of has two
developer-specified invariants.  The first invariant states the behavior of the
loop if a fault has yet to occur in the program. In this case, \code{next_x}
is the same in both the original and relaxed executions (i.e.,
\code{eq(next_x)}). 

The second invariant is a key step towards one main proof goal:
\code{bounded\_diff(N,~next\_x)}, which states that if no fault has occurred
previously, then the error in each position of \texttt{next_x} is bounded if a
fault occurs on this iteration. Because the next step of the algorithm
directly assigns \code{next_x} to \code{x}, this must be an invariant of the
middle loop.

\vspace{-.15cm} \paragraph{\bf Inner Loop.} The second invariant of the middle
loop is critical for verifying the innermost loop.
The invariant \code{sig(sum)} verifies that if an error
occurs on this inner loop iteration, then \code{sum<r>} is within \code{E} of
\code{sum<o>}. Otherwise, \code{eq(sum)}. This property is true because if a
fault did not occur on the previous calculation of \code{x} in the outer loops
(as provided by the third invariant of the middle loop) then at most one fault
may happen during the calculation of each \code{delta} that contributes to
\code{sum} and the error in each \code{delta} is bounded by \code{E}.  Leto
verifies that the unreliable multiplication in combination with the
conservative check on \cref{code:dyn-check} establishes this fact.

\section{Evaluation}
\label{sec:case-studies}

We next present our results from using {\TOOL} to implement and verify several
self-stabilizing and self-correcting algorithms.
 
\subsection{Benchmarks and Properties} \cref{fig:effort} presents for
each benchmark (Column 1) the execution model we verified under (Column 2), and
the number of lines of code it contains (Column 3).

\paragraph{\bf Jacobi Iterative Method.} We verify the Jacobi benchmark as
presented in \cref{sec:example} under a multiplicative (SEU) error model (\cref{fig:mult-seu}).   

\paragraph{\bf Self-Stabilizing Steepest Descent (SS-SD)} SS-SD is another
iterative linear system of equations solver that employs a periodic, reliable
correction step \cite{sao2013self} to repair the state of the program.  We
verify that a developer can correctly implement the correction step using instruction
duplication (i.e., dual modular redundancy (DMR))  under an unbounded SEU
execution model (\cref{fig:seu}).  We present the full benchmark in
\cref{sec:sd}.

\paragraph{\bf Self-Stabilizing Conjugate Gradient Descent (SS-CG)} SS-CG is
an iterative linear system of equations solver that employs a periodic,
reliable correction step to repair the program state in the presence of
faults~\cite{sao2013self}.  We verify under an additive SEU error model
(\cref{fig:add-seu}) that errors are sufficiently small such
that the algorithm does not diverge. We also verify that the correction step
can be correctly implemented using instruction duplication.
We present a full description of the SS-CG in \cref{sec:cg}.

\paragraph{\bf Self-Correcting Connected-Components (SC-CC)} SC-CC is an
iterative algorithm for computing the connected subgraphs in a graph where each
iteration consists of a faulty initial computation step followed by a
correction step \cite{sao2016self}.  We verify that each iteration
computes the correct result under a Rowhammer~\cite{rowhammer} error model that
allows for an unbounded number of faulty writes to storage. We specifically
verify that the implementation detects and corrects all errors.
We present the full benchmark in \cref{sec:cc}.

\begin{figure}
  \begin{small}
\begin{tabular}{|c|c|c||c|c|} \hline
    \textbf{Benchmark} & \textbf{Execution Model} & \textbf{LOC}         & \textbf{Manual Annotations}      & \textbf{Invariants Inferred} \\\hline
    Jacobi & Multiplicative SEU   &  51 & 16 & 30 \\\hline
    SS-CG  & Additive SEU         & 167 & 22 & 36 \\\hline
    SS-SD  & Unbounded SEU        &  57 &  9 &  0 \\\hline
    SC-CC  & Switchable Rowhammer &  89 & 38 & 42 \\\hline
\end{tabular}
  \end{small}
\caption{
  Benchmark Verification Effort
}
  \vspace{-2ex}
\label{fig:effort}
\end{figure}

\subsection{Verification Effort}

\cref{fig:effort} also presents the annotation
burden \TOOL imposes on the programmer.  For each benchmark, we present the
number of manual annotations (Column 4) and the number of automatically
inferred loop invariants (Column 5).  Manual annotations include loop
invariants, assertions, and function requirements.  We consider each conjunct a
separate annotation when counting inferred invariants and manual annotations.

\paragraph{\bf Results.} We significantly reduce the number of
invariants we must provide using inference in all but one benchmark.  In half
of the cases we infer more invariants than we provide.  We infer no invariants
for SS-SD as Z3 very quickly runs out of memory on our
machine and therefore we must disable inference on all loops in that benchmark.
We believe that we could resolve this issue by monitoring the memory usage of
the Z3 subprocess, killing the process if it consumes too much, and falling
back on our weak inference algorithm.

\begin{figure*}
\begin{center}
\begin{tabular}{|c|c|c|c|} \hline
  \textbf{Benchmark} & \textbf{Time (s)} & \textbf{Memory Usage (kbytes)} & \textbf{Constraints Generated} \\\hline
    Jacobi &       37.79 &   36132 & 12473 \\\hline
    SS-CG  &        4.24 &   37836 & 11707 \\\hline
    SS-SD  &        0.19 &   25440 &   420 \\\hline
    SC-CC  &  \worstTime &  199428 &  4200 \\\hline
\end{tabular}
\caption{
  Benchmark Runtime Characteristics
}
  \vspace{-2ex}
\label{fig:perf}
\end{center}
\end{figure*}

\paragraph{\bf Runtime Characteristics.} \cref{fig:perf}
presents also the runtime performance characteristics of the \TOOL C++
implementation.  We ran our experiments on an Intel i5-5200U CPU clocked at
2.20GHz with 8 GB of RAM. For each benchmark we present the time it took to run
in seconds (Column 2), the maximum memory usage in kilobytes (Column 3), and
the number of constraints generated for use with Z3 (Column 4).

\section{Related Work}

\newcommand{\most}{\algoref,\softref,\misc,\concern}

\paragraph{\bf Analyzing Approximate Computation.}

Researchers have developed a number of programming systems that enable
developers to reason about {\em approximate computations}: computations for
which the underlying execution substrate (e.g., the programming system and/or
hardware system) augments the behavior of the application to produce approximate
results. For example, EnerJ~\cite{sampson11} and FlexJava~\cite{park15} enable
developers to demonstrate non-interference between approximate computations and
critical parts of the computation that should not be modified.  Meola and
Walker propose a sub-structural logic for reasoning about fault tolerant
programs~\cite{meola10}. Their logic enables the proof system to count the
number of faults that have occurred and therefore reason about properties that
may hold for one model but not another.   In
contrast to all of these approaches, {\TOOL} provides a more expressive and
unconstrained logic that supports verifying complicated relational properties
of the application.

The relaxed programming model~\cite{pldi12} enables developers to prove both
safety and accuracy programs for relaxed computations by hand using a Coq
library.  In contrast, {\TOOL} automates many aspects of the proof. Further,
{\TOOL}'s first-class execution models enable the developer to automatically
weave in an execution model---which may impact many operations in the
program---whereas the relaxed programming framework requires that operations be
modelled individually by hand.

He et al.~\citep{he2016verifying} leverage the symdiff framework~\cite{symdiff}
to verify instances of approximate programs.
However, it does not enable a developer to integrate an execution model
without directly changing the application's logic.
\TOOL enables modular specification of execution models.

\vspace{-.15cm}
\paragraph{\bf Relational Hoare Logic.} Researchers have proposed a number of
relational Hoare Logics and verification systems to support verifying
relational properties of programs~\cite{benton04, barthe11, pldi12, lahiri12,
sousa16}.   The verification algorithms produced by Sousa
and Dillig~\cite{sousa16} and Lahiri et al.~\cite{lahiri12} demonstrate that it
is possible to automatically compose proofs for relational verification.
{\TOOL}'s verification system differs from that of CHL in that 1)~the semantics
of the two program executions are asymmetric and 2) {\TOOL} attempts to verify
with a specific program composition strategy that matches the types of proofs
that are seen in practice for approximate and unreliably executed programs.
Namely, although the semantics of the two executions of the program differ,
their structure is typically identical and therefore \code{assert} and
\code{assumes} can often be matched to enable maximum reuse of assumed
properties of the reliable execution during the verification of the relaxed.

\vspace{-.15cm}
\paragraph{\bf Type Systems for Self-Stabilization.}  
Self-Stabilizing Java provides developers with a type system and analysis that
enables a developer to prove that any corrupted state of the program exits
the system in a finite amount of time.    {\TOOL}'s logic (versus the information-flow type system of Self-Stabilizing Java)
enables developers to specify the richer invariants that need to be true of
emerging algorithms for self-stability. For example, instead of verifying that
corrupted state leaves the system within bounded time, {\TOOL} enables a
developer to verify that the corruption in the program's state is small enough
that the algorithm's correction steps will work as designed.

\vspace{-.15cm}
\paragraph{\bf Fault Rate Analysis}

We have driven the design and implementation of Leto by the anticipated fault
rates, abstract fault models, and resilience tools exported by the computer
architecture, high-performance computing, and fault tolerance communities.
Specifically, soft fault rates have led major organizations -- such as
Intel~\citep{kurd2010westmere,mitra2005robust,mitra2006combinational,mukherjee2003systematic},
Google~\citep{yim2014characterization}, NASA~\citep{johnston2000scaling},
DOE~\cite{exascale14}, and DARPA~\citep{amarasinghe09} -- to express concern
over such faults.  Leto is the first system -- to our knowledge -- to enable
automated verification for these faults.

The assumption of instruction-level arithmetic errors is the most common model
for building 1) application-specific fault analyses and
mechanisms~\citep{\algoref}, 2) software-level fault tolerance analyses and
mechanisms~\citep{\softref}, 3) micro-architectural resilience analyses and
mechanisms ~\cite{meixner2007argus,lu1982watchdog,austin1999diva}, and 4)
circuit-level resilience
analyses/mechanisms~\cite{kelin2010leap,lilja2013single,quinn2015frequency,quinn2015heavy,turowski201532nm,bowman2009energy,bowman201145}.

Mitra et al. have found that combinational logic faults account for 11\%
of all soft errors~\citep{mitra2005robust}. In addition, soft error rates,
including combinational faults, are expected to increase as chips
continue grow in the number of
transistors~\cite{shivakumar2002modeling,mitra2005robust}. These trends have inspired
a variety of different contributions, including modeling the propagation
of transient faults~\citep{chen2012efficient,omana2003model}, analyzing the
rate of combinational soft
faults~\citep{buchner1997comparison,rao2007computing,wang2011soft,zhang2006soft}, analyzing the impact of
combinational soft faults~\citep{rajaraman2006seat}, and correcting
combinational logic faults~\citep{mitra2006combinational}.

\section{Conclusion}

\begin{sloppypar}
Emerging computational platforms are increasingly vulnerable to errors. Future
computations designed to execute on these platforms must therefore be designed
to be {\em fault tolerant} and naturally resilient to error. We present a
verification system, \TOOL, that facilitates the verification of
application-specific fault tolerance mechanisms under programmer-specified
execution models. As these proofs frequently relate a faulty execution to a
fault-free one, \TOOL provides assertions that enable the developer to specify
and verify expressions that relate the semantics of both executions.
First-class execution models permit developers to convey information about the
class of faults they expect their computational platforms to experience.  By
giving developers tools to verify relational invariants under first-class
execution models, we enable developers to verify the self-stability  of their
programs.
\end{sloppypar}

\bibliography{references,demonstrate-concern,misc,use-assumption,validate-instruction-level}

\clearpage

\appendix

\section{Additional Proof Rules}
\label{sec:rules}

Figures~\ref{fig:additional-relaxed-logic}, \ref{fig:control-logic},
and~\ref{fig:additional-lockstep-logic} expand on the proof rules we presented
in \cref{sec:logic}.

\begin{figure*}
\begin{footnotesize}
\begin{mathpar}
 
\inferrule[read-l]
{ }
{
   \ljudge{ Q_r^*[\injo{x} / \injo{r}]}{\assign{r}{x}}{Q_r^*}
}

\inferrule[read-r]
{
   \fresh{r'} \\ 
   \exists ([x_\textit{mem}], X, P^*_w, P^*_e) \in \mu(\oplus) \cdot \injr{P^*_w[x / x_\textit{mem}]} \\
   Q'^*_r = \left(
   \bigvee_{([x_{\textit{mem}}], X, P^*_w, P^*_e) \in \mu(\texttt{read})}
   \injr{P^*_w[x / x_{\textit{mem}}]} \rightarrow
   \injr{P^*_e[x / x_{\textit{mem}}]
              [\forall x_m \in X \cdot \fresh{x_m'} / x_m]
              [ r' / \textit{result}]} \right)
}
{
   \reljudge[\mu]{Q'^*_r \land Q_r^*[\injr{r'} / \injr{r}]}{\assign{r}{x}}{Q_r^*}
}

\inferrule[write-l]
{ }
{
   \ljudge{Q_r^*[\injo{r} / \injo{x}]}{\assign{x}{r}}{Q_r^*}
}

\inferrule[write-r]
{
   \fresh{r'} \\ 
   \exists ([x_1, x_2], X, P^*_w, P^*_e) \in \mu(\oplus) \cdot  \injr{P^*_w[x / x_1][r / x_2]} \\
   Q'^*_r = \left( \bigvee_{([x_1, x_2], X,
   P^*_w, P^*_e) \in \mu(\texttt{write})}
   \injr{P^*_w[x / x_1][ r / x_2]} \rightarrow
   \injr{P^*_e[r' / x_1][r / x_2][\forall x_m \in X \cdot \fresh{x_m'} / x_m]} \right)
}
{
   \reljudge[\mu]{Q'^*_r \land Q_r^*[ \injr{r'} / \injr{x}]}{\assign{x}{r}}{Q_r^*}
}

\inferrule[relational-assert-l]
{ }
{\ljudge{\true}{\relassert{r}}{\true}}

\inferrule[relational-assert-r]
{ }
{\reljudge[\mu]{r}{\relassert{r}}{r}}
\end{mathpar}
\end{footnotesize}
  \caption{Left and Right Rules for Primitive Statements}
\label{fig:additional-relaxed-logic}
\end{figure*}

\subsection{Additional Left and Right Rules}
\cref{fig:additional-relaxed-logic} presents the remainder of our left and
right rules.

\paragraph{\bf Read.} The rules [\textsc{read-l}] and [\textsc{read-r}] give the semantics of reads from memory. The left rule [\textsc{read-l}] mimics the behavior of [\textsc{read-l}] with the primary differing being that it substitutes the value of a register, $r$, for a local variable $x$. The right rule,  [\textsc{read-r}], on the other hand more closely resembles  [\textsc{binop-r}] in that it models the potentially unreliable execution of the read from memory. 

\paragraph{\bf Write. } The rules [\textsc{write-l}] and [\textsc{write-r}] give the semantics of writes to memory. These rules are analogous to [\textsc{read-l}] and [\textsc{read-r}], except modified in their exact implementation to captures writes to memory.

\paragraph{\bf Relational Assert}
The rules [\textsc{relational-assert-l}] and [\textsc{relational-assert-r}]
give the semantics of relational assertion statements.
These statements are similar to my normal assert statements, but they take
relational predicates as arguments rather than standard predicates.
The logic requires that the condition of a relational assertion is verified in
relaxed execution, but in reliable execution the logic treats these statements
as no-ops.

\begin{figure*}
\centering
\begin{mathpar}
 
\inferrule[seq-l]
{
 \ljudge{P_r^*}{s_1}{R_r^*} \\
 \ljudge{R_r^*}{s_2}{Q_r^*}
}
{
   \ljudge{P_r^*}{\seq{s_1}{s_2}}{Q_r^*}
}

\inferrule[seq-r]
{
 \reljudge[\mu]{P_r^*}{s_1}{R_r^*} \\
 \reljudge[\mu]{R_r^*}{s_2}{Q_r^*}
}
{
 \reljudge[\mu]{P_r^*}{\seq{s_1}{s_2}}{Q_r^*}
}

\inferrule[if-l]
{
   b \equiv \assign{\injo{r}}{\true} \\
   \ljudge{P_r^* \land b}{s}{Q_r^*} \\
   \ljudge{P_r^* \land \neg b}{s}{Q_r^*}
}
{
   \ljudge{P_r^*}{\ifthen{r}{s_1}{s_2}}{Q_r^*}
}

\inferrule[if-r]
{
   b \equiv \assign{\injo{r}}{\true} \\
   \reljudge[\mu]{P_r^* \land b}{s}{Q_r^*} \\
   \reljudge[\mu]{P_r^* \land \neg b}{s}{Q_r^*}
}
{
   \reljudge[\mu]{P_r^*}{\ifthen{r}{s_1}{s_2}}{Q_r^*}
}

\inferrule[while-l]
{
   b \equiv \assign{\injo{r}}{\true} \\
}
{
   \ljudge{\true}{\whileR{r}{P_r^*}{s}}{P_r^* \land \neg b}
}

\inferrule[while-r]
{
   b \equiv \assign{\injo{r}}{\true} \\
   \reljudge[\mu]{P_r^* \land b}{s}{P_r^*} \\
}
{
   \reljudge[\mu]{P_r^*}{\whileR{r}{P_r^*}{s}}{P_r^* \land \neg b}
}

\end{mathpar}
\caption{Left and Right Rules for Structure and Control Flow}
\label{fig:control-logic}
\end{figure*}
\subsection{Left and Right Rules for Control Flow}
Figure~\ref{fig:control-logic} presents the left and right rules for control
flow statements.
With the exception of [\textsc{while-l}], the rules adhere to the standard
formalization as seen in traditional Hoare logic.
The only distinction between these rules and their standard implementation is
that they operate over relational predicates.

The rule [\textsc{while-l}] assumes that \(P_r^* \land \lnot b\) holds after the
while loop and does not place any constraints on the environment prior to the
loop.
In other words, I do not verify the invariants on loops that run under left
semantics.

\begin{figure*}
\centering
\begin{mathpar}
 
\inferrule[weak]
{
  P_r^* \models P'^*_r \\
  \lockjudge[\mu]{P'^*_r}{s}{Q'^*_r}  \\
  Q'^*_r \models Q_r^*
}
{
   \lockjudge[\mu]{P_r^*}{s}{Q_r^*}
}

\inferrule[while]
{
   b \equiv \assign{r}{\true} \\
   \lockjudge{R^* \land \injo{b} \land \injr{b}}{s}{R^*} \\
   \ljudge{R^* \land \injo{b} \land \neg\injr{b}}{s}{R^*} \\
   \reljudge[\mu]{R^* \land \neg\injo{b} \land \injr{b}}{ s}{R^*}
}
{
   \lockjudge[\mu]{R^*}{\whileR{r}{R^*}{s}}{R^* \land \neg\injo{b} \land \neg\injr{b}}
}
\end{mathpar}
\caption{Lockstep Control Flow and Structural Rules}
\label{fig:additional-lockstep-logic}
\end{figure*}

\subsection{Additional Lockstep Rules}
\cref{fig:additional-lockstep-logic} presents the remainder of our lockstep
control flow and structure rules.

\paragraph{\bf Weakening.} The rule  [\textsc{weak}] gives the standard semantics for weakening as found in the standard Hoare logic with the distinction that it operates on relational predicates.

\paragraph{\bf While.}  The rule [\textsc{while}] gives the semantics of
\code{while} statements. The rule is similar in design to the rule for
\code{if} statements in that it must also consider cases in which the control
flow of the two executions diverge. The rule first considers the case where the
two executions proceed in lockstep by both executing an iteration of the loop.
The next two cases leverage the left and right rules to consider the cases
when:
\begin{itemize}
  \item The relaxed execution halts, but the reliable execution executes an
    additional iteration.
  \item The reliable execution halts, but the relaxed execution executes an additional iteration, respectively.
\end{itemize}

\newcommand{\failed}[1]{\textit{failed}(#1)}

\section{Proofs of Properties}
\label{sec:proofs}

{\TOOL}'s program logic ensures two basic properties of {\TOOL} programs: {\em preservation} and {\em progress}.

\begin{sloppypar}
\begin{lemma}[Left Preservation]\ \\
  \label{lem:left-pres}
  If $\ljudge{P_r^*}{s}{Q_r^*}$ and \((\env_1, \env_2) \models P_r^*\) and
  $\tuple{s, \env_1} \judge \env_1'$,  then $(\env_1', \env_2) \models Q_r^*$.
\end{lemma}
  \TOOL's left preservation property states that given a proof in the left
  rules, for a pair of environments satisfying the precondition of the proof,
  then if a reliable execution of $s$ terminates, then the resulting
  environment pair satisfies the proof's postcondition.

\begin{proof}
  By induction on the lemma statement:
  \begin{itemize}
    \item \textsc{Assign-l}.
      Stepping \(\tuple{s, \env_1}\) produces an environment \(\env_1'\) which
      differs from \(\env_1\) only in that \(r\) is mapped to \(n\) in
      \(\frame\).
      As \(\env_1 \models Q_r^*[n / \injo{r}]\), \(\env_1'\) trivially satisfies
      \(Q_r^*\).
      Additionally, as the precondition contains no substitutions on
      \(\injr{r}\), \(\env_2\) trivially satisfies \(Q_r^*\).
    \item \textsc{binop-l}.
      Similar to \textsc{assign-l}.
    \item \textsc{read-l}.
      Similar to \textsc{assign-l}.
    \item \textsc{write-l}.
      Similar to \textsc{assign-l}.
    \item \textsc{assert-l}.
      Stepping \(\tuple{s, \env_1}\) by definition adds \(r\) to \(\env_1\) to
      form \(\env_1'\).
      Therefore, \((\env_1', \env_2) \models \injo{r}\).
    \item \textsc{assume-l}.
      Similar to \textsc{assert-l}.
    \item \textsc{relational-assert-l}.
      Relational assertions in left mode do nothing.
    \item \textsc{seq-l}.
      We start with inversion on \(\ljudge{P_r^*}{s}{Q_r^*}\), which yields
      \(\ljudge{P_r^*}{s_1}{R_r^*}\) and \(\ljudge{R_r^*}{s_2}{Q_r^*}\).
      Applying the induction hypothesis to \(\ljudge{P_r^*}{s_1}{R_r^*}\) yields
      \((\env_1'', \env_2) \models R_r^*\).
      Applying the induction hypothesis to \(\ljudge{R_r^*}{s_2}{Q_r^*}\) yields
      \((\env_1', \env_2) \models Q_r^*\).
    \item \textsc{while-l}.
      Stepping \(\tuple{s, \env_1}\) by definition adds \(P_r^*\) to \(\env_1\) to
      form \(\env_1'\).
      Therefore, \((\env_1', \env_2) \models P_r^*\).
    \item \textsc{if-l}.
      First, we perform inversion on \(s\).
      Then, we destruct \(b\).
      In the case where \(b\) is \(\true\), we apply the induction hypothesis
      to \(\ljudge{P_r^* \land b}{s}{Q_r^*}\).
      In the case where \(b\) is \(\false\), we apply the induction hypothesis
      to \(\ljudge{P_r^* \land \lnot b}{s}{Q_r^*}\).
  \end{itemize}
\end{proof}

\begin{lemma}[Right Preservation]\ \\
  \label{lem:right-pres}
  If $\reljudge[\mu]{P_r^*}{s}{Q_r^*}$ and \((\env_1, \env_2) \models P_r^*\) and
  $\tuple{s, \env_2} \judge_\mu \env_2'$,  then $(\env_1, \env_2') \models Q_r^*$.
\end{lemma}
  \TOOL's right preservation property states that given a proof in the right
  rules, for a pair of environments satisfying the precondition of the proof,
  then if execution of $s$ under the execution model specification \(\mu\)
  terminates, then the resulting environment pair satisfies the proof's
  postcondition.

\begin{proof}
  By induction on the lemma statement:
  \begin{itemize}
    \item \textsc{assign-r}.
      Stepping \(\tuple{s, \env_2}\) produces an environment \(\env_2'\) which
      differs from \(\env_2\) only in that \(r\) is mapped to \(n\) in
      \(\frame\).
      As \(\env_1 \models Q_r^*[n / \injo{r}]\), \(\env_2'\) trivially satisfies
      \(Q_r^*\).
      Additionally, as the precondition contains no substitutions on
      \(\injr{r}\), \(\env_1\) trivially satisfies \(Q_r^*\).
    \item \textsc{binop-r}.
      Stepping \(\tuple{s, \env_2}\) produces an environment \(\env_2'\) which
      differs from \(\env_2\) only in that \(r\) is mapped to \(r'\) in
      \(\frame\).
      The restrictions placed on \(r'\) in \(Q'^*_r\) codify the operator
      substitution routine from \textsc{f-binop} in the operational semantics.
      Therefore, the runtime always sets \(r\) in a way such that the
      resulting environment satisfies \(Q_r^*\).
      Additionally, as the precondition contains no substitutions on
      \(\injr{r}\), \(\env_1\) trivially satisfies \(Q_r^*\).
    \item \textsc{read-r}.
      Similar to \textsc{binop-r}.
    \item \textsc{write-r}.
      Similar to \textsc{binop-r}.
    \item \textsc{assert-r}.
      Stepping \(\tuple{s, \env_2}\) does not modify \(\env_2\).
      Therefore, \(\env_2 = \env_2'\).
      Since the precondition and postcondition are identical, \((\env_1,
      \env_2') \models r\).
    \item \textsc{assume-r}.
      Similar to \textsc{assert-r}.
    \item \textsc{relational-assert-r}.
      Similar to \textsc{assert-r}.
    \item \textsc{seq-r}.
      We start with inversion on \(\reljudge{P_r^*}{s}{Q_r^*}\), which yields
      \(\reljudge{P_r^*}{s_1}{R_r^*}\) and \(\reljudge{R_r^*}{s_2}{Q_r^*}\).
      Applying the induction hypothesis to \(\reljudge{P_r^*}{s_1}{R_r^*}\) yields
      \((\env_1, \env_2'') \models R_r^*\).
      Applying the induction hypothesis to \(\reljudge{R_r^*}{s_2}{Q_r^*}\) yields
      \((\env_1, \env_2') \models Q_r^*\).
    \item \textsc{while-r}.
      First, we perform inversion on \(s\).
      Then, we destruct \(b\).
      In the case where \(b\) is \(\true\), we apply the induction hypothesis to
      \(\reljudge{P_r^* \land b}{s}{P_r^*}\).
      This proves that \(P_r^*\) holds after each loop iteration.
      Upon exiting the loop, \(\lnot b\) trivially holds.
      In the case where \(b\) is \(\false\), the loop does not run and
      therefore \((\env_1, \env_2')\) trivially satisfies \(P_r^* \land \lnot
      b\).
    \item \textsc{if-r}.
      First, we perform inversion on \(s\).
      Then, we destruct \(b\).
      In the case where \(b\) is \(\true\), we apply the induction hypothesis to
      \(\reljudge{P_r^* \land b}{s}{Q_r^*}\).
      In the case where \(b\) is \(\false\), we apply the induction hypothesis
      to \(\reljudge{P_r^* \land \lnot b}{s}{Q_r^*}\).
  \end{itemize}
\end{proof}

\begin{theorem}[Preservation]\ \\
If $\lockjudge[\mu]{P_r^*}{s}{Q_r^*}$ and $ (\env_1, \env_2) \models P_r^*$ 
and $\tuple{s, \env_1} \judge \env_1'$ and $\tuple{s, \env_2} \judge_{\mu} \env_2'$,  then $(\env_1', \env_2') \models Q_r^*$
\end{theorem}

{\TOOL}'s preservation property states that given a proof in the program logic of a program $s$, for all pairs of environments $(\env_1, \env_2)$ that satisfy the
proof's precondition, if the executions of $s$ under both the reliable semantics and the relaxed semantics terminate in a pair of environments $(\env_1', \env_2')$, then
this pair of environments satisfies the proof's postcondition. Note that this states the partial correctness of the logic and does not establish termination (and therefore total correctness).

\begin{proof}
  By induction on the theorem statement:
  \begin{itemize}
    \item
      \textsc{stage}.
      We first perform inversion on \(s\).
      Then, we apply \autoref{lem:left-pres} to \(\ljudge{P_r^*}{s_1}{R_r^*}\) and
      \autoref{lem:right-pres} to \(\reljudge{R_r^*}{s_2}{Q_r^*}\).
    \item
      \textsc{inverse-stage}.
      We first perform inversion on \(s\).
      Then, we apply \autoref{lem:right-pres} to \(\reljudge{P_r^*}{s_2}{R_r^*}\) and
      \autoref{lem:left-pres} to \(\ljudge{R_r^*}{s_1}{Q_r^*}\).
    \item \textsc{split}.
      We first perform inversion on \(s\),
      followed by inversion on \(\lockjudge[\mu]{P_r^*}{s \sim s}{Q_r^*}\).
      Then, we apply \autoref{lem:left-pres} to \(\ljudge{P_r^*}{s}{R_r^*}\) and
      \autoref{lem:right-pres} to \(\reljudge{R_r^*}{s}{Q_r^*}\).
    \item \textsc{seq}.
      We start with inversion on \(\lockjudge[\mu]{P^*}{s}{Q^*}\), which yields
      \(\lockjudge[\mu]{P^*}{s_1}{R^*}\) and \(\lockjudge[\mu]{R^*}{s_2}{Q^*}\).
      Applying the induction hypothesis to \(\lockjudge[\mu]{P^*}{s_1}{R^*}\) yields
      \((\env_1'', \env_2'') \models R^*\).
      Applying the induction hypothesis to \(\lockjudge[\mu]{R^*}{s_2}{Q^*}\) yields
      \((\env_1', \env_2') \models Q^*\).
    \item \textsc{weak}.
      We start with inversion on \(\lockjudge[\mu]{P_r^*}{s}{Q_r^*}\).
      Applying the induction hypothesis to \(\lockjudge[\mu]{P'^*_r}{s}{Q'^*_r}\) yields
      \((\env_1', \env_2') \models Q'^*_r\).
      Since \(Q'^*_r \models Q_r^*\), \((\env_1', \env_2') \vdash Q_r^*\).
    \item \textsc{if}.
      We begin with inversion on \(s\).
      Then, we destruct \(b\).
      In the case where \(\injo{b} \land \injr{b}\), we apply the induction
      hypothesis to \(\lockjudge[\mu]{P^* \land \injo{b} \land \injr{b}}{s_1}{Q^*}\).
      In the case where \(\lnot\injo{b} \land \injr{b}\), we apply the induction
      hypothesis to \[\lockjudge[\mu]{P^* \land \lnot\injo{b} \land \injr{b}}{s_2
      \sim_r s_1}{Q^*}.\]
      In the case where \(\injo{b} \land \lnot\injr{b}\), we apply the induction
      hypothesis to \[\lockjudge[\mu]{P^* \land \injo{b} \land \lnot\injr{b}}{s_1
      \sim_r s_2}{Q^*}.\]
      In the case where \(\lnot\injo{b} \land \lnot\injr{b}\), we apply the induction
      hypothesis to \[\lockjudge[\mu]{P^* \land \lnot\injo{b} \land
      \lnot\injr{b}}{s_2}{Q^*}.\]
    \item \textsc{while}.
      First, we perform inversion on \(s\).
      Then, we destruct \(b\).
      In the case where \(\injo{b} \land \injr{b}\), we apply the induction
      hypothesis to \(\lockjudge[\mu]{R^* \land \injo{b} \land \injr{b}}{s}{R^*}\).
      This proves that \(R^*\) holds after each loop  iteration.
      Upon exiting the loop, \(\lnot\injo{b} \land \lnot\injr{b}\) trivially
      holds.
      The cases where \(\lnot\injo{b} \land \injr{b}\) and \(\injo{b} \land
      \lnot\injr{b}\) are similar to the previous case.
      In the case where \(\lnot\injo{b} \land \lnot\injr{b}\), the loop does
      not run in either execution and therefore \((\env_1', \env_2')\)
      trivially satisfies the postcondition.
  \end{itemize}
\end{proof}

\begin{lemma}[Right Progress]\ \\
  \label{lem:right-prog}
If $\reljudge[\mu]{P_r^*}{s}{Q_r^*}$ and $ (\env_1, \env_2) \models P_r^*$
and $\tuple{s, \env_2} \judge_{\mu} \env_2'$, then
  $\neg \textit{failed}(\env_2')$ where
  $\textit{failed}(\tuple{\code{fail}, \env}) = \textit{true}$
\end{lemma}

{\TOOL}'s right progress property states that given a proof in the right rules
of a program $s$, for all pairs of environments $(\env_1, \env_2)$ that satisfy
the proof's precondition, then if the relaxed execution of $s$ under $\mu$
terminates, then it does not terminate in an error.
The right progress property establishes that a {\TOOL}'s program of right rules
satisfies all of its \code{assert} and \code{assume} statements.

\begin{proof}
  By induction on the theorem statement:
  \begin{itemize}
    \item \textsc{assign-r}.
      Assignment cannot fail.
    \item \textsc{binop-r}.
      We start with inversion on \(s\),
      giving us the fact that the dynamic semantics is always able to
      perform an operator substitution
      (\(\exists ([x_1, x_2], X, P^*_w, P^*_e) \in \mu(\oplus) \cdot
      \injr{P^*_w[r_1 / x_1][r_2 / x_2]}\)), so binary operations cannot fail.
    \item \textsc{read-r}.
      We start with inversion on \(s\),
      giving us the fact that the dynamic semantics is always able to
      perform an operator substitution
      (\(\exists ([x_\textit{mem}], X, P^*_w, P^*_e) \in \mu(\oplus) \cdot
       \injr{P^*_w[x / x_\textit{mem}]}\)),
      so reads cannot fail.
    \item \textsc{write-r}.
      We start with inversion on \(s\),
      giving us the fact that the dynamic semantics is always able to
      perform an operator substitution
      (\(\exists ([x_1, x_2], X, P^*_w, P^*_e) \in \mu(\oplus) \cdot
      \injr{P^*_w[x / x_1][r / x_2]}\)),
      so writes cannot fail.
    \item \textsc{assert-r}.
      Stepping \(\tuple{s, \env_2}\) produces the environment \(\env_2'\) where
      \(\env_2 = \env_2'\).
      Since \(\lnot\failed{\env_2}\), \(\lnot\failed{\env_2'}\) trivially
      holds.
    \item \textsc{assume-r}.
      Similar to \textsc{assert-r}.
    \item \textsc{relational-assert-r}.
      Similar to \textsc{assert-r}.
    \item
      \begin{sloppypar}
        \textsc{seq-r}.
        We start with inversion on \(\reljudge{P_r^*}{s}{Q_r^*}\), which yields
        \(\reljudge{P_r^*}{s_1}{R_r^*}\) and \(\reljudge{R_r^*}{s_2}{Q_r^*}\).
        Applying \autoref{lem:right-pres} to \(\reljudge{P_r^*}{s_1}{R_r^*}\) yields
        \((\env_1, \env_2'') \models R_r^*\).
        Applying the induction hypothesis to \(\reljudge{P_r^*}{s_1}{R_r^*}\) provides
        \(\lnot\failed{\env_2''}\).
        Applying the induction hypothesis to \(\reljudge{R_r^*}{s_2}{Q_r^*}\) yields
        \(\lnot\failed{\env_2'}\).
      \end{sloppypar}
    \item \textsc{while-r}.
      First, we perform inversion on \(s\).
      Then, we destruct \(b\).
      In the case where \(b\) is \(\true\), we apply the induction hypothesis to
      \(\reljudge{P_r^* \land b}{s}{P_r^*}\).

      In the case where \(b\) is \(\false\), the loop does not run and
      therefore \((\env_2 = \env_2')\) trivially satisfies
      \(\lnot\failed{\env_2'}\).

      Lastly, since \((\env_1, \env_2) \models P_r^*\), the invariant check before
      the loop cannot fail.
    \item \textsc{if-r}.
      First, we perform inversion on \(s\).
      Then, we destruct \(b\).
      In the case where \(b\) is \(\true\), we apply the induction hypothesis to
      \(\reljudge{P_r^* \land b}{s}{Q_r^*}\).
      In the case where \(b\) is \(\false\), we apply the induction hypothesis
      to \(\reljudge{P_r^* \land \lnot b}{s}{Q_r^*}\).
  \end{itemize}
\end{proof}

\begin{theorem}[Progress]\ \\
If $\lockjudge[\mu]{P_r^*}{s}{Q_r^*}$ and $ (\env_1, \env_2) \models P_r^*$
and $\tuple{s, \env_1} \judge \env_1'$ and $\tuple{s, \env_2} \judge_{\mu}
  \env_2'$, then $\neg \textit{failed}(\env_2')$
where $\textit{failed}(\tuple{\code{fail}, \env}) = \textit{true}$
\end{theorem}

{\TOOL}'s progress property states that given a proof in the program logic of a
program $s$, for all pairs of environments $(\env_1, \env_2)$ that satisfy the
proof's precondition, if the reliable execution of $s$ terminates successfully,
then if the relaxed execution of $s$ under $\mu$ terminates, then it does not
terminate in an error.  The progress property establishes that a {\TOOL}'s
program satisfies all of its \code{assert} and \code{assume} statements --
provided that all reliable executions of the program also satisfy the program's
\code{assert} and \code{assume} statements.

\begin{proof}
  By induction on the theorem statement:
  \begin{itemize}
    \item
      \textsc{stage}.
      We first perform inversion on \(s\).
      Then, we apply \autoref{lem:left-pres} to \(\ljudge{P_r^*}{s_1}{R_r^*}\) and
      \autoref{lem:right-prog} to \(\reljudge{R_r^*}{s_2}{Q_r^*}\).
    \item
      \textsc{inverse-stage}.
      We first perform inversion on \(s\).
      Then, we apply \autoref{lem:right-prog} to \(\reljudge{P_r^*}{s_2}{R_r^*}\) and
      \autoref{lem:left-pres} to \(\ljudge{R_r^*}{s_1}{Q_r^*}\).
    \item
      \textsc{split}.
      We first perform inversion on \(s\), followed by
      inversion on \(\lockjudge[\mu]{P_r^*}{s \sim s}{Q_r^*}\).
      Then, we apply \autoref{lem:left-pres} to \(\ljudge{P_r^*}{s_1}{R_r^*}\).
      Finally, we apply \autoref{lem:right-prog} to \(\reljudge{R_r^*}{s_2}{Q_r^*}\).
    \item \textsc{seq}.
      We start with inversion on \(\lockjudge[\mu]{P^*}{s}{Q^*}\), which yields
      \(\lockjudge[\mu]{P^*}{s_1}{R^*}\) and \(\lockjudge[\mu]{R^*}{s_2}{Q^*}\).
      Applying the induction hypothesis to \(\lockjudge[\mu]{P^*}{s_1}{R^*}\) yields
      \(\lnot\textit{failed}(\env_2'')\).
      Applying the induction hypothesis to \(\lockjudge[\mu]{R^*}{s_2}{Q^*}\) yields
      \(\lnot\textit{failed}(\env_2')\).
    \item \textsc{weak}.
      We start with inversion on \(\lockjudge[\mu]{P_r^*}{s}{Q_r^*}\).
      Applying the induction hypothesis to \(\lockjudge[\mu]{P'^*_r}{s}{Q'^*_r}\) yields
      \(\lnot\textit{failed}(\env_2'')\).
      Since \(P_r^* \models P'^*_r\) and \(Q'^*_r \models Q_r^*\),
      \(\lnot\textit{failed}(\env_2'') \limp \lnot\textit{failed}(\env_2')\).
    \item \textsc{if}.
      We begin with inversion on \(s\).
      Then, we destruct \(b\).
      In the case where \(\injo{b} \land \injr{b}\), we apply the induction
      hypothesis to \(\lockjudge[\mu]{P^* \land \injo{b} \land \injr{b}}{s_1}{Q^*}\).
      In the case where \(\lnot\injo{b} \land \injr{b}\), we apply the induction
      hypothesis to \[\lockjudge[\mu]{P^* \land \lnot\injo{b} \land \injr{b}}{s_2
      \sim_r s_1}{Q^*}.\]
      In the case where \(\injo{b} \land \lnot\injr{b}\), we apply the induction
      hypothesis to \[\lockjudge[\mu]{P^* \land \injo{b} \land \lnot\injr{b}}{s_1
      \sim_r s_2}{Q^*}.\]
      In the case where \(\lnot\injo{b} \land \lnot\injr{b}\), we apply the induction
      hypothesis to \[\lockjudge[\mu]{P^* \land \lnot\injo{b} \land
      \lnot\injr{b}}{s_2}{Q^*}.\]
    \item \textsc{while}.
      First, we perform inversion on \(s\).
      Then, we destruct \(b\).
      In the case where \(\injo{b} \land \injr{b}\), we apply the induction
      hypothesis to \(\lockjudge[\mu]{R^* \land \injo{b} \land \injr{b}}{s}{R^*}\).

      In the case where \(\injo{b} \land \lnot\injr{b}\), taking a step from
      \(\ljudge{R^* \land \injo{b} \land \lnot\injr{b}}{s}{R^*}\) yeilds
      \(\env_2'\) where \(\env_2' = \env_2\).
      Since \(\lnot\textit{failed}(\env_2)\), \(\lnot\textit{failed}(\env_2')\)
      holds as well.

      In the \(\lnot\injo{b} \land \injr{b}\) case, we apply
      \autoref{lem:right-prog} to
      \(
         \reljudge[\mu]{R^* \land \neg\injo{b} \land \injr{b}}{ s}{R^*}
      \).

      In the case where \(\lnot\injo{b} \land \lnot\injr{b}\), the loop doesn't
      run and therefore \(\env_2 = \env_2'\), so
      \(\lnot\textit{failed}(\env_2')\) trivially holds.

      Lastly, since \((\env_1, \env_2) \models R^*\), the invariant check before
      the loop cannot fail.
  \end{itemize}
\end{proof}
\end{sloppypar}

\newcommand{\modTuple}[2]{\langle {#1}, \; {#2} \rangle}
\newcommand{\dmenv}{\modTuple{h}{o}}

\begin{figure}
\begin{small}
\begin{mathpar}
 
\inferrule[f-binop]
{
	\mu(\oplus, [x_1, x_2], X, P_w, P_e) \\
  \forall x_m \in X \cdot \fresh{x_m'} \\
	m[x_1 \mapsto n_1][x_2 \mapsto n_2] \models P_w \\
	m'[x_1 \mapsto n_1]
    [x_2 \mapsto n_2]
    [\forall x_m \in X \cdot x_m \mapsto x_m']
    [\textit{result} \mapsto n_3] \models P_e \\
	\mathrm{dom}(m) = \mathrm{dom}(m')
}
{
	\tuple{m, \oplus, (n_1, n_2)} \judge_{\mu} \tuple{n_3, \; m'}
}
   \\

    \inferrule[binop]
    {
        n_1 = \frame(r_1) \\
        n_1 = \frame(r_2) \\
        \tuple{m, \oplus, (n_1, n_2)} \judge_{\mu} \tuple{n_3, \; m'} \\
    }
    {
        \step{\tuple{\assign{r}{r_1 \oplus r_2}, \denv}}{\tuple{\assign{r}{n_3}, \envTuple{\frame}{\heap}{\region}{m'}}}
    }
    
    \inferrule[read]
    {
        a = \frame(x) \\
        n = h(a) \\
        q = \region(a) \\
        \tuple{m, \code{read}, (n, q)} \judge_{\mu} \tuple{n', \; m'}
    }
    {
        \step{\tuple{\assign{r}{x}, \denv}}{\tuple{\assign{r}{n'},
        \envTuple{\frame}{\heap}{\region}{m'}}}
    }

    \inferrule[write]
    {
         a = \frame(x) \\
         n_{\textit{old}} = h(a) \\
         n_{\textit{new}} = \frame(r) \\
         q = \region(a) \\
        \tuple{m, \code{write}, (n_{\textit{old}}, n_{\textit{new}}, q)} \judge_{\mu} \tuple{n_r, \; m'} \\
    }
    {
        \step{\tuple{\assign{x}{r}, \denv}}
             {\tuple{\texttt{skip},  \envTuple{\frame}
                                              {h[ a \mapsto n_r]}
                                              {\region}
                                              {m'}}}
    }
    
    \inferrule[assign]
    {
    }
    {
        \step{\tuple{\assign{r}{n}, \denv}}{\tuple{\texttt{skip},
        \envTuple{\frame[r \mapsto n]}{\heap}{\region}{m}}}
    }
    
    \inferrule[assert-t]
    {
        \piframe(\env)(r) = \true
    }
    {
        \step{\tuple{\assert{r}, \env}}{\tuple{\texttt{skip}, \env}}
    }

     \inferrule[assert-f]
    {
        \piframe(\env)(r) = \false
    }
    {
        \step{\tuple{\assert{r}, \env}}{\tuple{\texttt{fail}, \env}}
    }

    \inferrule[assume]
    {
    }
    {
        \step{\tuple{\assume{r}, \env}}{\tuple{\assert{r}, \env}}
    }

    \inferrule[if-t]
    {
         \piframe(\env)(r) = \true
    }
    {
        \step{\tuple{\ifthen{r}{s_1}{s_2}, \env}}{\tuple{s_1, \env}}
    }

    \inferrule[if-f]
    {
       \piframe(\env)(r) = \false
    }
    {
         \step{\tuple{\ifthen{r}{s_1}{s_2}, \env}}{\tuple{s_2, \env}}
    }

    \inferrule[seq1]
    {
        \step{\tuple{s_1, \env}}{\tuple{s_1', \env'}} \\
    }
    {
        \step{\tuple{s_1 \; ; s_2, \env}}{\tuple{s_1' \;  ; s_2, \; \env'}}
    }
    \\
    \inferrule[seq2]
    {
      \\
    }
    {
        \step{\tuple{\seq{\texttt{skip}}{s_2}, \env}}{\tuple{s_2, \env}}
    }

    \inferrule[while-f]
    {
        \piframe(\env)(r) = \false
    }
    {
        \step{\tuple{\while{r}{s}, \env}}{\tuple{\code{skip}, \env}}
    }
    
    \inferrule[while-t]
    {
           \piframe(\env)(r) = \true 
    }
    {
         \step{\tuple{\while{r}{s}, \env}}{\tuple{\seq{s}{\while{r}{s}}, \env}}
    }
\end{mathpar}
\end{small}
\caption{Semantics of Execution Model and Instructions}
\label{fig:main-operational}
\end{figure}

\section{Dynamic Semantics}
\label{sec:dyn-semantics}

Figure~\ref{fig:main-operational} presents an abbreviated dynamic semantics of
{\TOOL}'s language.
We present an extended semantics in \autoref{sec:full-semantics}.
{\TOOL}'s semantics models an abstract machine that includes a {\em frame}, a
{\em heap}, and an {\em execution model state}. {\TOOL} allocates memory for program variables (both scalar and array) in the heap. A frame serves two roles:
1) a frame maps a program variable to the address of the memory region
    allocated for that variable in the heap, and
2) a frame maps a register to its current value in the program.
The model state stores the values for state variables within the execution model.

\subsection{Preliminaries}

\paragraph{\bf Frames, Heaps, Model States, Environments.} A {\em frame},
\(
\frame \in \Frame =  \mathrm{Var} \cup \mathrm{Reg} \rightarrow \Int
\),
is a finite map from variables
and registers to N-bit integers. A {\em heap},
\(
  \heap \in \Heap = \Location \rightarrow \Int
\),
is a finite map from locations ($n \in \Location \subset \Int$) to N-bit
integer values. A {\em region map},
\(
  \region \in \Region = \Location \rightarrow \mathrm{Region}
\),
is a finite map from locations to memory regions.
A {\em model state},
\(
  m \in M = \mathrm{Var} \rightarrow \Int
\),
is a finite map from model state variables to N-bit integer values. An {\em
environment},
\(
  \env \in \Env = \Frame \times \Heap \times \Region \times M
\),
is a tuple consisting of a frame, a heap, a region map, and a model state. An {\em execution model specification},
\(
  \mu \subseteq \textit{Op} \times \textit{list}(\mathrm{Var}) \times \textit{set}(\mathrm{Var}) \times P \times P
\),
is a relation consisting of tuples of an operation $\textit{op} \in
\textit{Op}$, a list of input variables, a set of modified variables, and two
unary logical predicates representing the when and ensures clauses of the
operation.

\paragraph{\bf Initialization.} For clarity of presentation, we assume a compilation and execution model in which memory locations for program variables are allocated and the corresponding mapping in the frame are done prior to execution of the program (similar in form to C-style declarations).

\subsection{Execution Model Semantics}
\cref{fig:main-operational} provides an abbreviated presentation of the execution model relation
\(
  \tuple{m, \textit{op}, (\textit{args})} \judge_{\mu} \tuple{n, \; m'}.
\)
The relation states that given the arguments, $\textit{args}$ to an operation
$op$, evaluation of the operation from the model state $m$ yields a result $n$
and a new model state $m'$ under the execution model specification $\mu$. 
\vspace{.05cm}
\begin{small}
\begin{mathpar}
\inferrule[f-binop]
{
	\mu(\oplus, [x_1, x_2], X, P_w, P_e) \\
  \forall x_m \in X \cdot \fresh{x_m'} \\
	m[x_1 \mapsto n_1][x_2 \mapsto n_2] \models P_w \\
	m'[x_1 \mapsto n_1]
    [x_2 \mapsto n_2]
    [\forall x_m \in X \cdot x_m \mapsto x_m']
    [\textit{result} \mapsto n_3] \models P_e \\
	\mathrm{dom}(m) = \mathrm{dom}(m')
}
{
	\tuple{m, \oplus, (n_1, n_2)} \judge_{\mu} \tuple{n_3, \; m'}
}
\end{mathpar}
\end{small}
\vspace{-.15cm}

The [\textsc{f-binop}] rule specifies the meaning of this relation for binary
operations. This relation states that the value of an operation $\oplus$ given
a tuple of input values $(n_1,
n_2)$ and an execution model state $m$ evaluates to value $n_3$ and a new
model state $m'$. The rule relies on the relation $\mu(\textit{op},
$\textit{vlist}$, X, P_w, P_e)$ which specifies the list of argument names,
$\textit{vlist}$, the set of modified variables \(X\), the {\em precondition}
$P_w$, and the {\em postcondition} $P_e$ for the operation $op$ in the
developer-provided execution model.
The set of modified variables is the union of the \texttt{modifies} clauses in
the operation's specification.
The precondition of an operation is the conjunction of the \texttt{when}
clauses in the operation's specification.
The postcondition of an operation is the conjunction of the \texttt{ensures}
clauses in the operation's specification.

The semantics of the model relation non-deterministically selects an operation
specification, result value, and output model state subject to the constraint
that:
1) the current model state satisfies the precondition (after the inputs to
    the operation have been appropriately assigned into the model state),
2) the output model state satisfies the postcondition (after the inputs,
   modified variables, and result value have been appropriately assigned into
   the model state), and
3) the domains of the input and output state are the same.

Because of the uniformity of the execution model specification, the semantics
for other operations (e.g., reads and writes) is similar with the sole
distinction being the number of arguments passed to the operation. For clarity
of presentation, we elide the presentation of rules for those operations, but
we provide them in \cref{sec:full-semantics}, \cref{fig:ext-model-semantics}.

\subsection{Language Semantics}
Figure~\ref{fig:main-operational} presents the non-deterministic small-step
transition relation $\step{\tuple{s, \env}}{\tuple{s', \env'}}$ of a {\TOOL}
program. The relation states that execution of statement $s$ from the
environment $\env$ takes one step yielding the statement $s'$ and environment
$\env'$ under the execution model specification $\mu$. The semantics of the
statements are largely similar to that of traditional approaches except for the
ability to the statements to encounter faults. Broadly, we categorize {\TOOL}'s
instructions into four categories: {\em register instructions}, {\em memory
instructions}, {\em assertions}, and {\em control flow}.

\paragraph{\bf Register Instructions.} The rules [\textsc{assign}] and [\textsc{binop}] specify the semantics of two of {\TOOL}'s register manipulation instructions. [\textsc{assign}] defines the semantics of assigning an integer value to a register, $\assign{r}{n}$. This has the expected semantics updating the value of $r$ within the current frame with the value $n$. Of note is that register assignment executes fully reliably without faults. 

[\textsc{binop}] specifies the semantics of a register only binary operation,
$\assign{r}{r_1 \; \oplus \; r_2}$. Note that reads of the input registers
execute fully reliably. The result of the operation is $n_3$, which is the
value of the operation given the semantics of that operation's execution  model
when executed from the model state $m$ on parameters $n_1$ and $n_2$. Executing
the execution  model may change the values of the execution  model's state
variables. Therefore, the instruction evaluates to an instruction that assigns
$n_3$ to the destination register and evaluates with a environment that
consists of the unmodified frame, the unmodified heap, and the modified
execution  model state. Note that by virtue of the fact that both the frame and heap are unmodified, faults in register instructions cannot modify the contents or organization of memory. This modeling choice is consistent with standard fault modeling approaches. 

\paragraph{\bf Memory Instructions.}
The rules [\textsc{read}] and [\textsc{write}] specify the semantics of two of
{\TOOL}'s memory manipulation instructions. [\textsc{read}] defines the
semantics of reading the value of a program variable $x$ from it's
corresponding memory location: $\assign{r}{x}$. The rule fetches the program
variable's memory address from the frame, reads the value of the memory
location $n = h(a)$ and the region the memory location belongs to $q =
\region(a)$ and then executes the execution  model with the program variable's
current value in memory and the memory region it resides in as a parameters.
The execution  model non-deterministically yields a result $n'$ that the rule uses to complete its implementing by issuing an assignment to the register. 

[\textsc{write}] defines the semantics of writing the value of a register to
memory. The rule reads the value of the memory location to record the old value of
the memory location, reads the value of the input register, fetches the region
the memory location corresponds to, and then executes
the execution  model with these values as parameters. The execution  model yields a new
value $n_r$ that the rule then assigns to the value the program variable.

\paragraph{\bf Assertions.} The rules [\textsc{assert-t}] and [\textsc{assume-t}] specify the semantics of \texttt{assert} and \texttt{assume} statements, respectively. These statements have standard semantics, yielding a \texttt{skip} and continuing the execution of the program if their conditions are satisfied. For either of these statements, if their conditions evaluate to false, then execution yields \texttt{fail} denoting that execution has failed and become stuck in error.

\paragraph{\bf Control Flow.}
The rules for control flow ([\textsc{if-t}], [\textsc{if-f}], [\textsc{seq1}], [\textsc{seq2}], [\textsc{while-f}], and [\textsc{while-t}]) have standard semantics.
An important note is that the semantics of these statements is such that the transfer of control from one instruction to another always executes reliably and, therefore, faults do not introduce control flow errors into the program. This modeling assumption is consistent with standard fault injection and reliability analysis models~\cite{vulfi16}.

\section{Full Semantics}
\label{sec:full-semantics}

In this appendix we present the full dynamic semantics of the \TOOL language.
We have elided the rules presented in Figure~\ref{fig:main-operational} as they
remain unchanged.
A preprocessing pass performs the following actions:
\begin{itemize}
  \item It places all variables without a label into a reliable memory region.
  \item It flattens multidimensional vectors into single dimensional vectors.
  \item It inlines function calls.
\end{itemize}

The \textrm{alloc} function we use in \textsc{declare} and
\textsc{declare-array} takes a mapping from variables to addresses \(\sigma\)
and an integer \(n\) and returns the first addresses in a contiguous block of
\(n\) unmapped addresses in \(\sigma\).

Figures~\ref{fig:ext-semantics} and \ref{fig:ext-model-semantics} expand on the
operational semantics we present in \autoref{fig:main-operational}.

\begin{figure*}
  \begin{footnotesize}
\begin{mathpar}
    \inferrule[read-array]
    {
        a = \frame(x) \\
        n_o = h(o(r_2)) \\
        n = h(a + n_o) \\
        q = \region(a) \\
        \tuple{m, \code{read}, (n, q)} \judge_{\mu} \tuple{n', \; m'}
    }
    {
        \step{\tuple{\assign{r_1}{x[r_2]}, \denv}}{\tuple{\assign{r_1}{n'},
        \envTuple{\frame}{\heap}{\region}{m'}}}
    }

    \inferrule[write-array]
    {
         a = \frame(x) \\
         n_o = h(o(r_2)) \\
         n_{\textit{old}} = h(a + n_o) \\
         n_{\textit{new}} = \frame(r_1) \\
         q = \region(a) \\
        \tuple{m, \code{write}, (n_{\textit{old}}, n_{\textit{new}}, q)} \judge_{\mu} \tuple{n_r, \; m'} \\
    }
    {
        \step{\tuple{\assign{x[r_2]}{r_1}, \denv}}
             {\tuple{\texttt{skip},  \envTuple{\frame}
                                              {h[ (a + n_o) \mapsto n_r]}
                                              {\region}
                                              {m'}}}
    }

    \inferrule[relational-assert]
    { }
    {\step{\tuple{\relassert{r}, \env}}{\tuple{\code{skip}, \env}}}

    \inferrule[declare]
    { a = \mathrm{alloc}(\frame, 1) }
    {\step{\tuple{\code{@region}(q) \; \tau \; x, \denv}}
          {\tuple{\code{skip}, \envTuple{\frame[x \mapsto a]}
                                        {\heap[a \mapsto 0]}
                                        {\region[a \mapsto q]}
                                        {m}}}}

    \inferrule[declare-array]
    {
      a = \mathrm{alloc}(\frame, n) \\
      h' = \heap[a \mapsto 0]\ldots[(a + n-1) \mapsto 0]
    }
    {\step{\tuple{\code{@region}(q) \; \tau \; x(n), \denv}}
          {\tuple{\code{skip}, \envTuple{\frame[x \mapsto a]}
                                        {h'}
                                        {\region[a \mapsto q]}
                                        {m}}}}

    \inferrule[forall-t]
    {
      a = \mathrm{alloc}(\frame, 1) \\
      \forall n. \tuple{b, \envTuple{\frame[x \mapsto a]}{\heap[a \mapsto
      n]}{\region[a \mapsto \code{reliable}]}{m}} \judge_\mu 
      \tuple{\true, \envTuple{\frame[x \mapsto a]}{\heap[a \mapsto
      n]}{\region[a \mapsto \code{reliable}]}{m}}
    }
    {\tuple{\code{forall}(\tau\; x)(b), \denv} \judge_{\mu} \tuple{\true, \denv}}

    \inferrule[forall-f]
    {
      a = \mathrm{alloc}(\frame, 1) \\
      \exists n. \tuple{b, \envTuple{\frame[x \mapsto a]}{\heap[a \mapsto
      n]}{\region[a \mapsto \code{reliable}]}{m}} \judge_\mu 
      \tuple{\false, \envTuple{\frame[x \mapsto a]}{\heap[a \mapsto
      n]}{\region[a \mapsto \code{reliable}]}{m}}
    }
    {\tuple{\code{forall}(\tau\; x)(b), \denv} \judge_{\mu} \tuple{\false, \denv}}

    \inferrule[exists-t]
    {
      a = \mathrm{alloc}(\frame, 1) \\
      \exists n. \tuple{b, \envTuple{\frame[x \mapsto a]}{\heap[a \mapsto
      n]}{\region[a \mapsto \code{reliable}]}{m}} \judge_\mu 
      \tuple{\true, \envTuple{\frame[x \mapsto a]}{\heap[a \mapsto
      n]}{\region[a \mapsto \code{reliable}]}{m}}
    }
    {\tuple{\code{exists}(\tau\; x)(b), \denv} \judge_{\mu} \tuple{\true, \denv}}

    \inferrule[exists-f]
    {
      a = \mathrm{alloc}(\frame, 1) \\
      \forall n. \tuple{b, \envTuple{\frame[x \mapsto a]}{\heap[a \mapsto
      n]}{\region[a \mapsto \code{reliable}]}{m}} \judge_\mu 
      \tuple{\false, \envTuple{\frame[x \mapsto a]}{\heap[a \mapsto
      n]}{\region[a \mapsto \code{reliable}]}{m}}
    }
    {\tuple{\code{exists}(\tau\; x)(b), \denv} \judge_{\mu} \tuple{\false, \denv}}
\end{mathpar}
  \end{footnotesize}
  \caption{Extension to Dynamic Language Semantics}
  \label{fig:ext-semantics}
\end{figure*}

\begin{figure*}
\begin{mathpar}
  \inferrule[m-declare]
  { }
  {
    \tuple{\tau \; x, \dmenv} \sstep
    \tuple{\code{skip}, \modTuple{h[x \mapsto 0]}{o}}
  }

  \inferrule[m-assign]
  { \tuple{e, \dmenv} \bstep \tuple{n, \dmenv} }
  {\tuple{x = e, \dmenv} \sstep \tuple{\code{skip}, \modTuple{h[x \mapsto n]}{o}}}

  \inferrule[m-declare-op]
  { }
  {\tuple{\code{operator} \oplus x_1^* \; \code{when} \; p_w \; \code{modifies} \; x_2^* \; \code{ensures} \; p_e,
         \dmenv}
         \sstep
   \tuple{\code{skip}, \modTuple{h}{o::\tuple{\oplus, \; x_1^*, \; p_w, \; p_e}}}}

    \inferrule[f-read]
    {
        o = \mu(\code{read}, [x], X, P_w, P_e) \\
        o \in q \\
        \forall x_m \in X \cdot \fresh{x_m'} \\
        m[x_1 \mapsto n_1] \models P_w \\
        m'[x_1 \mapsto n_1]
          [\forall x_m \in X \cdot x_m \mapsto x_m']
          [\textit{result} \mapsto n_2] \models P_e \\
        \mathrm{dom}(m) = \mathrm{dom}(m')
    }
    {
      \tuple{\code{model.read}(n_1, q), \; m} \judge_{\mu} \tuple{n_2, \; m'}
    }

    \inferrule[f-write]
    {
        o = \mu(\code{write}, [x_1, x_2], X, P_w, P_e) \\
        o \in q \\
        \forall x_m \in X \cdot \fresh{x_m'} \\
        m[x_1 \mapsto n_\textit{old}][x_2 \mapsto n_\textit{new}] \models P_w \\
        m'[x_1 \mapsto n_3]
          [x_2 \mapsto n_\textit{new}]
          [\forall x_m \in X \cdot x_m \mapsto x_m']
          \models P_e \\
        \mathrm{dom}(m) = \mathrm{dom}(m')
    }
    {
      \tuple{\code{model.write}(n_\textit{old}, n_\textit{new}, q), \; m} \judge_{\mu} \tuple{n_3, \; m'}
    }

\end{mathpar}
  \caption{Extension to Dynamic Execution Model Semantics}
  \label{fig:ext-model-semantics}
\end{figure*}

\algnewcommand\algorithmicswitch{\textbf{match}}
\algnewcommand\algorithmiccase{\textbf{}}
\algnewcommand\algorithmicassert{\texttt{assert}}
\algnewcommand\Assert[1]{\State \algorithmicassert(#1)}%
\algnewcommand\Report[1]{\State \texttt{report}(#1)}%

\algdef{SE}[MATCH]{Match}{EndMatch}[1]{\algorithmicswitch\ #1\ \algorithmicdo}{\algorithmicend\ \algorithmicswitch}%
\algdef{SE}[CASE]{Case}{EndCase}[1]{\algorithmiccase\ #1:}{\algorithmicend\ \algorithmiccase}%
\algtext*{EndSwitch}%
\algtext*{EndCase}%

\newcommand{\model}{m}
\newcommand{\Model}{M}
 
\newcommand{\vars}{v}
\newcommand{\substs}{\omega}
\newcommand{\ops}{o}
\newcommand{\op}{\oplus}
\newcommand{\context}{\Gamma}
\newcommand{\solver}{\sigma}
\newcommand{\ir}{b_{ir}}
\newcommand{\io}{c}
\newcommand{\vs}{\Psi}
\newcommand{\vms}{\Delta}
\newcommand{\vme}{\delta}
\newcommand{\ims}{M}
\newcommand{\ime}{\mu}
\newcommand{\vr}{\textrm{vrel}}
\newcommand{\ve}{\textrm{vexp}}
\newcommand{\vb}{\textrm{vbexp}}
\newcommand{\ifcheck}{\textrm{Check}\xspace}

\section{Verification Algorithm}
\label{sec:verification}

Figures~\ref{fig:stmt}, and ~\ref{fig:verify-if} present the core of {\TOOL}'s
verification algorithm. The algorithm performs forward symbolic execution to
discharge verification conditions generated by \texttt{assert},
\texttt{assert_t}, \texttt{invariant}, and \texttt{invariant_r} statements in
the program. The algorithm directly implements the Hoare-style relational
program logic from Section~\ref{sec:logic}.

\paragraph{Preliminaries.} We denote Leto's verification algorithm by the function $\vs$, which takes as input a statement $s$, a logical predicate $\solver$, a model specification $m$, and a verification  mode $c$. The statement $s$ is the statement to be verified, $\solver$ is the symbolic context under which the verification algorithm is invoked, $m$ computes the symbolic representation for a operation in the execution model.

The control value $c \in C = \{\textit{lock}, \textit{left}, \textit{right}\}$ determines whether or not the algorithm is performing verification in {\em lockstep} mode, {\em left mode}, or {\em right mode}, respectively. When performing verification in lockstep, the algorithm models the original and relaxed execution as each executing 
 an instruction one at a time.  In this mode, the algorithm is able to demonstrate an easy correspondence between the two executions that therefore enables the algorithm
to, for example, transfer assumed properties of the original execution over to verify the relaxed execution. For both left mode and right mode,
the algorithm assumes the two executions have diverged and, therefore, that there is no simple correspondence between the two executions.
In left mode, the algorithm symbolically evaluates the original execution of the program, ignoring the verification conditions of the required for the relaxed execution. 
In right mode, the algorithm symbolically evalutes the relaxed execution of the program and checks the verification conditions that are required of the relaxed execution. 

The function $\ve(e, m, c)$ maps a standard unary expression $e$ to a constraint that represent the resulting value in either the original or relaxed execution. For example,
 $\ve$ maps a variable reference $x$ to either the variable reference
 $\origvar{x}$ or $\relvar{x}$ if $c$ equals \texttt{left} or \texttt{right},
 respectively. The function returns a constraint because the
 expression may reference an operation suffixed with a period, denoting that
 the operation has a custom semantics. The constraint characterizes the non-deterministic choice of the operation's implementation.
 The function uses the model specification \(m\) to compute the symbolic
 representation for these operations.
 
The function $\vb(b, m, c)$ maps a standard unary boolean expression $b$ to a constraint. Its operation is similar to that of $\ve$.

\begin{figure}
    \begin{small}
    \begin{algorithmic}[1]
        \Function{$\vs$}{$s, \solver, \model, \io$}
            \Match{s}
                \Case{$x \; \code{=} \; e$}   
                    \State $\solver_o \gets (\origvar{x} \; \texttt{=} \; \ve(e, m, \textit{left}))$
                    \State $\solver_r \gets (\relvar{x} \; \texttt{=} \;  \ve(e, m, \textit{right}))$
                    \State \Return $\solver \land \textrm{join}(\io, \solver_o, \solver_r)$
                \EndCase
             
                \Case{\assert{b_v}}
                \EndCase
                \Case{\assume{b_v}}
                    \State $\solver_o \gets \vb(b_v, m, \textit{left})$
                    \State $\solver_r \gets \vb(b_v, m, \textit{right})$
                    \If{$(\io == \textit{lock})$}
                        \State $\textrm{Verify}(\solver \land \solver_o, \solver_r)$
                    \ElsIf{$(\io == \textit{right})$}
                        \State $\textrm{Verify}(\solver, \solver_r)$
                    \EndIf
                    \State\Return $\solver \land \textrm{join}(\io, \solver_o, \solver_r)$
                \EndCase
                \Case{\code{assert_r} ($b_r$)}
                    \State $\textrm{Verify}(\solver, b_r)$
                    \State\Return $\solver \land b_r$
                \EndCase
                \Case{$s1 \code{;} s2$}
                    \State \Return \Call{$\vs$}{$s_2, \solver \land \vs(s_1, \solver, \model, \io), \model, \io$}
                \EndCase
            \EndMatch
        \EndFunction
    \end{algorithmic}
    \end{small}
    \caption{Verification Algorithm (sans Control Flow)}
    \label{fig:stmt}
\end{figure}

\paragraph{\bf Assignment.}
For an assignment statement $x \texttt{=} e$, the
algorithm maps the $e$ to an appropriate relational expression for both the
original and relaxed execution by creating the constraint that $x$ in the
original (relaxed) execution has the value $e$. The algorithm then uses the
\(\textrm{join}(\cdot)\) function to return a result.  The
\(\textrm{join}(\cdot)\) function joins two constraints into a conjunction depending on the value of $c$. If $c = \textit{lock}$ -- denoting the algorithm is modelling the lockstep execution of
both the original and relaxed executions -- then the \textit{join} includes both constraints. If $c = \textit{left}$ or $c = \textit{right}$ --  denoting
that the algorithm is modeling the original or relaxed execution, respectively -- then \textit{join} includes only the first or second constraint, respectively.

\paragraph{\bf Assume and Assert.}  The algorithm verifies both \texttt{assert} and \texttt{assume} using the same logical approach. The algorithm first generates the verification conditions for both the original and relaxed executions, namely that the statement's boolean expression $b_v$ is true ($\solver_o$ and $\solver_r$, respectively). The algorithm next considers two cases. In lockstep mode, the algorithm verifies that the current context $\solver$ extended with the {\em assumption} that the assertion or assumption is true in the original execution implies that the verification condition holds. The function $\textrm{Verify}(\sigma_1, \sigma_2)$ verifies that $\sigma_1$ implies $\sigma_2$ (Leto specifically uses an SMT solver to do so) and halts the execution of the algorithm if the implication does not hold or the solver is unable to demonstrate that it holds.  In right mode, the algorithm directly
verifies that the current context implies the verification condition. The insight is that unlike in lockstep mode, the algorithm must verify the relaxed execution independently of the original execution and, therefore, the algorithm cannot leverage the assumption that the assertion or assumption is valid in the original execution. In the last step,
the algorithm returns the join of the two verification conditions. 

\paragraph{\bf Relational Assert.}  The algorithm verifies relational assertions under the current context. If verification fails, then the verification procedure halts. If verification succeeds, then the algorithm appends the assertion to the context and returns the result.

\begin{figure}
    \begin{small}
       \begin{algorithmic}[1]
        \Function{$\vs$}{$\code{if} \; (b) \; \{ s_1 \} \; \code{else} \; \{ s_2 \}, \solver, \model, \io$} =
            
                    \State $\sigma_o \gets \vb(b, m, \textit{left}), \;\; \sigma_r \gets \vb(b, m, \textit{right})$
                    \If{$(\io == \textit{lock})$}
                      \If{$\ifcheck(\solver, \solver_o \land \solver_r)$}
                        \State \(\solver_1 \gets\) \Call{$\vs$}{$s_1, \solver \land \sigma_o \land \sigma_r,\textit{lock}$}
                      \Else \;
                        \(\solver_1 \gets \solver\)
                      \EndIf
                      \\
                      \If{$\ifcheck(\solver, \lnot\solver_o \land \lnot\solver_r)$} \;
                        \State \(\solver_2 \gets\) \Call{$\vs$}{$s_1, \solver \land
                        \lnot\sigma_o \land \lnot\sigma_r,\textit{lock}$}
                      \Else \;
                        \(\solver_2 \gets \solver\)
                      \EndIf
                        \\
                      \If{$\ifcheck(\solver, \solver_o \land \lnot\solver_r)$}
                        \State \(\solver_3 \gets\) \Call{$\vs$}{$s_1, \solver \land \sigma_o \land \neg \sigma_r, \textit{left}$}
                        \State \(\solver_4 \gets\) \Call{$\vs$}{$s_2, \solver \land \solver_3 \land \sigma_o \land \neg \sigma_r, \textit{right}$}
                      \Else \;
                        \(\solver_4 \gets \solver\)
                      \EndIf

                      \\
                    
                      \If{$\ifcheck(\solver, \lnot\solver_o \land \solver_r)$}
                        \State \(\solver_5 \gets\) \Call{$\vs$}{$s_2, \solver \land \neg \sigma_o \land \sigma_r, \textit{left}$}
                        \State \(\solver_6 \gets\) \Call{$\vs$}{$s_1, \solver \land \solver_5 \land \neg \sigma_o \land \sigma_r, \textit{right}$}
                      \Else \;
                        \(\solver_6 \gets \solver\)
                      \EndIf
                        \\
                    
                        \State \Return \((\solver_1 \land \solver_2 \land \solver_4 \land \solver_6)\)
                    \ElsIf{$(\io == \textit{left})$} \;
                      \If{$\ifcheck(\solver, \solver_o)$} \;
                        \(\solver_1 \gets\) \Call{$\vs$}{$s_1, \solver \land \sigma_o, \textit{left}$}
                      \Else \;
                        \(\solver_1 \gets \solver\)
                      \EndIf
                      \\
                      \If{$\ifcheck(\solver, \lnot\solver_o)$} \;
                        \(\solver_2 \gets\) \Call{$\vs$}{$s_1, \solver \land
                        \lnot\sigma_o, \textit{left}$}
                      \Else \;
                        \(\solver_2 \gets \solver\)
                      \EndIf
                      \\
                      \State \Return \((\solver_1 \land \solver_2)\)
                    \ElsIf{$(\io == \textit{right})$}
                      \If{$\ifcheck(\solver, \solver_r)$} \;
                        \(\solver_1 \gets\) \Call{$\vs$}{$s_1, \solver \land
                        \sigma_r, \textit{right}$}
                      \Else \;
                        \(\solver_1 \gets \solver\)
                      \EndIf
                      \\
                      \If{$\ifcheck(\solver, \lnot\solver_r)$} \;
                        \(\solver_2 \gets\) \Call{$\vs$}{$s_1, \solver \land
                        \lnot\sigma_r, \textit{right}$}
                      \Else \;
                        \(\solver_2 \gets \solver\)
                      \EndIf
                      \\
                      \State \Return \((\solver_1 \land \solver_2)\)
                    \EndIf
        \EndFunction
    \end{algorithmic}
    \end{small}
    \caption{If statement verification algorithm}
    \label{fig:verify-if}
\end{figure}

\paragraph{\bf If.} Figure~\ref{fig:verify-if} presents the algorithm's implementation for \texttt{if} statement verification. The algorithm has a different implementation for each of the verification modes:

\begin{itemize}
\item{\bf Lockstep}. In lockstep mode, the algorithm verifies and generates
a symbolic representation for four different scenarios: the case
when 1) the original execution and relaxed execution both take the
$\true$ branch of the statement, represented by $\sigma_1$, 2) the original execution
and relaxed execution both take the $\false$ branch of the statement, $\sigma_2$,
3) the original execution takes the $\true$ branch and the relaxed execution takes the $\false$ branch, $\sigma_4$, and 4) the original execution takes
the $\false$ branch and the relaxed execution takes the $\true$ branch, $\sigma_6$.

\item{\bf Left}. In left mode, the algorithm need only generate
a symbolic representation for the original execution. The algorithm
achieves this by conjoining the results of recursive calls to $\vs$ on
$s_1$ and $s_2$ given the current context.

\item{\bf Right}. In right mode, the algorithm need only generate a symbolic representation and discharge the verification conditions for the relaxed execution. Similar to that of left mode, algorithm achieves this by conjoining the results of recursive calls to $\vs$ on $s_1$ and $s_2$.
\end{itemize}

For performance, \TOOL only considers scenarios that are potentially viable.
The \(\ifcheck(\sigma_1, \sigma_2)\) function returns \(\true\) if a satisfying
assignment for \(\sigma_1 \limp \sigma_2\) may exist and \(\false\) otherwise.
If \ifcheck is able to prove that such an implication cannot exist, then
\TOOL does not recurse on that execution scenario.

\newcommand{\sinf}{\textsc{Inf}\xspace}
\newcommand{\winf}{\textsc{WeakInf}\xspace}
\newcommand{\cinv}{\sigma_i}
\newcommand{\cfalse}{b_{fv}}
\newcommand{\cfalser}{b_{fr}}
\newcommand{\infcheck}{\textrm{Check}\xspace}
\newcommand{\enum}{r}
\newcommand{\Enum}{R}
\newcommand{\sat}{\textit{sat}}
\newcommand{\unsat}{\textit{unsat}}
\newcommand{\unknown}{\textit{unknown}}
\newcommand{\vsinf}{\vs'}
\newcommand{\cand}{b_{cv}}
\newcommand{\candhead}{b_h}
\newcommand{\candtail}{b_{cv}'}
\newcommand{\candr}{b_{cr}}
\newcommand{\candrhead}{b_h}
\newcommand{\candrtail}{b_{cr}'}
\newcommand{\proginv}{b_{pv}}
\newcommand{\progrelinv}{b_{pr}}
\newcommand{\proginveval}{p_{pv}}

\section{Invariant Inference}
\label{sec:houdini}

\begin{figure*}
       \begin{algorithmic}[1]
        \Function{$\sinf$}{$\code{while} \; (b) \; (b_v) \; (b_r) \; \{ s_b \},
         \solver, \model, \io, \proginv, \progrelinv$} =
          \State $\sigma_o \gets \vb(b, m, \textit{left}), \;\; \sigma_r \gets \vb(b, m,\textit{right})$
          \If{$(\io == \textit{lock})$}
              \State $p_o \gets \vb(b_v, m, \textit{left}), \;\; p_r \gets
              \vb(b_v, m, \textit{right})$ \label{alg:strong-inf:begin-lock}
              \State $(\enum_1, \cfalse{}_1, \cfalser{}_1) \gets \infcheck(\solver::p_o,p_r::b_r)$
              \State $\solver_c \gets \solver::p_o::p_r::b_r::\vb(\proginv, m,
              \textit{left})::
              \vb(\progrelinv, m, \textit{right})$
              \State $(\enum_2, \cfalse{}_2, \cfalser{}_2) \gets \infcheck({\vsinf(s_b, \solver_c::\sigma_o::\sigma_r, \model, \textit{lock}), b_r::p_o::p_r)}$
              \State $(\enum_3, \cfalse{}_3, \cfalser{}_3) \gets
              \infcheck({\vsinf(s_b, \solver_c::\lnot\sigma_o::\sigma_r,
              \model, \textit{right}), b_r::p_r)}$
              \label{alg:strong-inf:end-lock-while}
              \If{$(\enum_1 == \unknown \lor
                    \enum_2 == \unknown \lor
                    \enum_3 == \unknown)$}
                \State \Return \Call{$\winf$}{$\code{while} \; (b) \; (\true) \;
                (\true) \; \{ s_b \}, \solver, \model, \io, p_o :: p_r ::
                b_r, \proginv, \progrelinv$} \label{alg:strong-inf:lock-weak}
              \ElsIf{$(\enum_1 == \sat \lor
                       \enum_2 == \sat \lor
                       \enum_3 == \sat)$}
                \State $\cfalse \gets \cfalse{}_1 \cup
                                      \cfalse{}_2 \cup
                                      \cfalse{}_3$
                                      \label{alg:strong-inf:lock-sat-begin}
                \State $\cfalser \gets \cfalser{}_1 \cup
                                       \cfalser{}_2 \cup
                                       \cfalser{}_3$
                \State \Return \Call{$\sinf$}{$\code{while} \; (b) \; (b_v
                \setminus \cfalse) \;
                (b_r \setminus \cfalser) \; \{ s_b \}, \solver, \model, \io,
                \proginv, \progrelinv$}
                \label{alg:strong-inf:lock-sat-end}
              \Else
                \State \Return $(b_v, b_r)$
                \label{alg:strong-inf:lock-unsat}
              \EndIf
          \ElsIf{$(\io == \textit{right})$}
              \State $\textit{p} \gets \vb(b_v, m, \textit{right}), \;\;
              \proginveval \gets \vb(\proginv, m, \textit{right})$
              \label{alg:strong-inf:begin-right}
              \State $(\enum_1, \cfalse{}_1, \cfalser{}_1) \gets \infcheck(\solver, p::b_r)$
              \State $(\enum_2, \cfalse{}_2, \cfalser{}_2) \gets \infcheck(\vsinf(s_b,
              \solver::p::b_r::\solver_r::\proginveval, \model, \textit{right}), \textit{p}::b_r)$
              \label{alg:strong-inf:end-right-while}
              \If{$(\enum_1 == \unknown \lor
                    \enum_2 == \unknown)$}
                \State \Return \Call{$\winf$}{$\code{while} \; (b) \; (\true) \;
                (\true) \; \{ s_b \}, \solver, \model, \io, p :: b_r, \proginv,
                \progrelinv$}
                \label{alg:strong-inf:right-weak}
              \ElsIf{$(\enum_1 == \sat \lor
                       \enum_2 == \sat)$}
                \State $\cfalse \gets \cfalse{}_1 \cup
                                      \cfalse{}_2 \cup$
                                      \label{alg:strong-inf:right-sat-begin}
                \State $\cfalser \gets \cfalser{}_1 \cup
                                       \cfalser{}_2 \cup$
                \State \Return \Call{$\sinf$}{$\code{while} \; (b) \; (b_v
                \setminus \cfalse) \;
                (b_r \setminus \cfalser) \; \{ s_b \}, \solver, \model, \io,
                \proginv, \progrelinv$}
                \label{alg:strong-inf:right-sat-end}
              \Else
                \State \Return $(b_v, b_r)$
                \label{alg:strong-inf:right-unsat}
              \EndIf
          \EndIf
        \EndFunction
    \end{algorithmic}
  \caption{Strong Inference Algorithm}
  \label{fig:strong-inf}
\end{figure*}

\begin{figure*}
  \begin{small}
       \begin{algorithmic}[1]
        \Function{$\winf$}{$\code{while} \; (b) \; (b_v) \; (b_r) \; \{ s_b \},
                            \solver, \model, \io, \cand, \candr, \proginv,
                            \progrelinv$} =
          \Match{$\cand$}
            \Case{$\candhead :: \candtail$}
              \If{$(\io == \textit{lock})$}
              \label{alg:weak-inf:start-stand}
                  \State $p_o \gets \vb(b_v, m, \textit{left})::\vb(\candhead,
                  m, \textit{left})$
                  \label{alg:weak-inf:begin-lock}

                  \State $p_r \gets \vb(b_v, m, \textit{right})::\vb(\candhead,
                  m, \textit{right})$

                  \State $(\enum_1, \cfalse{}_1, \cfalser{}_1) \gets \infcheck(\solver::p_o,p_r)$

                  \State $\solver_c \gets \solver::p_o::p_r::\vb(\proginv, m,
                  \textit{left})::
                  \vb(\progrelinv, m, \textit{right})$

                  \State $(\enum_2, \cfalse{}_2, \cfalser{}_2) \gets \infcheck({\vsinf(s_b, \solver_c::\sigma_o::\sigma_r, \model, \textit{lock}), p_o::p_r)}$

                  \State $(\enum_3, \cfalse{}_3, \cfalser{}_3) \gets \infcheck({\vsinf(s_b, \solver_c::\lnot\sigma_o::\sigma_r, \model, \textit{right}), p_r)}$ 
                  \label{alg:weak-inf:end-lock-while}

                  \If{$(\enum_1 == \unsat \land
                        \enum_2 == \unsat \land
                        \enum_3 == \unsat)$}
                    \State \Return \Call{$\winf$}{$\code{while} \; (b) \;
                    (\candhead :: b_v) \;
                    (b_r) \; \{ s_b \}, \solver, \model, \io, \candtail,
                    \candr, \proginv, \progrelinv$}
                    \label{alg:weak-inf:lock-unsat}
                  \Else \;
                    \Return \Call{$\winf$}{$\code{while} \; (b) \;
                    (b_v) \;
                    (b_r) \; \{ s_b \}, \solver, \model, \io, \candtail,
                    \candr, \proginv, \progrelinv$}
                    \label{alg:weak-inf:lock-not-unsat}
                  \EndIf
              \ElsIf{$(\io == \textit{right})$}
                  \State $\textit{p} \gets \vb(b_v, m, \textit{right}) ::
                  \vb(\candhead, m, \textit{right}), \;\;
                  \proginveval \gets \vb(\proginv, m, \textit{right})$
                  \label{alg:weak-inf:begin-right}

                  \State $(\enum_1, \cfalse{}_1, \cfalser{}_1) \gets \infcheck(\solver, p)$
                  \State $(\enum_2, \cfalse{}_2, \cfalser{}_2) \gets \infcheck(\vsinf(s_b,
                  \solver::p::\solver_r::\proginveval, \model, \textit{right}), \textit{p})$
                  \label{alg:weak-inf:end-right-while}
                  \If{$(\enum_1 == \unsat \land
                        \enum_2 == \unsat)$}
                    \State \Return \Call{$\winf$}{$\code{while} \; (b) \;
                    (\candhead :: b_v) \;
                    (b_r) \; \{ s_b \}, \solver, \model, \io, \candtail,
                    \candr, \proginv, \progrelinv$}
                    \label{alg:weak-inf:right-unsat}
                  \Else \;
                    \Return \Call{$\winf$}{$\code{while} \; (b) \;
                    (b_v) \;
                    (b_r) \; \{ s_b \}, \solver, \model, \io, \candtail,
                    \candr, \proginv, \progrelinv$}
                    \label{alg:weak-inf:right-not-unsat}
                  \EndIf
              \EndIf
              \label{alg:weak-inf:end-stand}
            \EndCase
            \Case{\code{[]}}
              \Match{$\candr$}
                \Case{\code{[]}}
                  \Return $(b_v, b_r)$
                  \label{alg:weak-inf:base}
                \EndCase
                \Case{$\candrhead :: \candrtail$}
                  \If{$(\io == \textit{lock})$}
                  \label{alg:weak-inf:start-rel}
                    \State $p_o \gets \vb(b_v, m, \textit{left}), \;\;
                            p_r \gets \vb(b_v, m, \textit{right})$

                    \State $(\enum_1, \cfalse{}_1, \cfalser{}_1) \gets
                    \infcheck(\solver::p_o,b_r::\candrhead::p_r)$

                    \State $\solver_c \gets \solver::p_o::p_r::b_r::\candrhead::\vb(\proginv, m,
                  \textit{left})::
                  \vb(\progrelinv, m, \textit{right})$

                    \State $(\enum_2, \cfalse{}_2, \cfalser{}_2) \gets
                    \infcheck({\vsinf(s_b, \solver_c::\sigma_o::\sigma_r, \model,
                    \textit{lock}), p_o::p_r::b_r::\candrhead)}$

                    \State $(\enum_3, \cfalse{}_3, \cfalser{}_3) \gets
                    \infcheck({\vsinf(s_b, \solver_c::\lnot\sigma_o::\sigma_r,
                    \model, \textit{right}), p_r::b_r::\candrhead)}$ 

                    \If{$(\enum_1 == \unsat \land
                          \enum_2 == \unsat \land
                          \enum_3 == \unsat)$}
                      \State \Return \Call{$\winf$}{$\code{while} \; (b) \;
                      (b_v) \;
                      (b_r :: \candrhead) \; \{ s_b \}, \solver, \model, \io,
                      \cand,
                      \candrtail, \proginv, \progrelinv$}
                    \Else \;
                      \Return \Call{$\winf$}{$\code{while} \; (b) \;
                      (b_v) \;
                      (b_r) \; \{ s_b \}, \solver, \model, \io, \cand,
                      \candrtail, \proginv, \progrelinv$}
                    \EndIf
              \ElsIf{$(\io == \textit{right})$}
                  \State $\textit{p} \gets \vb(b_v, m, \textit{right}), \;\;
                  \proginveval \gets \vb(\proginv, m, \textit{right})$
                  \State $(\enum_1, \cfalse{}_1, \cfalser{}_1) \gets \infcheck(\solver,
                  p::b_r::\candrhead)$
                  \State $(\enum_2, \cfalse{}_2, \cfalser{}_2) \gets \infcheck(\vsinf(s_b,
                  \solver::p::b_r::\candrhead::\solver_r::\proginveval, \model,
                  \textit{right}), \textit{p}::b_r::\candrhead)$
                  \If{$(\enum_1 == \unsat \land
                        \enum_2 == \unsat)$}
                      \State \Return \Call{$\winf$}{$\code{while} \; (b) \;
                      (b_v) \;
                      (b_r :: \candrhead) \; \{ s_b \}, \solver, \model, \io,
                      \cand,
                      \candrtail, \proginv, \progrelinv$}
                  \Else \;
                      \Return \Call{$\winf$}{$\code{while} \; (b) \;
                      (b_v) \;
                      (b_r) \; \{ s_b \}, \solver, \model, \io, \cand,
                      \candrtail, \proginv, \progrelinv$}
                  \EndIf
                  \EndIf
                  \label{alg:weak-inf:end-rel}
                \EndCase
              \EndMatch
            \EndCase
          \EndMatch
        \EndFunction
    \end{algorithmic}
  \end{small}
  \caption{Weak Inference Algorithm}
  \label{fig:weak-inf}
\end{figure*}

Figures~\ref{fig:strong-inf} and \ref{fig:weak-inf} present \TOOL's loop
invariant inference algorithm.
\TOOL uses Houdini-style loop invariant inference \cite{flanagan2001houdini} to
reduce the annotation burden on the programmer.

\paragraph{\bf Preliminaries}
We denote a modified version of \TOOL's verification algorithm by the function
\(\vsinf\), which is identical to \(\vs\) except that it ignores calls to the
Verify function.

The value \(\enum \in \Enum = \{\sat, \unsat, \unknown\}\) represents the
response from the SMT solver \TOOL uses.
\(\sat\) indicates that a satisfying assignment exists for all variables in the
predicate.
\(\unsat\) indicates that no satisfying assignment exists for all variables in
the predicate.
\(\unknown\) indicates that the SMT solver cannot determine whether a
satisfying assignment exists.

The function \infcheck is identical to the Verify function except that:
\begin{itemize}
  \item \(\infcheck(\solver_1, \solver_2)\) does not halt execution if
    \(\solver_1\) does not imply \(\solver_2\).
  \item \infcheck returns a tuple consisting of:
    \begin{itemize}
      \item An SMT result \(\enum \in \Enum\).
      \item A set of false conjuncts from the loop invariant \(b_v\).
      \item A set of false conjuncts from the relational invariant
        \(b_r\).
    \end{itemize}
\end{itemize}

\paragraph{\bf Strong Inference.}

\autoref{fig:strong-inf} presents the strong inference algorithm.
Before checking a loop, \TOOL assembles the candidate invariants
\begin{align*}
  b_v &\equiv p_v \\
  b_r &\equiv p_r \land \left( \bigwedge_{x \in \text{vars}} x_r = x_o \right)
\end{align*}
where \(p_v\) is the invariant for the immediate parent loop or function,
\(p_r\) is the relational invariant for the immediate parent loop or function,
and vars is the set of program variables currently in scope.

\TOOL replaces the programmer provided invariants in the loop with \(b_v\) and
\(b_r\) and invokes the \sinf function.
It also provides the \sinf function with the programmer provided invariant as
\(\proginv\) and the  programmer provided relational invariant as
\(\progrelinv\).
\TOOL uses these invariants as assumptions during the inference process.
The algorithm has a different behavior for each of the verification modes:
\begin{itemize}
  \item \textbf{Lockstep.}
    The beginning of the lockstep algorithm
    (Lines~\ref{alg:strong-inf:begin-lock} through
    ~\ref{alg:strong-inf:end-lock-while}) is similar to the lockstep case for
    while loop verification, but we've replaced all invocations of Verify with
    invocations of \infcheck and all applications of \(\vs\) with applications
    of \(\vsinf\).

    The algorithm proceeds in three possible ways based on the results of the
    \infcheck function:
    \begin{itemize}
      \item If any of the three \infcheck results is \unknown, then \TOOL falls
        back on its weak inference algorithm (\cref{alg:strong-inf:lock-weak}).
      \item If any of the three \infcheck results is \sat, the \sinf function
        recurses with false conjuncts removed from \(b_v\) and \(b_r\)
        (\crefrange{alg:strong-inf:lock-sat-begin}{alg:strong-inf:lock-sat-end}).
      \item If all three of the \infcheck results are \unsat, then the
        algorithm has converged on a set of invariants and returns \((b_v,
        b_r)\) (\cref{alg:strong-inf:lock-unsat}).
    \end{itemize}
  \item \textbf{Left.}
    In left mode \TOOL does no invariant inference due to the fact that \TOOL
    does not verify loop invariants in left mode.
  \item \textbf{Right.}
    The beginning of the right algorithm
    (Lines~\ref{alg:strong-inf:begin-right} through
    ~\ref{alg:strong-inf:end-right-while}) is similar to the right case for
    while loop verification, but we've replaced all invocations of Verify with
    invocations of \infcheck and all applications of \(\vs\) with applications
    of \(\vsinf\).

    The algorithm proceeds in three possible ways based on the results of the
    \infcheck function:
    \begin{itemize}
      \item If any of the two \infcheck results is \unknown, then \TOOL falls
        back on its weak inference algorithm (\cref{alg:strong-inf:right-weak}).
      \item If any of the two \infcheck results is \sat, the \sinf function
        recurses with false conjuncts removed from \(b_v\) and \(b_r\)
        (\crefrange{alg:strong-inf:right-sat-begin}{alg:strong-inf:right-sat-end}).
      \item If both of the \infcheck results are \unsat, then the
        algorithm has converged on a set a invariants and returns
        \((b_v, b_r)\) (\cref{alg:strong-inf:right-unsat}).
    \end{itemize}
\end{itemize}

\paragraph{\bf Weak Inference.}

\Cref{fig:weak-inf} presents the weak inference algorithm.
\TOOL falls back on this algorithm when any call to the SMT solver returns
\unknown.
While the strong inference algorithm iteratively prunes a set of candidate
invariants, the weak inference algorithm builds up a set of invariants one at a
time from a set of candidates.
This is inherently weaker than the strong inference algorithm as it cannot
always infer invariants that depend on other invariants.
The \winf function takes these candidates as parameters (\(\cand\) for standard
invariants and \(\candr\) for relational invariants) in addition to the loop to
perform inference over and the programmer provided invariants for that loop.

The weak inference algorithm operates in three stages:
\begin{itemize}
  \item \textbf{Standard invariant inference
    (\crefrange{alg:weak-inf:start-stand}{alg:weak-inf:end-stand}).}
    \TOOL adds the head of the standard candidate invariant list
    (\(b_h\)) to the loop invariant then proceeds differently depending on the
    verification mode:
    \begin{itemize}
      \item \textbf{Lockstep.}
        The beginning of the lockstep algorithm
        (\crefrange{alg:weak-inf:begin-lock}{alg:weak-inf:end-lock-while}) is
        similar to the lockstep case for while loop verification, but we've
        replaced all invocations of Verify with invocations of \infcheck and
        all applications of \(\vs\) with applications of \(\vsinf\) and added
        the head of the candidate invariant list at each step.

        The algorithm proceeds in two possible ways based on the results of the
        \infcheck function:
        \begin{itemize}
          \item If all three of the \infcheck results are \unsat, then the
            algorithm recurses with \(\candhead\) appended to \(b_v\) and the
            tail of \(\cand\) as the candidate invariant list
            (\cref{alg:weak-inf:lock-unsat}).
          \item If any of the three \infcheck results are not \unsat, then the
            algorithm discards the candidate invariant and recurses
            (\cref{alg:weak-inf:lock-not-unsat}).
        \end{itemize}
      \item \textbf{Left.}
        In left mode \TOOL does no invariant inference due to the fact that \TOOL
        does not verify loop invariants in left mode.
      \item \textbf{Right.}
        The beginning of the right algorithm
        (\crefrange{alg:weak-inf:begin-right}{alg:weak-inf:end-right-while}) is
        similar to the right case for while loop verification, but we've
        replaced all invocations of Verify with invocations of \infcheck and
        all applications of \(\vs\) with applications of \(\vsinf\) and added
        the head of the candidate invariant list at each step.

        The algorithm proceeds in two possible ways based on the results of the
        \infcheck function:
        \begin{itemize}
          \item If both of the \infcheck results are \unsat, then the
            algorithm recurses with \(\candhead\) appended to \(b_v\) and the
            tail of \(\cand\) as the candidate invariant list
            (\cref{alg:weak-inf:right-unsat}).
          \item If any of the two \infcheck results are not \unsat, then the
            algorithm discards the candidate invariant and recurses
            (\cref{alg:weak-inf:right-not-unsat}).
        \end{itemize}
    \end{itemize}
  \item \textbf{Relational invariant inference
    (\crefrange{alg:weak-inf:start-rel}{alg:weak-inf:end-rel}).}
    After exhausting the standard candidate invariant list, \TOOL iterates
    through the relational candidate invariant list.
    This process is identical to the previous stage but uses \(\candr\) in
    place of \(\cand\).
  \item \textbf{Base case (\cref{alg:weak-inf:base}).}
    When no candidate invariants remain, \winf returns the pair of invariants
    \((b_v, b_r)\).
\end{itemize}

\section{Self-correcting Connected Components}
\label{sec:cc}

\autoref{fig:cc} presents an implementation of self-correcting connected
components (SC-CC)~\cite{sao2016self}, an iterative algorithm that computes the
connected components of an input graph.
A connected component is a subgraph in which every pair of vertices in the
subgraph is connected through some path, but no vertex is connected to another
vertex that is not also in the subgraph.

The standard connected components algorithm begins by constructing a vector
\(CC^0\) and initializing this vector such that \(\forall v.\ CC^0[v] = v\).
Then, on iteration \(i\) for each node \(v\) the algorithm looks up the value
of each of \(v\)'s neighbors in \(CC^{i-1}\) and sets \(CC^{i}[v]\) to the minimum
of its neighbors and \(CC^{i-1}[v]\).  In other words,
\begin{equation}\label{eqn:cc}
  CC^i[v] = \min_{j \in \mathcal{N}(v)}CC^{i-1}[j]
\end{equation}
where \(\mathcal{N}(v)\) is the union of \(v\) and the neighbors of node \(v\).
The algorithm iterates this process until no elements in \(CC\) are updated at
which point it has converged.

Self-correcting connected components adds an additional step of checking \(CC^i\) after each iteration to verify that it is valid and has not been corrupted by memory errors.
If SC-CC detects an error at \(CC^i[v]\), it repeats the computation for node
\(v\) with reliably backed storage.

Our implementation allows errors when writing \(CC^i\) so long as the errors are sufficiently large.
Therefore, we consider \(CC^i\) to be valid if \(\forall v.\ 0 \leq CC^i[v] <
|V|\) and in all other cases SC-CC corrects the invalid positions.
When this property holds, then after each iteration \code{CC<o>~==~CC<r>}, even though intermediate values may differ during faulty execution.

The original SC-CC algorithm described by Sao et al.\ contains an additional
data structure \code{P*} and permits a larger class of errors than this
implementation does.  However, this flexibility comes at a cost: the original
algorithm is not guaranteed to converge.
As such, we modified the algorithm to prove strong convergence properties that
the oringal does not provide.

\paragraph{\bf Self Correction.}
SC-CC is \emph{self correcting}.
This means that given some valid state, SC-CC can correct errors encountered
during each iteration.
In this case, if an error occurs at iteration \(i\), SC-CC can correct \(CC^i\)
using data from \(CC^{i-1}\).
Therefore, SC-CC always stores \(CC\) for the previous iteration correctly.
This is weaker than self-stabilizing algorithms which may correct themselves
from any state and do not rely on certain state elements remaining uncorrupted.

\subsection{SC-CC Implementation}
\begin{figure}
\begin{wraplst}
  \begin{lstlisting}[basicstyle=\ttfamily\footnotesize]
requires N < max_N ~\label{cc:n-bound}~ requires_r eq(adj) ~\label{cc:req-eq-adj}~
vector<uint> cc(uint N, matrix<uint> adj(N, N)) {
  vector<uint> CC(N);                            ~\label{cc:begin-decls}~
  @region(unreliable) vector<uint> next_CC(N);   ~\label{cc:end-decls}~

  for (uint v = 0; v < N; ++v) invariant_r vec_bound(CC, v) {CC[v] = v;} ~\label{cc:init-cc}~ 
  uint N_s = N;

  @noinf while (0 < N_s)                         ~\label{cc:outer}~
      invariant N < max_N ~\label{cc:n-bound-while}~
      invariant_r vec_bound(CC, N)
      invariant_r eq(N) && eq(adj) && eq(N_s) && eq(CC) ~\label{cc:outer-eq}~ {

    next_CC = CC; ~\label{cc:faulty-init}~N_s = 0; ~\label{cc:ns1}~model.reliable = false; ~\label{cc:unreliable}~
    for (uint v = 0; v < N; ++v) ~\label{cc:faulty-outer}~
        invariant_r vec_bound(next_CC, N)
        invariant_r large_error_r(next_CC, N) ~\label{cc:faulty-large-err}~
        invariant_r ~$\forall$~(uint fi)((v<o><=fi<N<o>) -> next_CC<o>[fi]==CC<o>[fi])
        invariant_r outer_spec(v<o>, N<o>, next_CC<o>, CC<o>, adj<o>) { ~\label{cc:faulty-outer-spec}~

      for (uint j = 0; j < N; ++j) ~\label{cc:faulty-inner}~
          invariant v < N && N < max_N ~\label{cc:bounds-inner}~
          invariant_r ~$\forall$~(uint fi)((v<o><fi<N<o>) -> next_CC<o>[fi]==CC<o>[fi]) ~\label{cc:faulty-inner-spec-begin}~
          invariant_r inner_spec(j<o>, v<o>, next_CC<o>, CC<o>, adj<o>) {               ~\label{cc:faulty-inner-spec}~
        if (CC[j] < next_CC[v] && next_CC[v] <= v && adj[v][j] == 1) { ~\label{cc:faulty-cond}~
          next_CC[v] = CC[j];                    ~\label{cc:faulty-copy}~
        }
      }
    }
    model.reliable = true; ~\label{cc:reliable}~
    @noinf @label(outer_correction) for (uint v = 0; v < N; ++v)          ~\label{cc:corr-outer}~
        invariant_r outer_spec(v<r>, N<r>, next_CC<r>, CC<r>, adj<r>)      ~\label{cc:corr-outer-spec}~
        invariant_r eq(N) && eq(CC) && eq(adj) && eq(v) && eq(N_s) ~\label{cc:corr-eq-ns}~
        invariant_r forall(uint fi)(((fi < v<r>) -> ~\label{cc:corr-eq}~
            (next_CC<r>[fi] == next_CC<o>[fi])))
        invariant_r vec_bound(next_CC, N)
        invariant_r model.reliable
        invariant_r large_error_r_inclusive(next_CC, v, N) ~\label{cc:corr-large-err}~
        invariant_r outer_spec(N<o>, N<o>, next_CC<o>, CC<o>, adj<o>) { ~\label{cc:corr-next-spec}~
      if (v < corrected_next_CC[v]) {  ~\label{cc:detect}~
        next_CC[v] = CC[v];
        for (uint j = 0; j < N; ++j)   ~\label{cc:corr-inner}~
            invariant v < N && v < outer_correction[next_CC[v]]
            invariant_r inner_spec(j<r> , v<r>, next_CC<r> CC<r>, adj<r>) ~\label{cc:corr-inner-spec}~
            invariant_r large_error_r_exclusive(next_CC, v, N) {
          if (CC[j] < next_CC[v] && adj[v][j] == 1) { next_CC[v] = CC[j]; }
        }
      }
      if (next_CC[v] < CC[v]) {++N_s;} ~\label{cc:corr-ns}~
    }
    CC = corrected_next_CC;                      ~\label{cc:cc-update}~
  }
  return CC; ~\label{cc:return}~
}

    \end{lstlisting}
    \end{wraplst}
    \caption{Self-Correcting Connected Components}
    \label{fig:cc}
\end{figure}

The overall structure of the SC-CC implementation is as follows:
\begin{itemize}
  \item \textbf{Initialization.}
    The \code{cc} function takes a description of a graph in the form of an
    adjacency matrix (\code{adj}).
    It then declares and initializes \code{CC}, which holds the result of the previous
    iteration.
    It also declares \code{next_CC}, which holds the result of the current
    iteration, in the \code{unreliable} memory region.
  \item \textbf{Outer while loop (Line~\ref{cc:outer})}
    The outer while loop computes the next iteration of \code{CC}.
    It converges when the algorithm makes no changes to \code{CC} over the
    course of a single iteration.
  \item \textbf{Faulty step (Line~\ref{cc:faulty-outer})}
    The faulty step computes \autoref{eqn:cc} element-wise over \code{next_CC}.
    The inner loop allows errors during writes to \code{next_CC}, which SC-CC
    will correct in the correction step.
    Prior to the entrance of the outer loop, SC-CC sets the
    \code{model.reliable} to false to permit errors (\cref{cc:unreliable}).
  \item \textbf{Correction step (Line~\ref{cc:corr-outer})}
    The correction step detects and corrects errors in \code{next_CC}.
    First, it sets the \code{model.reliable} flag to true to prevent further
    errors in this iteration (\cref{cc:reliable}).
    If \code{next_CC} contains an error at index \code{v}, the implementation
    reliably computes \code{next_CC[v]} using \autoref{eqn:cc}.
    After correcting \code{next_CC}, the implementation sets \code{CC}
    equal to \code{next_CC} and begins the next iteration.
\end{itemize}

\paragraph{\bf Constants and Properties.}
\begin{figure}
\begin{wraplst}
    \begin{lstlisting}
const real max_N = ..; ~\label{cc-prop:max-n}~

property_r vec_bound(vector<uint> V, uint i) : ~\label{cc-prop:vec-bound}~
    forall(uint j)((j < i<o>) -> (V<o>[j] <= j));

property_r large_error_r(vector<uint> V, uint i) : ~\label{cc-prop:large-err}~
  forall(uint j)((j < i<r>) ->
      (V<r>[j] == V<o>[j] || j < V<r>[j]));

property_r large_error_r_inclusive(vector<uint> V, uint from, uint to) : ~\label{cc-prop:large-err-inc}~
  forall(uint j)((from<r> <= j < to<r>) ->
      (V<r>[j] == V<o>[j] || j < V<r>[j]));

property_r large_error_r_exclusive(vector<uint> V, uint from, uint to) : ~\label{cc-prop:large-err-ex}~
  forall(uint j)((from<r> < j < to<r>) ->
      (V<r>[j] == V<o>[j] || j < V<r>[j]));

property_r outer_spec(uint to, uint N, vector<uint> next_CC, ~\label{cc-prop:outer}~
                      vector<uint> CC, matrix<uint> adj) :
  forall(uint fi)((fi < to) ->
      (forall(uint fj)((fj < N && adj[fi][fj] == 1) ->
          next_CC[fi] <= CC[fj]) &&
       next_CC[fi] <= CC[fi] &&
       (exists(uint ej)(next_CC[fi] == CC[ej] && ej < N &&
            adj[fi][ej] == 1) ||
        next_CC[fi] == CC[fi])));

property_r inner_spec(uint to, uint v, uint N, vector<uint> next_CC, ~\label{cc-prop:inner}~
                      vector<uint> CC, matrix<uint> adj) :
  forall(uint fi)((fi < to && adj[v][fi] == 1) ->
      next_CC[v] <= CC[fi]) &&
  next_CC[v] <= CC[v] &&
  (exists(uint ei)(next_CC[v] == CC[ei] && ei < N &&
       adj[v][ei] == 1) ||
   next_CC[v] == CC[v]);
    \end{lstlisting}
    \end{wraplst}
    \caption{Constant and Properties for Self-Correcting Connected Components}
    \label{fig:cc-props}
\end{figure}

SC-CC uses the following constants and properties, found in
\autoref{fig:cc-props}:
\begin{itemize}
  \item \textbf{\code{max\_N} (\autoref{cc-prop:max-n}).}
    This constant bounds the maximum number of nodes an input graph may
    contain.
  \item \textbf{\code{vec\_bound} (\autoref{cc-prop:vec-bound}).}
    This property takes a vector \code{V} and an index \code{i} and stipulates
    that
    \(
      \forall \; \code{j} < \code{i<o>}. \; \code{V<o>[j]} \leq \code{j}.
    \)
\item \textbf{\code{large_error_r} (\autoref{cc-prop:large-err}).}
  This property takes a vector \code{V} and an index \code{i} and asserts that
    \(
      \forall \; \code{j} < \code{i<r>}. \; \code{V<r>[j]} = \code{V<o>[j]}
      \lor
      \code{j} < \code{V<r>[j]}.
    \)
\item \textbf{\code{large_error_r_inclusive} (\autoref{cc-prop:large-err-inc}).}
  This property takes a vector \code{V}, an index \code{from}, an index
    \code{to} and asserts that
    \(
      \forall \; \code{i}. \; (\code{from<r>} \leq \code{i} < \code{to<r>}) \limp  \; \code{V<r>[i]} = \code{V<o>[i]}
      \lor
      \code{i} < \code{V<r>[i]}.
    \)
\item \textbf{\code{large_error_r_exclusive} (\autoref{cc-prop:large-err-ex}).}
  This property takes a vector \code{V}, an index \code{from}, an index
    \code{to} and asserts that
    \(
      \forall \; \code{i}. \; (\code{from<r>} < \code{i} < \code{to<r>}) \limp  \; \code{V<r>[i]} = \code{V<o>[i]}
      \lor
      \code{i} < \code{V<r>[i]}.
    \)
\item \textbf{\code{outer\_spec} (\autoref{cc-prop:outer}).}
    This property takes an index \code{to}, a size \code{N}, a vector
    \code{next_CC}, a vector \code{CC}, and an adjacency matrix \code{adj}.
    It ensures that every element of \code{next_CC} from index 0 to index
    \code{to} (exclusive) satisfies \autoref{eqn:cc}.
  \item \textbf{\code{inner\_spec} (\autoref{cc-prop:inner}).}
    This property takes an index \code{to}, an index \code{v}, a size \code{N},
    a vector \code{next_CC}, a vector \code{CC}, and an adjacency matrix
    \code{adj}.
    It ensures that \code{next_CC[v]} satisfies \autoref{eqn:cc} up to the
    neighbor at index \code{to}.
    That is,
    \[
      \code{next_CC[v]} = \min_{j \in \mathcal{N}(\code{v, to})} \code{CC[j]}
    \]
    where \(\mathcal{N}(\code{v, to})\) is the union of \code{v} and the
    neighbors of node \code{v} up to (but not including) nodes with the id
    \code{to}.
\end{itemize}

\paragraph{\bf Error Model.}
I evaluate SC-CC under the switchable rowhammer fault model from \autoref{fig:big-err}.
This model provides two implementations for the write operator in the memory region \code{unreliable}.
The first implementation is fully reliable while the second allows for errors so long as they are larger than the programmer specified constant \code{min_error}.

I use this fault model that allows only faulty writes because it captures all
possible rowhammer attacks over high order bits in the elements of
\code{next_CC}.
A rowhammer attack allows an attacker to selectively flip bits in DRAM by
issuing frequent reads on DRAM rows surrounding the row under
attack~\cite{rowhammer}.
This drains capacitors in the attacked row and therefore permanently flips
bits in that row.
As empty capacitors may indicate a 0 or 1 depending on location on the
chip, this attack does not always flip bits from 1 to 0.
Although researchers have devised protections to address rowhammer
attacks~\cite{kim2015architectural, rowhammer}, the JDEC Solid State Technology
Association did not include these protections in the DDR4
standard~\cite{ddr4spec} and Mark Lanteigne has demonstrated rowhammer attacks
on some DDR4 memory~\cite{ddr4rowhammer}.

Given that some regions of memory may be more vulnerable to rowhammer attacks
than others~\cite{rowhammer}, I place \code{next_CC} in an unreliable region
prone to attacks (\autoref{cc:end-decls}) and all other variables in reliable
memory.

\subsection{Specification}
We use \TOOL's specification abilities to verify the error detection,
self-correction, and convergence properties of SC-CC.

\paragraph{\bf Perfect Error Detection.}
To verify that this SC-CC implementation detects all errors, the developer must
verify that \code{N < max_N}, where \code{N} is the number of nodes in the
input graph and \code{max_N} is a programmer specified value bounding the
maximum graph size a calling function may provide.
The developer must also ensure that \code{max_N} is less than the magnitude of
the largest error they expect to encounter.
This property ensures that any error will be larger than \code{N}, which in
turn ensures that SC-CC detects all errors as errors result in trivially
invalid values.
SC-CC specifies this property as a unary prerequisite to calling the \code{cc}
function on Line~\ref{cc:n-bound}.
It also specifies this property as unary invariants that must hold before and
after each loop iteration on Lines~\ref{cc:n-bound-while} and
\ref{cc:bounds-inner}.
Although not explicitly stated, \TOOL infers this invariant for the loop on
Line~\ref{cc:faulty-outer}.

Error detection also requires that impact of errors on \code{next_CC<r>} is
large enough to be detected.
Specifically, it is necessary that
\[
  \forall i.\ \code{next_CC<r>}[i] > i \lor \code{next_CC<r> == next_CC<o>}.
\]
This property ensures that for every element \(e\) of \code{next\_CC<r>} at
index \(i\), \(e\) is in one of two states:
\begin{itemize}
  \item \textbf{Detectable Error.}
    When \(e > i\), \(e\) violates the property that every element of \(CC\)
    does not increase from its initialization value.
    The check in the conditional on \cref{cc:detect} of the correction step
    trivially detects this error.
  \item \textbf{Equality.}
    When \(e = \code{next\_CC<o>}[i]\), \(e\) is correctly computed.
\end{itemize}
Note that \(e\) can not be in a state where it contains an undetectable error.
Loop invariants on Lines~\ref{cc:faulty-large-err} and
\ref{cc:corr-large-err} (\code{large_error_r(next_CC, N)}) ensure this
property.

The final property SC-CC requires to ensure that it has perfect error
detection is \(\code{eq(adj)}\).
This property asserts that the graphs both executions operate over are
equivalent, and unable to experience errors.
SC-CC enforces this property by specifying it as a requirement to call the \code{cc}
function on Line~\ref{cc:req-eq-adj}.
On all loops it either explicitly specify \code{eq(adj)}, or \TOOL infers it.

\paragraph{\bf Self-Correction.}
To verify that SC-CC is self correcting, the implementation first verifies that
every element of \code{next_CC<o>} satisfies Equation~\ref{eqn:cc}.
It capture this with the \code{outer_spec} property application on
Line~\ref{cc:faulty-outer-spec} and the \code{inner_spec} property application
on Line~\ref{cc:faulty-inner-spec}.
Both of these properties capture the semantics of the min operator.
After exiting inner faulty loop (\code{j == N}), \code{inner_spec(j<o>, v<o>,
next_CC<o>, CC<o>, adj<o>)} is equivalent to Equation~\ref{eqn:cc} at index
\(v\).
Similarly, after exiting the outer loop (\code{v == N}), the application
\code{outer_spec(v<o>, N<o>, next_CC<o>, CC<o>, adj<o>)} is equivalent to
Equation~\ref{eqn:cc} at all indices, or
\[
  \forall v. \; CC^i[v] = \min_{j \in \mathcal{N}(v)} CC^{i-1}[j].
\]

The correction step applies the same two properties (\code{outer_spec} and
\code{inner_spec}) in the same fashion to specify that \code{next_CC<r>} also
satisfies Equation~\ref{eqn:cc} at all indices.

SC-CC passes the specification of \code{next_CC<o>} into the correction loop on
Line~\ref{cc:corr-next-spec} where \TOOL combines it with the specification of
\code{next_CC<r>} to prove that
\[
  \forall \code{i} < \code{v<r>}. \; \code{next_CC<r>[i] ==
  next_CC<o>[i]},
\]
stated on Line~\ref{cc:corr-eq}.

Finally, with the assignment \code{CC = next_CC} Leto can verify the
outer loop invariant \code{eq(CC)} (Line~\ref{cc:outer-eq}) thus proving that
SC-CC is self correcting.

\paragraph{\bf Convergence Equality.}
Given that SC-CC is self correcting and detects all errors,
it is trivial to see that both executions converge in the same number of
iterations.
That is, the outer while loop (Line~\ref{cc:outer}) must run in lockstep.
To demonstrate this fact to \TOOL SC-CC uses the \code{eq(N_s)} invariant on
Lines~\ref{cc:outer-eq} and \ref{cc:corr-eq-ns}.

\code{N_s} itself is updated in two places:
\begin{itemize}
  \item \textbf{Line~\ref{cc:ns1}}. SC-CC sets \code{N_s} to 0 at the top of
    the outer loop.
    As \code{N_s} is stored in reliable memory, it is clear that \code{N_s<o> ==
    N_s<r>}.
  \item \textbf{Line~\ref{cc:corr-ns}}.
    SC-CC increments \code{N_s} at this location if \code{next_CC[v] <
    CC[v]}.
    From the surrounding loop invariants \TOOL knows that
    \[
      \code{next_cc<o>[v<o>] == next_CC<r>[v<r>]}
    \]
    and
    \[
      \code{CC<o>[v<o>] == CC<r>[v<r>]}
    \]
    so the if statement must execute in lockstep.
    Therefore, if \code{N_s} was equal across both executions before the if
    statement, then it is equal afterwards as the implementations performs the
    increment to \code{N_s} reliably.
\end{itemize}

\TOOL realizes that \code{N_s} is equal across both executions after both
assignments and therefore \code{eq(N_s)} must hold in both loop invariants.
This forces the outer while loop into a lockstep execution and proves that
convergence time is equal under relaxed and reliable execution semantics.

\subsection{Verification Approach}
I next demonstrate how the developer works with \TOOL to verify that the
implementation meets these specifications.

\paragraph{\bf Precondition.}
The verification algorithm begins with the preconditions on \code{cc}.
The precondition stipulates that \code{N} be less than \code{max_N} and that
the graph be equal across both executions (\code{eq(adj)}).
\TOOL adds these preconditions as assumptions to its context.

\paragraph{\bf Initialization}

On Line~\ref{cc:init-cc} SC-CC initializes the \code{CC} vector such that \(\forall
\code{v}.\ \code{CC[v]} = \code{v}\).
The loop invariant verifies that all elements with index \code{v} in \code{CC} are between 0 and \code{v}.
This property is critical in the detection of errors and must hold for \code{CC} after each iteration of the connected components algorithm.

\paragraph{\bf Outer Loop}
The loop on \autoref{cc:outer} runs the iterative portion of the algorithm.
The loop enforces the critical invariants \code{vec_bound(CC, N)}, which
ensures that \(\forall \code{v}.\ \code{CC[v]} \leq \code{v}\), and
\code{eq(CC)}.
It also contains the invariant \code{eq(N_s)}, which ensures that the loop runs
in lockstep.
Lastly, it enforces the invariants \code{eq(N)} and \code{eq(adj)} which ensure
that the input graph is identical across both executions.

The loop also sports the \code{@noinf} annotation, which disables inference
over this loop.
SC-CC disables inference on this loop because the inference algorithm's time
complexity is exponential in the depth of nested loops.

\paragraph{\bf Faulty Middle Loop.}
Verification then proceeds to the faulty middle loop on
\autoref{cc:faulty-outer}.
This loop contains the following invariants:
\begin{itemize}
  \item \textbf{\code{vec_bound(next_CC, N).}}
    This invariant enforces that elements of \code{next_CC<o>} are bounded by
    their respective indices.
    This fact is important to pass on to the correction step as it implies that
    the original execution never runs the inner correction loop.
  \item \textbf{\code{large_error_r(next_CC, N)}.}
    This invariant enforces that errors to \code{next_CC<r>} are large enough
    to be detected.
    It enables the implementation to detect and correct all errors during the
    correction step.
  \item \textbf{\code{forall(uint fi)((v<o> <= fi < N<o>) -> next_CC<o>[fi] ==
    CC<o>[fi])}.}\\
    This invariant states that elements in \code{next_CC} that the
    implementation hasn't yet updated are still equal to \code{CC}.
    This is necessary as it communicates to \TOOL that if an element is not
    updated on this iteration, then it is already the minimum of its neighbors.
  \item \textbf{\code{outer_spec(v<o>, N<o>, next_CC<o>, CC<o>, adj<o>)}.}
    This invariant specifies the contents of \code{next_CC<o>}.
    I use it to pass information about \code{next_CC<o>} on to the correction
    step.
\end{itemize}

\paragraph{\bf Faulty Inner Loop.}
Verification then continues to the loop on \autoref{cc:faulty-inner}, where it
encounters the following invariants:
\begin{itemize}
  \item \textbf{\code{v < N}.}
    This invariant bounds \code{v} so that \TOOL knows that the vector accesses
    within this loop are in bounds.
  \item \textbf{\code{N < max_N}.}
    This invariant bounds the maximum size of the graph so that \TOOL knows
    that errors are larger than the maximum graph size and are therefore
    detectable.
  \item \textbf{\code{forall(uint fi)((v<o> < fi < N<o>) -> next_CC<o>[fi] ==
    CC<o>[fi])}.}\\
    This invariant serves to pass on that although the implementation may have
    altered \code{next_CC<o>[v<o>]}, the other elements that \TOOL previously
    knew were equal to \code{CC<o>} in the middle faulty loop are still equal
    to \code{CC<o>}.
  \item \textbf{\code{inner_spec(j<o>, v<o>, next_CC<o>, CC<o>, adj<o>)}.}
    This invariant specifies the contents of \code{next_CC<o>[v<o>]}.
    It serves only to pass this information on to the outer loop so
    \TOOL may verify the \code{outer_spec} invariant in that loop.
\end{itemize}

\paragraph{\bf Correction Middle Loop.}
Verification next proceeds to the loop on \autoref{cc:corr-outer}.
This loop contains the following notable invariants:
\begin{itemize}
  \item \textbf{\code{outer_spec(v<r>, N<r>, next_CC<r>, CC<r>,
    adj<r>)}.}
    This invariant specifies the contents of \code{next_CC<r>}.
    \TOOL combines it with the other \code{outer_spec} invariant to verify
    \code{next_CC<o> == next_CC<r>} after exiting the loop.
  \item \textbf{\code{eq(v)}.}
    This invariant states that \code{v<o> == v<r>}.
    It informs \TOOL that this loop must run in lockstep.
  \item \textbf{\code{forall(uint fi)(((0 <= fi < v<r>) ->
    (next_CC<r>[fi] == next_CC<o>[fi])))}.}
    This invariant states that elements in \code{next_CC} that the
    implementation has updated are equal across both executions.
    This allows \TOOL to prove \code{eq(CC)} at the end of the outer while
    loop.
  \item \textbf{\code{vec_bound(next_CC, N)}.}
    This invariant enforces that elements of \code{next_CC<o>} are bounded by
    their respective indices.
    This implies that the original execution never runs the inner correction
    loop.
    Therefore, \TOOL prunes any paths in which the original execution runs
    through the inner loop.
  \item \textbf{\code{model.reliable}.}
    This invariant enforces that the correction step runs reliably.
  \item \textbf{\code{large_error_r_inclusive(next_CC, v, N)}.}
    This invariant enforces that errors to \code{next_CC<r>} are large enough
    to be detected.
    It enables the implementation to detect and correct all errors during the
    course of this loop.
  \item \textbf{\code{outer_spec(N<o>, N<o>, next_CC<o>, CC<o>, adj<o>)}.}
    This invariant passes in the specification for \code{next_CC<o>} from the
    faulty loop.
    \TOOL combines it with specifications over \code{next_CC<r>} to
    prove that \code{next_CC<o>[v<o>]} is equivalent to
    \code{next_CC<r>[v<r>]} at the end of the current loop
    iteration.
\end{itemize}

\paragraph{\bf Correction Inner Loop.}
Verification then continues to the loop on \autoref{cc:corr-inner} which
contains three developer specified invariants:
\begin{itemize}
  \item \textbf{\code{v < N}.}
    This invariant bounds \code{v} so that \TOOL knows that the vector accesses
    within this loop are in bounds.
  \item \textbf{\code{v < outer_correction[next_CC[v]]}.}
    The syntax \code{outer_correction[next_CC[v]]} refers to the value of
    \code{next_CC[v]} at the top of the loop labeled \code{outer_correction}.
    Therefore, this invariant states that \code{v} is less than
    \code{next_CC[v]} at the top of the outer correction loop.
    The conditional that contains this loop implies this invariant.
  \item \textbf{\code{large_error_r_exclusive(next_CC, v, N)}.}
    This invariant enforces that errors to \code{next_CC<r>} are large enough
    to be detected.
    It enables the implementation to detect and correct all errors during the
    course of this loop.
  \item \textbf{\code{inner_spec(j<r> , v<r>, corrected_next_CC<r> CC<r>, adj<r>)}.}
    This invariant specifies the contents of \code{corrected_next_CC<r>[v<r>]}.
    It serves only to pass this information on to the outer loop so that
    \TOOL may verify the \code{outer_spec} in that loop.

\end{itemize}

\section{Self-stabilizing Conjugate Gradient Descent}
\label{sec:cg}

We use \TOOL to verify an implementation of self-stabilizing conjugate gradient
(SS-CG)~\cite{sao2013self} and present relevant snippets required for verification
below.  Conjugate gradient descent is another method for solving linear systems
of equations. However, unlike Jacobi, the standard conjugate gradient method is
sensitive to errors that may corrupt internal state variables and, therefore,
is not naturally self-stabilizing. SS-CG employs a periodic correction step to
recalculate appropriate values for internal state variables from the current
estimated solution vector \code{x} and the input matrix \code{A}.

The standard conjugate gradient descent algorithm computes the next iteration's
variables as follows:
\begin{align}
  q_i &= A p_i \label{eqn:faulty-matvec} \\
  \alpha_i &= \frac{r_i^Tr_i}{p_i^T q} \\
  x_{i+1} &= x_i + \alpha  p_i \\
  r_{i+1} &= r_i - \alpha  q_i \\
  \beta &= \frac{r_{i+1}^T r_{i+1}}{r_i^T r_i} \\
  p_{i+1} &= r_{i+1} + \beta p_i
\end{align}

SS-CG adds a periodic correction step to repair state variables that may have
been corrupted by errors.
Unlike the previous steps, the repair step must be computed reliably.
This repair step computes:
\begin{align}
  r_i' &= A x_i \label{eqn:matvec1} \\
  q_i &= A p_i \label{eqn:matvec2} \\
  r_i &= b - r_i' \\
  \alpha_i &= \frac{r_i^T p_i}{p_i^T q} \\
  x_{i+1} &= x_i + \alpha p_i \\
  r_{i+1} &= r_i - \alpha q \\
  \beta &= -\frac{r_{i+1}^T q_i}{p_i^T q} \\
  p_{i+1} &= r_{i+1} + \beta p_i
\end{align}

We use \TOOL to verify two properties of SS-CG that are necessary for self-stability under
the SEU model we present in \cref{fig:seu}:
\begin{itemize}
  \item \textbf{Reliable Correction Step.}
    We verify that it is possible to correct errors even when the matrix vector
    products in Equations~\ref{eqn:matvec1} and \ref{eqn:matvec2} may
    experience faults.
    To accomplish this, the implementation uses dual modular redundancy (DMR)
    to duplicate arithmetic instructions and repeat the correction step until
    the result of both sets of instructions agree with each other.
  \item \textbf{Correctable Errors.}
    We verify that errors in the matrix vector product from
    \autoref{eqn:faulty-matvec} are sufficiently small.
    Specifically, SS-CG requires that if an element of \(q\) is corrupted by
    \(\epsilon\), then
    \[
      \epsilon^2 < \max_{i, j}(A[i][j])^2.
    \]
\end{itemize}

We present the code snippet for the reliable correction step in
\autoref{fig:cg-correct} and the snippet for correctable errors in
\autoref{fig:cg-small}.
We have included the full implementation in~\autoref{sec:cg-full}.

\subsection{Implementation}

\subsubsection{Reliable Correction Step}
\begin{figure}
\begin{wraplst}
\begin{lstlisting}
property_r dmr_eq(vector<real> x1, vector<real> x2, vector<real> sx) : ~\label{cgc:eq}~
  !model.upset -> (x1<r> == sx && x2<r> == sx);

property_r dmr_imp(vector<real> x1, vector<real> x2, vector<real> sx) : ~\label{cgc:imp}~
  (x1<r> == x2<r>) -> (x1<r> == sx);

vector<real> r2(N), q2(N); ~\label{cgc:init}~
specvar vector<real> spec_r(N), spec_q(N);
r = r2 = spec_r = q = q2 = spec_q = zeros;
bool not_run = true;

@noinf while (not_run || r != r2 || q != q2) ~\label{cgc:while}~
             invariant_r dmr_eq(r, r2, spec_r) ~\label{cgc:while-begin-invs}~
             invariant_r dmr_eq(q, q2, spec_q)
             invariant_r dmr_imp(r, r2, spec_r)
             invariant_r dmr_imp(q, q2, spec_q) ~\label{cgc:while-end-invs}~
{
  not_run = false;

  for (uint i = 0; i < N; ++i) ~\label{cgc:middle-for}~
  {
    for (uint j = 0; j < N; ++j) ~\label{cgc:inner-for}~
    {
      real tmp = A[i][j] *. x[j];
      real tmp2 = A[i][j] *. x[j];
      specvar real spec_tmp = A[i][j] * x[j];
      r[i] = r[i] +. tmp;
      r2[i] = r2[i] +. tmp2;
      spec_r[i] = spec_r[i] + spec_tmp;

      tmp = A[i][j] *. p[j];
      tmp2 = A[i][j] *. p[j];
      spec_tmp = A[i][j] * p[j];
      q[i] = q[i] +. tmp;
      q2[i] = q2[i] +. tmp2;
      spec_q[i] = spec_q[i] + spec_tmp;
    }
  }
}

assert_r (!outer_while[model.upset] -> (r<r> == spec_r)); ~\label{cgc:assert-r}~
assert_r (!outer_while[model.upset] -> (q<r> == spec_q)); ~\label{cgc:assert-q}~
\end{lstlisting}
\end{wraplst}
\caption{SS-CG Correction Step}
\label{fig:cg-correct}
\end{figure}

The SS-CG correction step we present in \autoref{fig:cg-correct} operates
over the following pre-existing variables:
\begin{itemize}
  \item \textbf{\code{A}.}
    \code{A} is a matrix of coefficients.
  \item \textbf{\code{x}.}
    \code{x} is a solution vector.
  \item \textbf{\code{r}.}
    \code{r} holds the residual of the current iteration.
  \item \textbf{\code{p} and \code{q}}.
    \code{p} and \code{q} are vectors of loop carried state.
\end{itemize}

The overall structure of the SS-CG correction step is as follows:
\begin{itemize}
  \item \textbf{Initialization (\autoref{cgc:init}).}
    Initialization declares the following variables:
    \begin{itemize}
      \item \textbf{\code{r2} and \code{q2}.}
        The algorithm computes \code{r2} according to \autoref{eqn:matvec1} and
        \code{q2} according to \autoref{eqn:matvec2}.
        It then uses \code{r2} and \code{q2} to verify that it has correctly
        computed \code{r} and \code{q} respectively.
      \item \textbf{\code{spec\_r} and \code{spec\_q}.}
        \code{spec\_r} and \code{spec\_q} are specification variables that also
        compute \autoref{eqn:matvec1} and \autoref{eqn:matvec2} respectively.
        Unlike \code{r}, \code{r2}, \code{q}, and \code{q2}, the implementation
        computes these specification variables reliably.
      \item \textbf{Outer while loop (\autoref{cgc:while}).}
        The outer while loop repeats the correction step until \code{r == r2}
        and \code{q == q2}.
      \item \textbf{Middle for loop (\autoref{cgc:middle-for}).}
        The middle for loop computes \autoref{eqn:matvec1} element-wise for
        \code{r}, \code{r2}, and \code{spec\_r} and \autoref{eqn:matvec2}
        element-wise for \code{q}, \code{q2}, and \code{spec\_q}.
      \item \textbf{Inner for loop (\autoref{cgc:inner-for}).}
        The inner for loop computes the matrix vector products:
        \begin{align*}
          \code{r} &= \code{A * x} \\
          \code{q} &= \code{A * p}
        \end{align*}
        It computes \code{r2} and \code{spec\_r} similarly to \code{r}, and
        \code{q2} and \code{spec\_q} similarly to \code{q}.
        The algorithm permits errors in the computation of \code{r}, \code{r2},
        \code{q}, and \code{q2}, but not in \code{spec\_r} or \code{spec\_q}.
    \end{itemize}
\end{itemize}

\paragraph{\bf Properties.}
The SS-CG correction step implementation uses the following properties, found in
\autoref{fig:cg-correct}:
\begin{itemize}
  \item \textbf{\code{dmr\_eq} (\autoref{cgc:eq}).}
    This property asserts that if an upset has not occurred, then \code{x1} and
    \code{x2} are both equal to the specification variable \code{sx}.
  \item \textbf{\code{dmr\_imp} (\autoref{cgc:imp}).}
    This property asserts that if \code{x1} and \code{x2} are equal, then
    \code{x1} is equal to the specification variable \code{sx}.
\end{itemize}

\subsubsection{Faulty Matrix Vector Product}
\begin{figure}
\begin{wraplst}
\begin{lstlisting}
const real M = ...; ~\label{cgs:m}~

property_r sqr_lt(vector<real> v, uint i) : ~\label{cgs:sqrlt}~
  ((v<r>[i<r>] - v<o>[i<o>]) * (v<r>[i<r>] - v<o>[i<o>])) < M;

for (uint i = 0; i < N; ++i) ~\label{cgs:outer-for}~
{

  q[i] = 0;
  @label(inner_err)
  for (int j = 0; j < N; ++j) ~\label{cgs:inner-for}~
      invariant_r (!model.upset && eq(p)) -> q<r>[i<r>] == q<o>[i<o>] ~\label{cgs:eq}~
  {
    real tmp = A[i][j] *. p[j];
    q[i] = q[i] +. tmp;

    assert_r((!inner_err[model.upset] && eq(p)) -> sqr_lt(q, i)); ~\label{cgs:assert}~
  }
}
\end{lstlisting}
\end{wraplst}
\caption{SS-CG Faulty Matrix Vector Product}
\label{fig:cg-small}
\end{figure}

The SS-CG faulty matrix vector product I present in \autoref{fig:cg-small}
operates over the following pre-existing variables:
\begin{itemize}
  \item \textbf{\code{A}.}
    \code{A} is a matrix of coefficients.
  \item \textbf{\code{p} and \code{q}}.
    \code{p} and \code{q} are vectors of loop carried state.
\end{itemize}

The structure of the SS-CG faulty matrix vector product is as follows:
\begin{itemize}
  \item \textbf{Outer for loop (\autoref{cgs:outer-for}).}
    The outer for loop computes \autoref{eqn:faulty-matvec} element-wise over
    \code{q}.
  \item \textbf{Inner for loop (\autoref{cgs:inner-for}).}
    The inner for loop computes the unreliable matrix vector product
    \(
      \code{q} = \code{A * p}.
    \)
\end{itemize}

\paragraph{\bf Constants and Properties}
The SS-CG faulty matrix vector product uses the following constants properties,
found in \autoref{fig:cg-small}:
\begin{itemize}
  \item \textbf{\code{M} (\autoref{cgs:m}).}
    \code{M} represents the maximum square error permissible in a single
    element of \code{q}.
    The developer must set it according to the formula
    \[
      \code{M} < \min_{a \in A}\left(\max_{(i, j)}(a[i][j])^2\right)
    \]
    where \(A\) is the set of \code{A} input matrices the developer expects to
    run our implementation over.
  \item \textbf{\code{sqr\_lt} (\autoref{cgs:sqrlt}).}
    \code{sqr\_lt} takes a vector \code{v}, and index \code{i}, and ensures
    that the square error of \code{v[i]} is strictly less than \code{M}.
    In other words, it mandates that
    \[
      (\code{v<r>[i<r>]} - \code{v<o>[i<o>]})^2 < \code{M}.
    \]
\end{itemize}

\subsection{Specification}
We use Leto's specification abilities to verify the error correction and small
errors properties of SS-CG.

\paragraph{\bf Error Correction}
Using DMR, the correction step corrects \code{r} and \code{q} even in the
presence of errors.
SS-CG enforces this property through the assertions on Lines~\ref{cgc:assert-r}
and \ref{cgc:assert-q} of \autoref{fig:cg-correct}.
As the algorithm computes \code{spec\_r} and \code{spec\_q} correctly, \TOOL
knows that if \code{r<r> == spec\_r} and \code{q<r> == spec\_q}, then the
system has computed \code{r<r>} and \code{q<r>} correctly.

The \code{dmr\_imp} and \code{dmr\_eq} property applications in the outer while
loop invariants (Lines~\ref{cgc:while-begin-invs} through
\ref{cgc:while-end-invs}) pass the information \TOOL needs to verify this
assertion out of the loop.
Leto infers these invariants for the inner loops, thus allowing the outer loop
to verify that the invariant holds after the modifications the inner loop
performs.

\paragraph{\bf Correctable Errors}
Under SEU, SS-CG requires that if an element of \code{q} is corrupted by
\(\epsilon\), then
\begin{equation}
  \label{eqn:cgs-eps}
  \epsilon^2 < \max_{(i, j)}(A[i][j])^2.
\end{equation}
SS-CG enforces this property through the assertion on \autoref{cgs:assert} of
\autoref{fig:cg-small}.
The invariant on \autoref{cgs:eq} enforces the complementary invariant that if
no error occurred, then \code{q<r>[i<r>]} is equal to \code{q<o>[i<o>]}.
\code{eq(p)} guards both of these constraints because the implementation
verifies this section in isolation without specifying the global properties of
\code{p}.
However, if no upset occurred prior to the start of this section then
\code{eq(p)} trivially holds before and throughout as this snippet does not
modify \code{p}.

\subsection{Verification Approach}
Next, we demonstrate how the developer works with \TOOL to verify that the
implementation meets these specifications.

\subsubsection{Correction Step}
\paragraph{\bf Outer Loop.}
Verification begins with the loop on \autoref{cgc:while} of
\autoref{fig:cg-correct}.
This loop contains the following invariants:
\begin{itemize}
  \item \textbf{\code{dmr_eq(r, r2, spec\_r)}.}
    This invariant enforces that in the absence of errors during a loop
    iteration, \code{r<r>} is equal to \code{spec\_r}.
    That is, if no errors occur then \code{r} is correct.
    This fact is important to pass on so that \TOOL may verify the assertions
    on \cref{cgc:assert-r}.
  \item \textbf{\code{dmr_eq(q, q2, spec\_q)}.}
    This invariant enforces that in the absence of errors during a loop
    iteration, \code{q<r>} is equal to \code{spec\_q}.
    That is, if no errors occur then \code{q} is correct.
    This fact is important to pass on so that \TOOL may verify the assertions
    on \cref{cgc:assert-q}.
  \item \textbf{\code{dmr\_imp(r, r2, spec\_r)}.}
    This invariant states that if the duplicated \code{r} variables are equal
    to each other, then \code{r} is also equal to \code{spec\_r}.
    Combining this with the loop condition, \TOOL knows that \code{r<r> ==
    spec\_r} after exiting the loop.
  \item \textbf{\code{dmr\_imp(q, q2, spec\_q)}.}
    This invariant states that if the duplicated \code{q} variables are equal
    to each other, then \code{q} is also equal to \code{spec\_q}.
    Combining this with the loop condition, \TOOL knows that \code{q<r> ==
    spec\_q} after exiting the loop.
\end{itemize}

\paragraph{\bf Assertions.}
Verification concludes with the assertions on Lines~\ref{cgc:assert-r} and
\ref{cgc:assert-q}:
\begin{itemize}
  \item \textbf{\code{!outer\_while[model.upset] - > (r<r> == spec\_r)}.}
    This assertion verifies that if no upset occurred prior to entering the
    loop containing the outer while loop, then the correction step computed
    \code{r} correctly.
  \item \textbf{\code{!outer\_while[model.upset] - > (q<r> == spec\_q)}.}
    This assertion verifies that if no upset occurred prior to entering the
    loop containing the outer while loop, then the correction step computed
    \code{q} correctly.
\end{itemize}

\subsubsection{Faulty Matrix Vector Product}
\paragraph{\bf Outer Loop}
Verification begins with the loop on \autoref{cgs:outer-for} of
\autoref{fig:cg-small}.
This loop contains no invariants.

\paragraph{\bf Inner Loop.}
Verification then proceeds to the inner loop on \autoref{cgs:inner-for}.
This loop contains the following invariants:
\begin{itemize}
  \item \textbf{\code{(!model.upset \&\& eq(p)) -> q<r>[i<r>] == q<o>[i<o>]}.}
    This invariant ensures that if no upset occurred and \code{p<o> ==
    p<r>}, then \code{q[i]} is equal across both executions.
\end{itemize}

\paragraph{\bf Assertion.}
Verification concludes with the assertion on \autoref{cgs:assert}.
This assertion verifies that if no error had occurred prior to the top of the
inner loop, and \code{p<o> == p<r>}, then the square difference between
\code{q<o>} and \code{q<r>} is less than \code{M}.
This ensures that errors are sufficiently small to be correctable.
That is, it ensures that the error in \code{q} satisfies \autoref{eqn:cgs-eps}.

\section{Self-Stabilizing Steepest Descent Correction Step}
\label{sec:sd}

\begin{figure}
\begin{wraplst}
\begin{lstlisting}
vector<real> correct_sd(int N, matrix<real> A(N, N),
                        vector<real> b(N), vector<real> x(N))
{
  vector<real> zeros(N);
  @noinf for (int i = 0; i < N; ++i)  { zeros[i] = 0; }

  vector<real> r(N), r2(N);
  specvar vector<real> spec_r(N);
  spec_r = r;
  bool run = false;

  @noinf @label(outer) while (run == false || r != r2) ~\label{sd:outer}~
        invariant_r r<r> == r2<r> -> r<r> == spec_r
  {
    model.upset = false; ~\label{sd:set-upset}~
    run = true;

    vector<real> Ax(N), Ax2(N);
    specvar vector<real> spec_Ax(N);
    Ax = Ax2 = spec_Ax = r = r2 = spec_r = zeros;

    @noinf for (int i = 0; i < N; ++i) ~\label{sd:middle}~
        invariant_r upset(r, r2, spec_r, Ax, Ax2, spec_Ax) ~\label{sd:middle-upset}~
    {

      @noinf for (int j = 0; j < N; ++j) ~\label{sd:inner}~
          invariant_r upset(r, r2, spec_r, Ax, Ax2, spec_Ax) ~\label{sd:inner-upset}~
      {

        real tmp = A[i][j] *. x[j];
        real tmp2 = A[i][j] *. x[j];
        specvar real spec_tmp = A[i][j] * x[j];

        Ax[i] = Ax[i] +. tmp;
        Ax2[i] = Ax2[i] +. tmp2;
        spec_Ax[i] = spec_Ax[i] + tmp;
      }

      r[i] = b[i] -. Ax[i];
      r2[i] = b[i] -. Ax2[i];
      spec_r[i] = b[i] - spec_Ax[i];
    }
  }

  assert_r (r<r> == spec_r); ~\label{sd:assert}~

  return r;
}
\end{lstlisting}
\end{wraplst}
\caption{SS-SD Correction Step}
\label{fig:sd}
\end{figure}

\cref{fig:sd} presents an implementation of the correction step from
self-stabilizing steepest descent (SS-SD) \cite{sao2013self}.
SS-SD is an iterative algorithm that computes the solution to a linear system
of equations.
It takes as input a matrix of coefficients \code{A}, a vector \code{b} of
intercepts, and returns an approximate solution vector \code{x} such that
\(\code{A} * \code{x} \approx \code{b}\).
On each iteration, steepest descent uses \(r_i\), \(x_i\), and \(A\) to compute
\(r_{i+1}\) and \(x_{i+1}\) as follows:

\begin{align}
  q_i &= Ar_i \\ \label{eqn:sd-begin}
  \alpha_i &= \frac{r_i^T r_i}{r_i^Tq_i} \\
  x_{i+1} &= x_i + \alpha_i r_i \\
  r_{i+1} &= r_i - \alpha_i q_i \label{eqn:sd-end}
\end{align}

SS-SD adds a periodic correction step to repair the residual \(r\) of any
errors it may have incurred.
This repair step computes
\begin{equation}
  r_i = b - Ax. \label{eqn:sd-correct}
\end{equation}
Unlike \autoref{eqn:sd-begin} through \autoref{eqn:sd-end}, SS-SD requires
that \autoref{eqn:sd-correct} is performed reliably.
Therefore, the implementation verifies that it is possible to correct errors
under the SEU execution model from \autoref{fig:seu} even when the error
correction step may experience errors.
To accomplish this, we use dual modular redundancy (DMR) to duplicate arithmetic
instructions and repeat the correction step until the result of both sets of
instructions agree with each other.

\subsection{SS-SD Correction Step Implementation}
The overall structure of the SS-SD correction step is as follows:
\begin{itemize}
  \item{\bf Initialization.}
    The \code{correct_sd} function takes a matrix of coefficients \code{A}, a
    vector of intercepts \code{b}, and a solution vector \code{x}.
    It then declares:
    \begin{itemize}
      \item \code{\bf r.} \code{r} holds the residual that will be returned.
      \item\code{\bf r2.} The function computes \code{r2} according to
        \autoref{eqn:sd-correct}.
        The algorithm uses \code{r2} to verify that it has correctly computed
        \code{r}.
      \item
        \code{\bf spec_r.} \code{spec_r}  is a specification variable that
        also computes \autoref{eqn:sd-correct}.
        Unlike \code{r} and \code{r2}, the implementation computes
        \code{spec_r} correctly.
    \end{itemize}
  \item{\bf Outer while loop (\autoref{sd:outer}).}
    The outer while loop repeats the correction step until \code{r~==~r2}.
    This loop also sets \code{model.upset} to false (\cref{sd:set-upset}) to
    permit at most one fault per correction iteration.
  \item{\bf Middle for loop (\autoref{sd:middle}).}
    The middle for loop computes \autoref{eqn:sd-correct} element-wise for
    \code{r}, \code{r2}, and \code{spec_r}.
    The algorithm permits errors in the computation of \code{r} and \code{r2}
    but not \code{spec_r}.
  \item{\bf Inner for loop (\autoref{sd:inner}).}
    The inner for loop computes the matrix vector product
    \(
      \code{Ax} = \code{A} * \code{x}.
    \)
    It computes \code{Ax2} and \code{spec_Ax} similarly.
    The algorithm permits errors in the computation of \code{Ax} and
    \code{Ax2}, but not in \code{spec_Ax}.
\end{itemize}

\begin{figure}
\begin{wraplst}
\begin{lstlisting}
property_r upset(vector<real> r, vector<real> r2, vector<real> spec_r, ~\label{sd:prop-upset}~
                 vector<real> Ax, vector<real> Ax2, vector<real> spec_Ax):
  (model.upset ->
    ((r<r> == spec_r && Ax<r> == spec_Ax) ||
     (r2<r> == spec_r && Ax2<r> == spec_Ax))) &&
  (!model.upset ->
    (r<r> == spec_r && r2<r> == spec_r &&
     Ax<r> == spec_Ax && Ax2<r> == spec_Ax));
\end{lstlisting}
\end{wraplst}
\caption{SS-SD Properties}
\label{fig:sd-props}
\end{figure}

\paragraph{\bf Properties.}
The SS-SD correction step implementation uses the following properties, found in
\autoref{fig:sd-props}:
\begin{itemize}
  \item \textbf{\code{upset} (\autoref{sd:prop-upset}).}
    This property consists of two conjuncts:
    \begin{itemize}
      \item The first conjunct states that if there was an upset since the
        start of the outer loop then at least one of the duplicated
        computations is correct.
        That is, \code{r<r> == spec_r \&\& Ax<r> == spec_Ax} or \code{r2<r> ==
        spec_r \&\& Ax2<r>~==~spec_Ax}.
      \item The second conjunct states that if no upset has occurred since the
        toop of the outer loop then both sets of duplicated instructions are
        correct.
    \end{itemize}
\end{itemize}

\subsection{Specification}
Leto's specification abilities verify the error correction property of SS-SD's
correction step.

\paragraph{\bf Error Correction.}
Using DMR, the correction step corrects \code{r} even in the presence of
errors.
SS-SD enforces this property through the assertion on \autoref{sd:assert}.
As the algorithm computes \code{spec_r} correctly, if \code{r<r> ==
spec_r}, then \code{r<r>} is the correct residual.

The invariant on the outer while loop passes the information needed to verify
this assertion out of the loop.
In turn, this invariant relies on the \code{upset} invariant
in the middle loop (\cref{sd:middle-upset}),
which iteself relies on the \code{upset} invariant in the inner loop
(\cref{sd:inner-upset}).

\subsection{\bf Verification Approach}
Next, we demonstrate how the developer works with Leto to verify that the
implementation meets these specifications.

\paragraph{\bf Outer Loop.}
Verification begins with the loop on \autoref{sd:outer}.
This loop contains the following invariants:
\begin{itemize}
  \item \textbf{\code{r<r> == r2<r> -> r<r> == spec_r}.}
    This invariant states that if the duplicated \code{r} variables are equal
    to each other, then \code{r} is also equal to \code{spec_r}.
    Therefore, \TOOL knows that \code{r<r>~==~spec_r} after exiting the loop.
\end{itemize}

\paragraph{\bf Middle Loop.}
Verification then proceeds to the middle loop on \autoref{sd:middle}.
This loop contains the following invariants:
\begin{itemize}
  \item \textbf{\code{upset(r, r2, spec_r, Ax, Ax2, spec_Ax)}.}
    This invariant enforces that in the event of an error during this outer
    loop iteration, the algorithm computes at least one of set of instructions
    (\code{r} and \code{Ax} or \code{r2} and \code{Ax2}) correctly.
    Otherwise, the algorithm computes both sets correctly.
    It is an invariant in the middle loop because it must hold at the top of
    the inner loop.
\end{itemize}

\paragraph{\bf Inner Loop.}
Verification then proceeds to the inner loop on \autoref{sd:inner}.
This loop contains the following invariants:
\begin{itemize}
  \item \textbf{\code{upset(r, r2, spec_r, Ax, Ax2, spec_Ax)}.}
    This invariant enforces that in the event of an error during this outer
    loop iteration, the algorithm computes at least one of set of instructions
    (\code{r} and \code{Ax} or \code{r2} and \code{Ax2}) correctly.
    Otherwise, the algorithm computes both sets correctly.
    This invariant captures information about the impacts of errors on
    variables  modified in the inner loop and passes this information
    back to the middle loop.
\end{itemize}

\paragraph{\bf Assertion.}
Verification concludes with the assertion on \autoref{sd:assert}.
This assertion verifies that \code{r<r>~==~spec_r}, which enforces that
\code{r<r>} is the correct residual and it satisfies \autoref{eqn:sd-correct}.

\section{Full Self-stabilizing Conjugate Gradient Descent Implementation}
\label{sec:cg-full}

We present our full implementation for self-stabilizing conjugate gradient
descent below.
We use the \code{@noinf} annotation regularly in this benchmark on loops that
do not contribute to the final verification restult to increase performance.

\begin{lstlisting}[xleftmargin=0.85cm, xrightmargin=0.5em]
const real SQR_MIN_MAX_AIJ = 2;

property_r sqr_lt(vector<real> v, uint i) :
  ((q<r>[i<r>] - q<o>[i<o>]) * (q<r>[i<r>] - q<o>[i<o>])) <
      SQR_MIN_MAX_AIJ;

property_r dmr_eq(vector<real> x1,
                  vector<real> x2,
                  vector<real> spec_x) :
  !model.upset -> x1<r> == spec_x && x2<r> == spec_x;

property_r dmr_imp(vector<real> x1,
                   vector<real> x2,
                   vector<real> spec_x) :
  (x1<r> == x2<r>) -> (x1<r> == spec_x);

r_requires eq(N) && eq(M) && eq(F) && eq(A)
vector<real> ss_cg(uint N,
                   uint M,
                   uint F,
                   matrix<real> A(N, N),
                   vector<real> b(N),
                   vector<real> x(N))
{
  vector<real> r(N), p(N), q(N), next_x(N), next_r(N), next_p(N)
  real alpha, beta, tmp, tmp2, num, denom;
  uint man_mod;

  vector<real> zeros(N);
  @noinf for (uint i = 0; i < N; ++i) { zeros[i] = 0 };

  uint it = 0;

  @noinf for (uint i = 0; i < N; ++i)
  {
    tmp = 0;
    @noinf for (uint j = 0 ; j < N; ++j)
    {
      tmp = tmp + A[i][j] * x[i];
    }
    r[i] = b[i] - tmp;
  }
  p = r;

  @noinf @label(outer_while)
  while (it < M)
        invariant_r eq(A) && eq(it) && eq(M) &&
                    eq(N) && eq(man_mod) && eq(F)
  {
    if (man_mod == F)
    {
      vector<real> r2(N), q2(N);
      specvar vector<real> spec_r(N), spec_q(N);

      r = r2 = spec_r = q = q2 = spec_q = zeros;
      bool not_run = true;
      @noinf while (not_run || r != r2 || q != q2)
                   invariant_r dmr_eq(r, r2, spec_r)
                   invariant_r dmr_eq(q, q2, spec_q)
                   invariant_r dmr_imp(r, r2, spec_r)
                   invariant_r dmr_imp(q, q2, spec_q)
      {
        not_run = false;

        for (uint i = 0; i < N; ++i)
        {
          for (uint j = 0; j < N; ++j)
          {
            tmp = A[i][j] *. x[j];
            tmp2 = A[i][j] *. x[j];
            specvar real spec_tmp = A[i][j] * x[j];
            r[i] = r[i] +. tmp;
            r2[i] = r2[i] +. tmp2;
            spec_r[i] = spec_r[i] + spec_tmp;

            tmp = A[i][j] *. p[j];
            tmp2 = A[i][j] *. p[j];
            spec_tmp = A[i][j] * p[j];
            q[i] = q[i] +. tmp;
            q2[i] = q2[i] +. tmp2;
            spec_q[i] = spec_q[i] + spec_tmp;
          }
        }
      }

      assert_r(!outer_while[model.upset] -> r<r> == spec_r);
      assert_r(!outer_while[model.upset] -> q<r> == spec_q);

      @noinf for (uint i = 0; i < N; ++i) { r[i] = b[i] - r[i]; }

      num = 0;
      denom = 0;
      @noinf for (uint i = 0; i < N; ++i)
      {
        tmp = r[i] * p[i];
        num = num + tmp;
        tmp = p[i] * q[i];
        denom = denom + tmp;
      }
      alpha = num / denom;

      @noinf for (i = 0; i  < N; ++i)
      {
        tmp = alpha * p[i];
        next_x[i] = x[i] + tmp;
        tmp = alpha * q[i];
        next_r[i] = r[i] - tmp;
      }

      num = 0;
      denom = 0;
      @noinf for (uint i = 0; i < N; ++i)
      {
        tmp = -next_r[i];
        tmp = tmp * q[i];
        num = num + tmp;

        tmp = p[i] * q[i];
        denom = denom + tmp;
      }
      beta = num / denom;

      @noinf for (i = 0; i < N; ++i)
      {
        tmp = beta * p[i];
        next_p[i] = next_r[i] + tmp;
      }
    } else {
      for (uint i = 0; i < N; ++i)
      {
        q[i] = 0;
        @label(inner_err)
        for (uint j = 0; j < N; ++j)
            invariant_r (model.upset == false && eq(p)) ->
                q<r>[i<r>] == q<o>[i<o>]
        {
          tmp = A[i][j] *. p[j];
          q[i] = q[i] +. tmp;

          assert_r((!inner_err[model.upset] && eq(p)) -> sqr_lt(q, i));
        }
      }

      num = 0;
      denom = 0;
      @noinf for (uint i = 0; i < N; ++i)
      {
        tmp = r[i] * r[i];
        num = num + tmp;
        tmp = p[i] * q[i];
        denom = denom + tmp;
      }
      alpha = num / denom;

      @noinf for (i = 0; i  < N; ++i)
      {
        tmp = alpha * p[i];
        next_x[i] = x[i] + tmp;
        tmp = alpha * q[i];
        next_r[i] = r[i] - tmp;
      }

      num = 0;
      denom = 0;
      @noinf for (uint i = 0; i < N; ++i)
      {
        tmp = next_r[i] * next_r[i];
        num = num + tmp;
        tmp = r[i] * r[i];
        denom = denom + tmp;
      }
      beta = num / denom;

      @noinf for (i = 0; i < N; ++i)
      {
        tmp = beta * p[i];
        next_p[i] = next_r[i] + tmp;
      }
    }
    ++it;

    p = next_p;
    x = next_x;
    r = next_r;


    ++man_mod;

    if (man_mod == M)
    {
      man_mod = 0;
    }
  }
  return x;
}
\end{lstlisting}

\end{document}